\documentclass[a4paper,fleqn,usenatbib]{mnras}
\usepackage{newtxtext,newtxmath}

\usepackage[T1]{fontenc}
\usepackage{ae,aecompl}

\usepackage{graphicx}	
\usepackage{amsmath}	
\usepackage{amssymb}	
\usepackage{subfig}
\usepackage{fixltx2e}

\newcommand{\MgII}{\ion{Mg}{ii}}
\newcommand{\MgI}{\ion{Mg}{i}}
\newcommand{\CIV}{\ion{C}{iv}}
\newcommand{\OII}{\ion{O}{ii}}
\newcommand{\CaII}{\ion{Ca}{ii}}
\newcommand{\NII}{\ion{N}{ii}}
\newcommand{\SII}{\ion{S}{ii}}
\newcommand{\FeII}{\ion{Fe}{ii}}
\newcommand{\OIII}{\ion{O}{iii}}
\newcommand{\HI}{\ion{H}{i}}
\newcommand{\II}{\ion{}{ii}}

\title[MUSE-ALMA Halos]{MUSE-ALMA Halos V: Physical properties and environment of $z \leq 1.4$ HI quasar absorbers}

\author[A. Hamanowicz et al.]{Aleksandra Hamanowicz,$^{1}$\thanks{E-mail: ahamanow@eso.org}
C\'eline P\'eroux,$^{1,2}$ Martin A. Zwaan,$^{1}$ Hadi Rahmani,$^{3}$\newauthor Max Pettini,$^{5}$ Donald G. York,$^{7}$ Anne Klitsch$^{1,4}$, Ramona Augustin,$^{2}$  
\newauthor Jens-Kristian Krogager,$^{9}$  Varsha Kulkarni,$^{6}$ Alejandra Fresco,$^{8}$ Andrew D. Biggs,$^{1}$, 
\newauthor Bruno Milliard,$^{2}$  Jo\"el D.R. Vernet$^{1}$
\\
$^{1}$ European Southern Observatory, Karl-Schwarzschild-Str. 2, 85748 Garching near Munich, Germany\\
$^{2}$ Space Telescope Science Institute, 3700 San Martin Drive, Baltimore, MD 21218, USA\\
$^{3}$ GEPI, Observatoire de Paris, PSL Research University, CNRS, Place Jules Janssen, F-92190 Meudon, France \\
$^{4}$ Department of Physics, Centre for Extragalactic Astronomy, Durham University, South Road, Durham DH1 3LE, UK\\
$^{5}$ Institute of Astronomy, University of Cambridge, Madingley Road, Cambridge CB3 0HA, UK \\
$^{6}$ Department of Physics and Astronomy, University of South Carolina, Columbia, SC 29208, USA \\
$^{7}$ Department of Astronomy and Astrophysics, The Enrico Fermi Institute, University of Chicago, 5640 S. Ellis Ave, Chicago, IL 60637, USA\\
$^{8}$ Max-Planck-Institut f\"ur extraterrestrische Physik (MPE), Giessenbachstrasse 1, D-85748 Garching bei Müchen, Germany \\
$^{9}$ Institut d'Astrophysique de Paris, CNRS-SU, UMR7095, 98bis bd Arago, F-75014 Paris, France
}

\date{Accepted XXX. Received YYY; in original form ZZZ}

\pubyear{2019}

\begin{document}
\label{firstpage}
\pagerange{\pageref{firstpage}--\pageref{lastpage}}
\maketitle
\begin{abstract}
 We present results of the MUSE-ALMA Halos, an ongoing study of the Circum-Galactic Medium (CGM) of low redshift galaxies ($z \leq 1.4$), currently comprising 14 strong \HI\ absorbers in five quasar fields. We detect 43 galaxies associated with absorbers down to star formation rate (SFR) limits of 0.01-0.1 M$_{\odot}$yr$^{-1}$, found within impact parameters ($b$) of 250 kpc from the quasar sightline. Excluding the targeted absorbers, we report a high detection rate of 89 per cent and find that most absorption systems are associated with pairs or groups of galaxies (three to eleven members). We note that galaxies with the smallest impact parameters are not necessarily the closest to the absorbing gas in velocity space. Using a multi-wavelength dataset (UVES/HIRES, HST, MUSE), we combine metal and \HI\ column densities, allowing for derivation of the lower limits of neutral gas metallicity as well as emission line diagnostics (SFR, metallicities) of the ionised gas in the galaxies. We find that groups of associated galaxies follow the canonical relations of N(\HI) -- $b$ and W$_{r}$(2796) -- $b$, defining a region in parameter space below which no absorbers are detected. The metallicity of the ISM of associated galaxies, when measured, is higher than the metallicity limits of the absorber. In summary, our findings suggest that the physical properties of the CGM of complex group environments would benefit from associating the kinematics of individual absorbing components with each galaxy member.
\end{abstract}
\begin{keywords}
galaxies: abundances -- galaxies: halos -- galaxies: absorption lines -- intergalactic medium
\end{keywords}
\section{Introduction}
 Galaxies grow, evolve and sustain their star formation thanks to the accretion of gaseous material from filaments of the cosmic web \mbox{\citep{rubin, martin}}. In turn, strong outflows from supernovae or AGN feedback expel enriched material out of the galaxies \mbox{\citep{shull}}, some of which comes back in the form of galactic fountains \mbox{\citep{fraternali}}, while some is carried away from the system by galactic winds. Most of the gas recycling and inflow happens within a few hundreds of kiloparsecs from the galaxy centre in a region dubbed the Circum-Galactic Medium (CGM). These extended gaseous halos surrounding galaxies are transition zones where inflowing Inter-Galactic Medium (IGM) material meets metal-enriched gas expelled from galaxies \mbox{\citep{tumlinson}}. 

Investigating the physical processes taking place within the CGM is key to understanding galactic evolution. However, studies of these extended regions in emission are challenging due to the low surface brightness of the gas \mbox{\citep{frank, corlies, augustin19}}. So far emission from the CGM has only been detected around extreme ionization sources, like luminous quasars, in the form of gigantic filamentary Ly$\alpha$ nebulae \citep{cantalupoNature,borisova, fabMUSE,lusso, umehata}.
and using deep MUSE exposures, around faint galaxies at redshift $z = 3-6$ \citep{wisotzki}. To study the CGM around typical star-forming galaxies, we must rely on distant bright point-like light sources such as quasars or Gamma Ray Bursts whose light gets attenuated by the material around a foreground galaxy \citep{bouche, peroux11, LisaGrb}. Various absorption lines, from the most abundant Ly$\alpha$, to numerous metal lines like \MgII, \FeII\ or \CIV, originating in the intervening systems can reveal information about kinematics and metallicities of the absorbing medium. Metals can, in principle, be used as tracers of gas flows in the CGM. \citet{lehner}, \citet{quiret} and more recently \citet{wotta}, observed a bimodality in the metallicity distribution of the low redshfit quasar absorbers, possibly a trace of metal-poor inflows and metal-rich outflows. 

Understanding the physical properties of the CGM is at present best constrained observationally by the identification of the host galaxies linked to the gas observed in absorption. Historically, the first associated galaxy candidates have been identified on wide-field images of the quasar field as the closest object to the quasar sight-line \citep{bergeron, berbo, Steidel, lebrun}. By using long-slit spectroscopy it was then possible to confirm the redshifts of these objects \citep{fynbo10, fynbo13, rahmani16}. Supplementary to the wide-field search was the method of using a narrow-band optical filter (about 10 \AA) centred around the expected [\OII] emission at the absorbers redshift \citep{yanny87, yanny89,yanny90, moller93}. Another detection method involved surveying the quasar field with a X-Shooter slit triangulation in search for associated galaxies \citep{moller04, krogager17}. Narrow filter integral field searches, using an array of optic fibers, for extended emission near QSOs with absorbers were also attempted \citep{yanny}.

Integral-field spectroscopy with instruments like SINFONI \citep{sinfoni}, MUSE \citep{muse} or OSIRIS \citep{osiris}, make it possible to obtain spectra of many sources in the field-of-view and efficiently classify galaxies associated with absorbers. Combined with high-resolution UV spectroscopy of quasars, these studies have revealed relevant features interpreted as galactic outflows \citep{schroetter, rahmani18wind, schro19}, warped accretion discs \citep{rahmani18disk} and galactic fountains \citep{bouche16}.\\
Alongside the detection of kinematic traces of gas inflows and outflows, new studies detected multiple galaxies associated with one absorber \citep{bielby17,peroux17, klitsch, nielsen18, peroux19, muzahid19}. These findings point towards a more complex view of the gaseous halos of galaxies: from the halo gas of single systems, through tidal streams and halo substructures \citep{whiting,kacprzak10} to intra-group gas \citep{bielby17, fossati}. All these studies have however focused on single absorber systems or on single-species absorbers such as OVI \citep[eg.][]{bielby19} or \MgII\ \citep{zabl19}, lacking the information about the \HI\, which is crucial for metallicity estimates. 
At higher redshifts, some damped Ly$\alpha$ absorbers (DLAs, with log[N(\HI)/cm$^{-2}$] $\geq $20.3) and Lyman Limit Systems, (LLS, log[N(\HI)/cm$^{-2}$] $\leq $19.0) have been found to be associated with multiple Ly$\alpha$ Emitters \citep{mackenzie, magg}.
In this work, we undertake a statistical analysis of HI-selected absorbers. We present a MUSE study of five quasar fields with 14 \HI\ absorbers at $z \leqslant$ 1.4, to perform a statistical study of the properties of the absorbers and their associated galaxies, including their metallicities. Each of these fields also have associated ALMA observations, targeted to detect CO(2-1) and CO(3-2) emission lines from galaxies associated to \HI\ rich quasar absorbers. Measuring the molecular content of these systems makes it possible to constrain the physical properties and kinematics of molecular gas, known to be the fuel of star formation (forthcoming publication). 

This paper is organized as follows: in Section 2 we describe the data and the reduction process, in Section 3 we present the analysis details and the results. In section 4 we discuss our findings comparing them to previous studies, and we summarize the study in Section 5. Images and spectra of all absorbers and associated galaxies can be found in the Appendix. \\
We adopt the following cosmology: $H_{0}$ = 70 km s$^{-1}$ Mpc$^{-1}$, $\Omega_{\rm M}$ = 0.3, $\Omega_{\rm \Lambda}$  = 0.7.
\begin{figure*}
\subfloat{\includegraphics[width=\columnwidth]{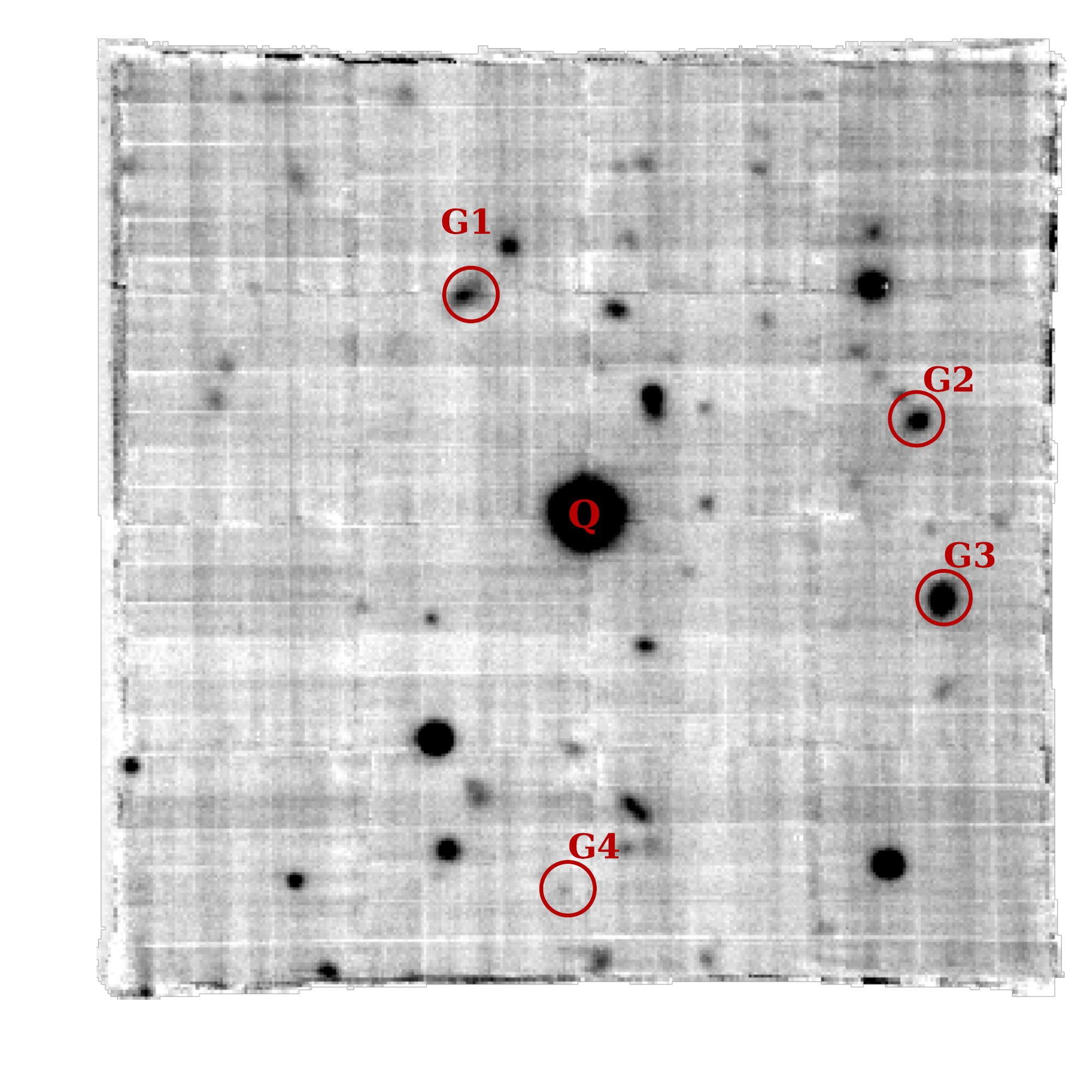}}
\subfloat{\includegraphics[width=\columnwidth]{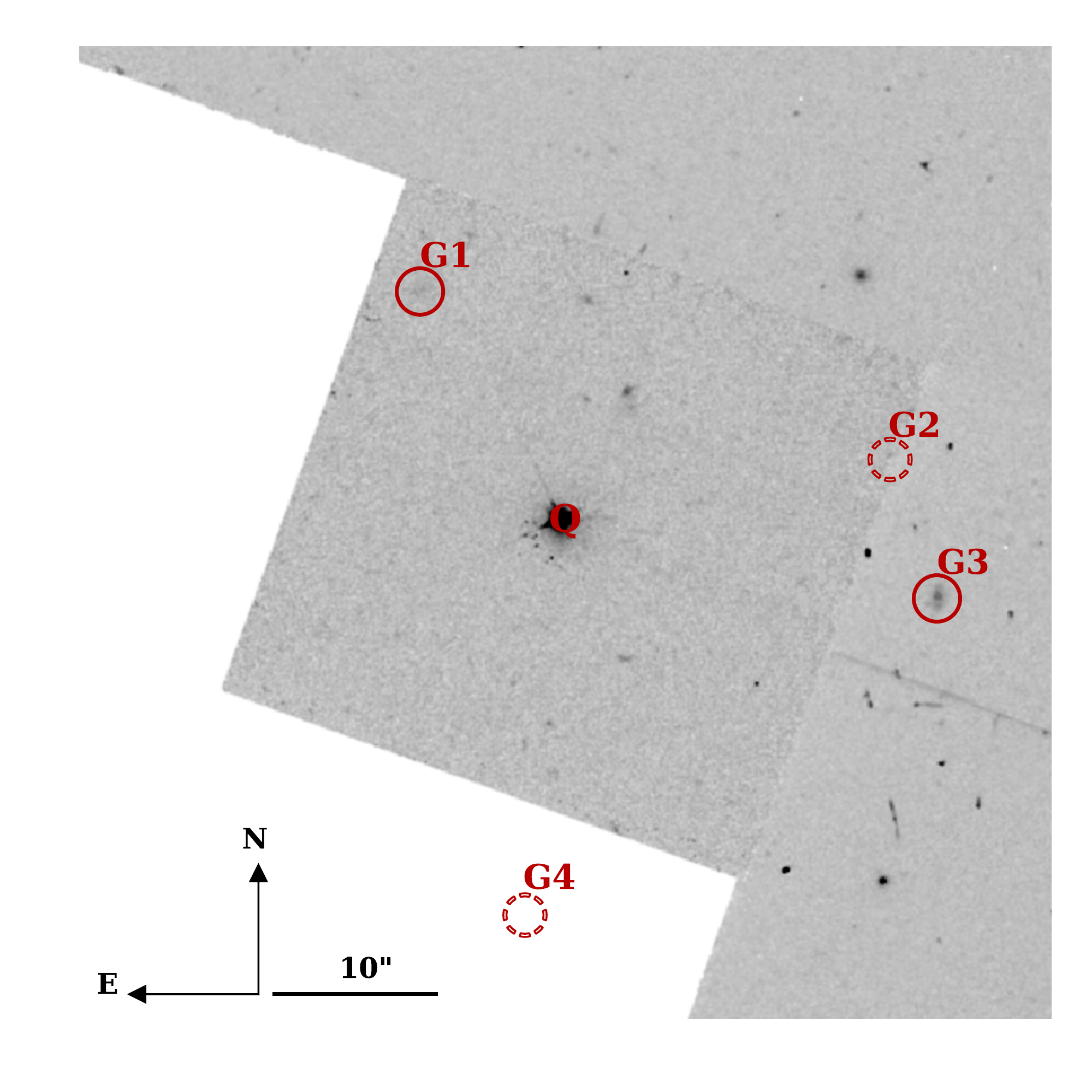}}
\caption{MUSE white light (left) and HST F702W (right) images of the Q1232z083 absorber field. Galaxies associated with this absorber are marked with red circles. Non-detections are marked with dashed circles: both galaxies G4 and G2 are undetected in the HST image, G4 because it falls outside the field of view of the WFPC2 camera, and G2 because it is at the edge between two camera chips. The MUSE and the HST images in the figure are both centred around the background Q1232-0224 (zQSO=1.05). This figure illustrates the power of IFU observations to identify galaxies associated with absorbers. Images of the remaining fields from the sample with galaxies associated with the absorbers marked on them are presented in the Appendix.}
\label{fig:example}
\end{figure*}

\section{MUSE Observations and ancillary data}
The MUSE-ALMA Halos sample of quasar absorbers is based on five quasar observed with MUSE and archival high-resolution UV-spectroscopy: Q0152-2001, Q1130-1449, Q1211+1030, Q1229-0207, Q2131-1207; quasars coordinates and redshifts are summarized in Table \ref{tab:quasars}. Through the paper we refer to these quasars with their first four right ascension numbers (e.g. Q1130). Similarly, we adopt the naming scheme for the absorbers in our sample: first four right ascension numbers + z + the redshift of the absorber (up to second decimal point (e.g.Q1229z076). 

The MUSE quasar fields were selected (proposal 096.A-0303(A), PI: P\'eroux) for the presence of known \HI\ absorbers, with column density $\log [N(\HI)/{\rm cm}^{-2}] > 18.5$ at redshift z $\sim$ 0.4, for which MUSE covers all prominent galactic emission lines: H$\beta$, [\OII], [\OIII], H$\alpha$, [\NII], [\SII]. Furthermore, for all selected quasars, at least one galaxy was known to be associated with this \HI\ absorber, identified with photometry or low-resolution spectroscopy (Q0152 - \citet{rao}, Q1130 - \citet{kacprzak11}, Q1211 - \citet{kacprzak11}, Q1232 - \citet{lebrun}, Q2131 - \citet{kacprzak11}). 

\begin{table}

\caption{Parameters of the quasars in MUSE-ALMA Halos sample: (1) reference name used in this paper (2) full name of the quasar, (3) right ascension, (4) declination, (5) quasar redshift}
\begin{tabular}{ c  c  c c c }
\hline
QSO name & alternative name &  RA & DEC  & z  \\
 & & [hh:mm:ss] & [dd:mm:ss] & \\
\hline\hline
Q0152-2001 & UM 657 & 01:52:27 & -20:01:07.1 & 2.06\\
Q1130-1449 & B1127-145 & 11:30:07 & -14:49:27.7 & 1.19 \\
Q1211+1030 & 1209+107 & 12:11:41 & +10:30:02.8 & 2.19\\
Q1232-0224 & 1229-021 & 12:32:00 & -02:24:04.6 &  1.05 \\
Q2131-1207 & Q2128-123 & 21:31:35 &  -12:07:04.8 & 0.43\\
\hline

\end{tabular}
\label{tab:quasars}
\end{table}

\subsection{High resolution UV/optical quasar spectroscopy}
We used UVES/VLT and HIRES/Keck archival high-resolution spectra for all five quasars in our sample to identify absorbers and their multiple metal absorption lines. 
Q0152-2001 and Q2131-1207 were observed with HIRES/Keck at resolution $R$ = 45,000 - 48,000.
The remaining quasar spectra (Q1130-1449, Q1211+1030 and Q1232-0224) we covered by the VLT/UVES spectrograph BLUE arm, with combined exposure times from 4 to 6 hours and spectral resolution up to $R$ = 80,000. The wavelength coverage differs between objects due to different initial observation setups, introducing gaps in the spectra. The data were uniformly reprocessed using the ESO UVES pipeline and individual exposures of each target were merged, weighted by their signal-to-noise-ratios \citep{zafar13}.  

We used also archival HST/FOS spectra covering the \HI\ absorption line for all the absorbers in our sample. We used the already reduced FOS UV spectra from HST archives for Q1232-0224 (PI: Bergeron ID:5351, FOS-G190H, $R$ = 1300, $T_{\rm exp}$ = 4580s) and  Q1130-1449 (PI: Deharveng, ID: 3483, FOS-G160L, $R$ = 250, $T_{\rm exp}$ = 1450s). For the remaining systems, we relied on the measurements of \HI\ column density from the literature (references included in Table \ref{tab:emmiters}). 

We note that the wavelength reference frame differs for MUSE (air) and UVES/HIRES (vacuum). To obtain precise redshift and velocity measurements we shifted the MUSE spectra to a vacuum wavelength reference system.

\subsection{MUSE and HST imaging of five quasar fields}
 The five quasar fields we present in this work were observed with VLT/MUSE IFU in period 96 (proposal 096.A-0303(A), PI: P\'eroux). All observations were conducted in good seeing conditions (< 0.85 arcsec) with an average of 1-2 hours per target for all but field Q1130-1449 \citep{peroux19}, which was significantly deeper (12 $\times$ 1200s). All fields were observed in "nominal mode", resulting in a spectral coverage of 4800 - 9300 \AA.  
 
 All raw MUSE observations of our quasar fields were reduced with the ESO default MUSE reduction pipeline v2.2 \citep{musepipe}. We applied bias, flat and wavelength calibration, and line spread functions as well as illumination correction frames to each individual exposure with \textit{scibasic} command, creating MUSE specific PIXTABLE output. The astrometry solution and the correction for geometry, together with flux calibrations, were all combined within the \textit{scipost} command. Finally, individual exposures were combined including field rotation. The detailed description of the reduction procedure can be found in \citet{peroux19}. We refrained from using the pipeline sky subtraction method, due to unsatisfactory results. Instead, removal of sky emission lines was performed on the final datacube with a Principal Component Analysis (PCA) algorithm \citep{brend}. The code creates PCA component of selected sky regions used further to remove sky residuals.
 
 In addition to MUSE observations we used readily available, reduced archival HST imaging for the galaxy identification: WFPC2 (Wide Field Planetary Camera 2) in filter F702W of quasars Q0152-2001 (PI: Steidel, ID:6557), Q1211+1030, Q1232-0224 (PI: Bergeron, ID:5351) and Q2131-1207 (PI: Maccheto, ID:5143) and WFC3 (Wide Field Camera) IR-F140W of Q1130-1449 (PI: Bielby, ID: 14594), with exposure times from 10 to 50 minutes. An example of MUSE and corresponding HST WFPC2 quasar field images for Q1232z083 is shown in Fig.\ref{fig:example}.
 
\begin{table*}
	\centering
\caption{Parameters of the absorbers and their associated galaxies. Each absorber row is followed by the rows of associated galaxies. Absorber parameters: (1) AbsorberID: first part of the quasar name +  the absorber redshift, selection method in brackets (2) ID of the quasar, (3) redshift of the quasar, (4) redshift of the absorber, (5) column density of \HI, (6) Equivalent width of \MgII $\lambda$2796, (7) absorber [Fe/H] (not dust corrected) (8) Corresponding figures in the appendix (position on the sky and spectra of associated galaxies). Each galaxy parameters row consists of: (1) Galaxy ID, in the order of increasing impact parameter, unless the nomenclature adapted from literature states differently (Q1130z031, Q1130z032), (2) redshift of the galaxy, (3) impact parameter in kpc and arcsec, (4) observed magnitude of the galaxy in the HST F702W filter if the galaxy was detected (unless marked otherwise), (5) star formation rate measured from [\OII] emission line flux (not dust corrected) \citep{kobulnicky}, (6-7) 12 + log(O/H) (not dust corrected): we report both metallicity branches following \citet{kobulnicky}; in case of literature values where only one metallicity was reported we leave the upper metallicity column blank.
Literature references: \textsuperscript{a} \citet{kacprzak11}, \textsuperscript{b} \citet{lane19}, \textsuperscript{c} \citet{muzahid16},  \textsuperscript{d} \citet{rao06}, \textsuperscript{ e} \citet{boise98}, \textsuperscript{ f} \citet{rahmani18wind},\textsuperscript{ g} \citet{peroux19}. \textsuperscript{1} Observed magnitudes in HST F814W from \citet{kacprzak10}}

\begin{tabular}{ c  c  c c  c  c c c c c }
\hline

\multicolumn{1}{c}{\textbf{Absorber ID}}&\multicolumn{1}{c}{\textbf{Quasar ID}} & \multicolumn{1}{c}{$\mathbf{z}_{\rm \textbf{QSO}}$} & \multicolumn{1}{c}{$\mathbf{z}_{\rm \textbf{abs}}$} &
\multicolumn{1}{c}{\textbf{log(N(\HI)/cm$^{\rm -2}$)}} &\multicolumn{1}{c}{\textbf{W$_{\rm r}$(2796) [\AA]} } & \multicolumn{1}{c}{\textbf{[Fe/H]}} & \multicolumn{1}{c}{\textbf{Figure}} \\
\hline
Galaxy &$z_{\rm gal}$ &  b [kpc/arcsec]& HST$_{\rm mag}$ F702W & SFR$_{\rm [OII]}$ [M$_{\odot}$yr$^{-1}$] & 12+log(O/H)$_{\rm lower}$ & 12+log(O/H)$_{\rm upper}$ & \\
\hline
\hline

\multicolumn{1}{c}{\textbf{Q0152z038 (\HI)}}&\multicolumn{1}{c}{\textbf{Q0152-2001}} &\multicolumn{1}{c}{\textbf{2.06}}& \multicolumn{1}{c}{\textbf{0.3887}} & 
\multicolumn{1}{c}{\textbf{< 18.8 \textsuperscript{ f}}} &\multicolumn{1}{c}{\textbf{0.58 $\pm$ 0.05\textsuperscript{ a}}} & \multicolumn{1}{c}{\textbf{> -1.36} }& \multicolumn{1}{c}{\textbf{\ref{fig:Q0152}, -}} \\
\hline

G1 & 0.3826 &  60 / 11.5 & - & 1.04 $\pm$ 0.03 & 8.65 $\pm$ 0.09 & - \\
G2 & 0.3802 &  80 / 15.4 & - & 0.33 $\pm$ 0.01 & 8.70 $\pm$ 0.07 & - \\
G3 & 0.3815 &  86 / 16.5 & - & 0.31 $\pm$ 0.01 & 8.42 $\pm$ 0.06 & - \\
G4 & 0.3814 & 123 / 23.7 & - & 0.12 $\pm$ 0.02 & 8.94 $\pm$ 0.08 & - \\
G5 & 0.3814 & 172 / 33.1 & - & < 0.02	        & -	                    & - \\
G6 & 0.3824 & 205 / 39.4 & - & 0.07 $\pm$ 0.01 & 7.82 $\pm$ 0.13 & - \\

\hline

\multicolumn{1}{c}{\textbf{Q0152z078 (\MgII)}}&\multicolumn{1}{c}{\textbf{Q0152-2001}} & \multicolumn{1}{c}{\textbf{2.06}} & \multicolumn{1}{c}{\textbf{0.7800}} & \multicolumn{1}{c}{\textbf{18.87$_{\rm \textbf{-0.14}}^{\rm \textbf{+0.11}}$\textsuperscript{ d}}} & \multicolumn{1}{c}{\textbf{0.36 $\pm$ 0.04 \textsuperscript{ d}}} & \multicolumn{1}{c}{\textbf{> -0.44 }} & \multicolumn{1}{c}{\textbf{ \ref{fig:Q0152}, - }}\\
\hline
G1 & 0.7803 & 54 / 7.6 & - &  23.00 $\pm$ 5  & 7.90 $\pm$ 0.20 &  8.60 $\pm$ 	0.20 \\
\hline
\hline

\multicolumn{1}{c}{\textbf{Q1130z019 (\MgII)}}&\multicolumn{1}{c}{\textbf{Q1130-1449}} & \multicolumn{1}{c}{\textbf{1.19}}  & \multicolumn{1}{c}{\textbf{0.1906}} & \multicolumn{1}{c}{\textbf{< 19.1}} &\multicolumn{1}{c}{\textbf{0.14 $\pm$ 0.01}}& \multicolumn{1}{c}{\textbf{ -} } & \multicolumn{1}{c}{ \textbf{\ref{fig:Q1130}, \ref{fig:Q1130z019spec} } }\\ 
\hline

G1 & 0.1905 & 17 / 5.3 & 22.26 & - & - & - \\

\hline
\multicolumn{1}{c}{\textbf{Q1130z031 (\HI)}}&\multicolumn{1}{c}{\textbf{Q1130-1449}} & \multicolumn{1}{c}{\textbf{1.19}} & \multicolumn{1}{c}{\textbf{0.3127}} & \multicolumn{1}{c}{\textbf{21.71 $\pm$ 0.07\textsuperscript{b}}}&\multicolumn{1}{c}{\textbf{2.21 $\pm$ 0.12\textsuperscript{d}}} & \multicolumn{1}{c}{\textbf{-1.94 $\pm$ 0.08 \textsuperscript{ g}}}& \multicolumn{1}{c}{\textbf{ \ref{fig:Q1130z031}, -  }} \\ 
\hline

G0  & 0.3131 &  11 /  2.3  & -     & 0.07 $ \pm$ 0.1 & 8.64 $\pm$ 0.14 & - \\
G1  & 0.3121 &  18 /  3.8  & 21.55\textsuperscript{ 1} & 2.80 $\pm$ 0.8  & 8.12 $\pm$ 0.08 & - \\
G2  & 0.3127 &  44 /  9.5  & 18.81\textsuperscript{ 1} & 0.44 $\pm$ 0.3  & 8.77 $\pm$ 0.05 & - \\
G4  & 0.3126 &  82 / 17.7  & 18.64\textsuperscript{ 1} & > 0.40          & < 8.65          & - \\	
G6  & 0.3115 &  98 / 21.3  & 19.79\textsuperscript{ 1} & 1.14 $\pm$ 0.7  & 8.94 $\pm$ 0.16 & - \\
G16 & 0.3133 &  21 /  4.5  & -     & 0.42 $\pm$ 0.1  & 7.96 $\pm$ 0.14 & - \\
G17 & 0.3136 &  35 /  7.6  & 22.50 & 3.18 $\pm$ 0.7  & 8.01 $\pm$ 0.17 & - \\
G18 & 0.3125 &  27 /  5.9  & -     & 0.30 $\pm$ 0.1  & 8.69 $\pm$ 0.18 & - \\
G19 & 0.3119 & 120 / 26.0  & 21.43\textsuperscript{ 1} & 1.09 $\pm$ 0.2  & 8.67 $\pm$ 0.16 & - \\
G20 & 0.3140 &  56 / 12.1  & 21.82\textsuperscript{ 1} & < 0.01          &  -              & - \\	
G21 & 0.3122 & 131 / 28.4  & 24.67\textsuperscript{ 1} & 0.01 $\pm$ 0.01 & 8.76 $\pm$ 0.17 & - \\

\hline
\multicolumn{1}{c}{\textbf{Q1130z032 (\MgII)}}&\multicolumn{1}{c}{\textbf{Q1130-1449}} & \multicolumn{1}{c}{\textbf{1.19}}&  \multicolumn{1}{c}{\textbf{0.3283}} & \multicolumn{1}{c}{ \textbf{< 18.9} } &\multicolumn{1}{c}{ \textbf{0.028 $\pm$ 0.003 \textsuperscript{ a}} } & \multicolumn{1}{c}{ \textbf{-} }& \multicolumn{1}{c}{ \textbf{\ref{fig:Q1130z031}, \ref{fig:Q1130z032spec} }} \\ 
\hline
G3 & 0.3282 & 74 / 15.3 & 20.12\textsuperscript{ 1} & 0.02 $\pm$ 0.01   & 7.52 $\pm$ 0.28 & 8.93 $\pm$ 0.09 \\
G5 & 0.3284	& 87 / 18.2	& 18.84\textsuperscript{ 1} & 0.006 $\pm$ 0.001 & -               & -               \\
\hline
\hline
\multicolumn{1}{c}{\textbf{Q1211z039 (\HI)}}&\multicolumn{1}{c}{\textbf{Q1211+1030}} &  \multicolumn{1}{c}{\textbf{2.19}}& \multicolumn{1}{c}{\textbf{0.3929}} & \multicolumn{1}{c}{\textbf{19.46 $\pm$ 0.08 \textsuperscript{ d}}} &\multicolumn{1}{c}{\textbf{1.19 $\pm$ 0.01 \textsuperscript{ a}}}& \multicolumn{1}{c}{\textbf{> -1.05}}& \multicolumn{1}{c}{ \textbf{\ref{fig:Q1211}, \ref{fig:Q1211z039spec}} } \\
\hline
G1 & 0.3928 &  37 /  6.8 & 21.93 & 4.71 $\pm$ 0.08 & 8.16 $\pm$ 0.01 & 8.48	$\pm$ 0.01 \\
\hline
\multicolumn{1}{c}{\textbf{Q1211z062 (\HI)}}&\multicolumn{1}{c}{\textbf{Q1211+1030}} &  \multicolumn{1}{c}{\textbf{2.19}}& \multicolumn{1}{c}{\textbf{0.6296}} & \multicolumn{1}{c}{\textbf{20.30$_{\rm \textbf{-0.30}}^{\rm \textbf{+0.18}}$ \textsuperscript{ d}}} &  \multicolumn{1}{c}{\textbf{2.92 $\pm$ 0.23 \textsuperscript{ d}}} & \multicolumn{1}{c}{\textbf{-0.98 $\pm$ 0.3 \textsuperscript{ e}}}& \multicolumn{1}{c}{\textbf{ \ref{fig:Q1211}, \ref{fig:Q1211z062spec} (Fig. \ref{fig:qsores})} } \\
\hline

G1 & 0.6283 & 12  /  1.7 & 21.59 & 1.04 $\pm$ 0.14	& - & - \\
G2 & 0.6303 & 132 / 19.1 & 24.53 & 0.59 $\pm$ 0.05  & - & - \\

\hline
\multicolumn{1}{c}{\textbf{Q1211z089 (\MgII)}}&\multicolumn{1}{c}{\textbf{Q1211+1030}} &  \multicolumn{1}{c}{\textbf{2.19}}& \multicolumn{1}{c}{\textbf{0.8999}} & \multicolumn{1}{c}{ \textbf{< 18.50} } & \multicolumn{1}{c}{\textbf{ 0.023 $\pm$ 0.005}} & \multicolumn{1}{c}{ \textbf{-} }& \multicolumn{1}{c}{\textbf{\ref{fig:Q1211}, \ref{fig:Q1211z089spec} }} \\
\hline
G1 & 0.8991 &  79 / 10.0 & 24.65 & 4.33 $\pm$ 0.43 & - & - \\
G2 & 0.8995 & 185 / 23.4 & 23.26 & 2.03 $\pm$ 0.19 & - & - \\
G3 & 0.8991 & 275 / 34.8 & -    & 0.57 $\pm$ 0.07 & - & - \\
G4 & 0.8953 & 149 / 18.9 & 24.37 & 3.29 $\pm$ 0.14 & - & - \\
G5 & 0.8960 & 197 / 24.9 & -    & 1.68 $\pm$ 0.21 & - & - \\
\hline
\multicolumn{1}{c}{\textbf{Q1211z105 (\MgII)}}&\multicolumn{1}{c}{\textbf{Q1211+1030}} &  \multicolumn{1}{c}{\textbf{2.19}}& \multicolumn{1}{c}{\textbf{1.0496}} & \multicolumn{1}{c}{ \textbf{< 18.90} } & \multicolumn{1}{c}{ \textbf{0.18 $\pm$ 0.01}}& \multicolumn{1}{c}{\textbf{> -1.69}}& \multicolumn{1}{c}{ \textbf{--} } \\
\hline 
\multicolumn{7}{c}{No galaxies found to match the absorber} \\
\hline
\hline

\end{tabular}
	\label{tab:emmiters}
\end{table*}

\begin{table*}
	\centering
		\contcaption{}
	\begin{tabular}{ c  c  c c  c  c c c c}
	\hline
\multicolumn{1}{c}{\textbf{Absorber ID}}&\multicolumn{1}{c}{\textbf{Quasar ID}} & \multicolumn{1}{c}{$\mathbf{z}_{\rm \textbf{QSO}}$} & \multicolumn{1}{c}{$\mathbf{z}_{\rm \textbf{abs}}$} &
\multicolumn{1}{c}{\textbf{log(N(\HI)/cm$^{\rm -2}$)}} &\multicolumn{1}{c}{\textbf{W$_{\rm r}$(2796) [\AA]} } & \multicolumn{1}{c}{\textbf{[Fe/H]}} & \multicolumn{1}{c}{\textbf{Figure}} \\
\hline
Galaxy &$z_{\rm gal}$ &  b [kpc/arcsec]& HST$_{\rm mag}$ F702W & SFR [M$_{\odot}$yr$^{-1}$] ( [OII] )& 12+log(O/H)$_{\rm lower}$ & 12+log(O/H)$_{\rm  upper}$\\
\hline
\multicolumn{1}{c}{\textbf{Q1232z039 (\HI)}}&\multicolumn{1}{c}{\textbf{Q1232-0224}} &  \multicolumn{1}{c}{\textbf{1.05}}& \multicolumn{1}{c}{\textbf{0.3950}} & \multicolumn{1}{c}{ \textbf{20.75 $\pm$ 0.07 \textsuperscript{ e}} } & \multicolumn{1}{c}{\textbf{2.894 $\pm$ 0.11}} &\multicolumn{1}{c}{\textbf{ < -1.31 \textsuperscript{ e}}}  & \multicolumn{1}{c}{ \textbf{\ref{fig:Q1229}, (Fig.\ref{fig:qsores})}}\\
\hline

G1 & 0.3953 & 8 / 1.5 & 23.00 & 0.67 $\pm$ 0.09 & 8.02 $\pm$ 0.06 & 8.66 $\pm$ 0.04 \\

\hline

\multicolumn{1}{c}{\textbf{Q1232z075 (\MgII)}}&\multicolumn{1}{c}{\textbf{Q1232-0224}}&  \multicolumn{1}{c}{\textbf{1.05}} & \multicolumn{1}{c}{\textbf{0.7572}} & \multicolumn{1}{c}{\textbf{18.36$_{\rm \textbf{-0.08}}^{\rm \textbf{+0.09}}$\textsuperscript{ d}}} &\multicolumn{1}{c}{\textbf{0.52 $\pm$ 0.07 \textsuperscript{ d}}} & \multicolumn{1}{c}{ \textbf{> -1.48 } } & \multicolumn{1}{c}{ \textbf{\ref{fig:Q1229},\ref{fig:Q1229z075spec}} }\\
\hline
G1 & 0.7566 & 68 / 9.1 & 22.89 & 2.58 $\pm$ 0.23 & 8.54 $\pm$ 0.19 & 8.19 $\pm$ 0.19 \\

\hline
\multicolumn{1}{c}{\textbf{Q1232z076 (\MgII)}}&\multicolumn{1}{c}{\textbf{Q1232-0224}} &  \multicolumn{1}{c}{\textbf{1.05}}& \multicolumn{1}{c}{\textbf{0.7691}} & \multicolumn{1}{c}{\textbf{18.11 $\pm$ 0.15 }} &\multicolumn{1}{c}{\textbf{0.053 $\pm$ 0.001 }} & \multicolumn{1}{c}{\textbf{> -2.34 }} & \multicolumn{1}{c}{\textbf{ \ref{fig:Q1229},\ref{fig:Q1229z076spec}} } \\
\hline
G1 & 0.7666 &  57 /  7.7 & 22.11 & 3.85 $\pm$ 0.14 & - & - \\	
G2 & 0.7685 & 131 / 17.5 & -     & 0.84 $\pm$ 0.07 & 8.06 $\pm$ 0.14 & 8.55 $\pm$ 0.10 \\
G3 & 0.7688 & 170 / 22.7 & 21.62 & <1.41 & - & - \\ 

\hline
\multicolumn{1}{c}{\textbf{Q1232z083 (\MgII)}}&\multicolumn{1}{c}{\textbf{Q1232-0224}} &  \multicolumn{1}{c}{\textbf{1.05}}& \multicolumn{1}{c}{\textbf{0.8311}} & \multicolumn{1}{c}{ \textbf{18.84 $\pm$ 0.10} } &\multicolumn{1}{c}{\textbf{0.238 $\pm$ 0.004}}& \multicolumn{1}{c}{\textbf{> -2.19 }}  & \multicolumn{1}{c}{\textbf{\ref{fig:Q1229}, \ref{fig:Q1229z083spec}}}  \\ 
\hline

G1 & 0.8309 & 122 / 15.8 & 22.73 & 3.10 $\pm$ 0.25 & - & - \\
G2 & 0.8323 & 166 / 21.6 & -     & < 0.70 & - & - \\
G3 & 0.8325 & 176 / 22.8 & 21.84 & 5.42 $\pm$ 0.29 & - & - \\
G4 & 0.8321 & 182 / 23.6 & -     & 0.61 $\pm$ 0.13 & - & - \\

\hline
\hline
\multicolumn{1}{c}{\textbf{Q2131z043 (\HI)}}&\multicolumn{1}{c}{\textbf{Q2131-1207}} &  \multicolumn{1}{c}{\textbf{0.43}}& \multicolumn{1}{c}{\textbf{0.4298}} & \multicolumn{1}{c}{\textbf{19.50 $\pm$ 0.15 \textsuperscript{ c}}} &\multicolumn{1}{c}{\textbf{0.41 $\pm$ 0.01 \textsuperscript{ d}}} & \multicolumn{1}{c}{\textbf{> -0.96}}  & \multicolumn{1}{c}{\textbf{\ref{fig:Q2131}, -}} \\  
\hline
G1 & 0.4301 &  52 /  9.2 & 20.54 & 2.00 $\pm$ 0.2 & 8.98 $\pm$ 0.02 & -\\
G2 & 0.4307 &  61 / 10.7 & -     & 0.20 $\pm$ 0.1 & 8.32 $\pm$ 0.16 & - \\
G3 & 0.4301 & 147 / 26.0 & -     & 0.08 $\pm$ 0.1 & - & -  \\
G4 & 0.4298 & 174 / 30.7 & -     & 0.10 $\pm$ 0.1 & - & - \\

\hline
\hline
\end{tabular}

\end{table*}

\begin{figure*} 
\includegraphics[width=2\columnwidth]{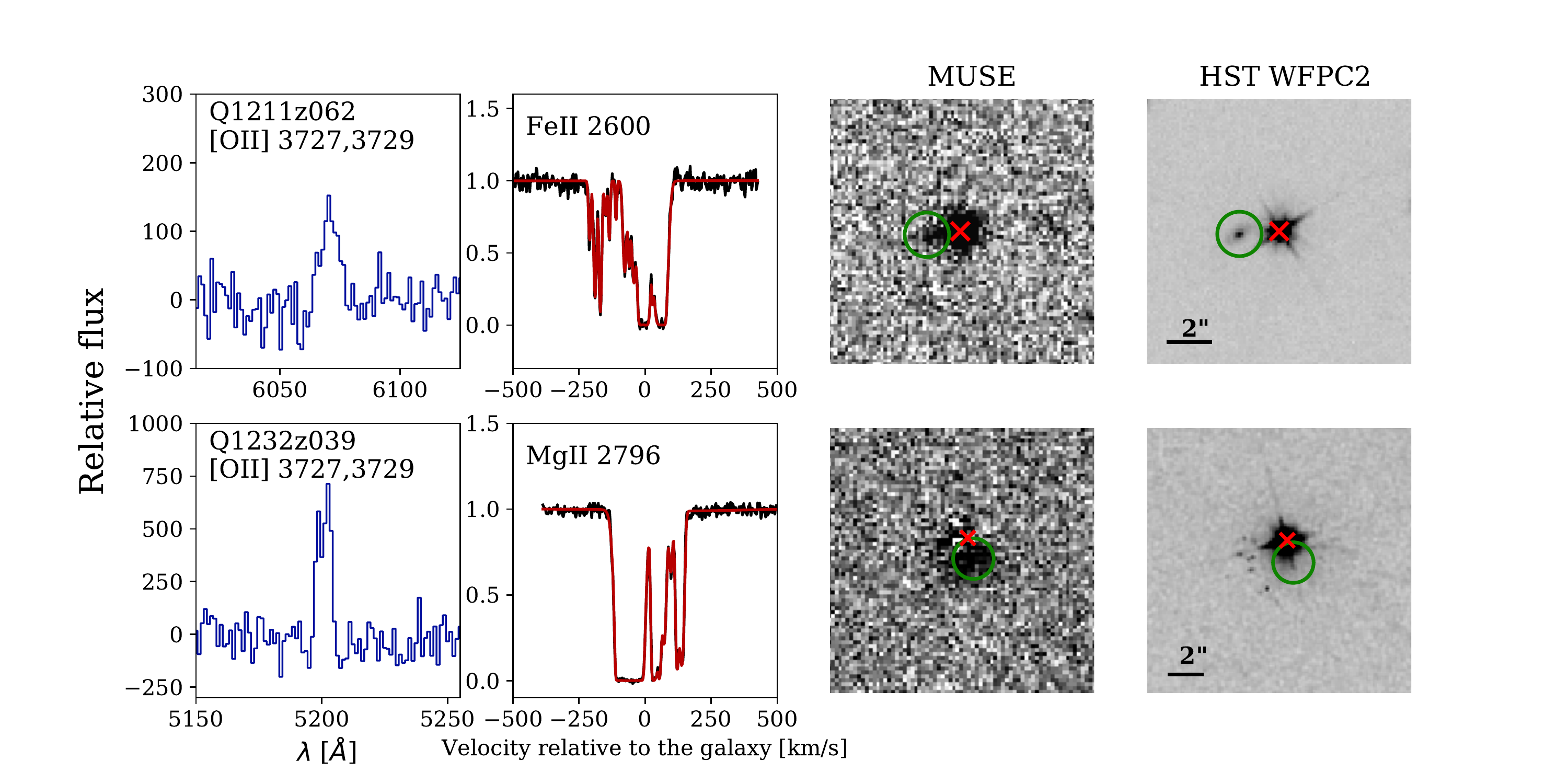}
\caption{Spectral quasar PSF subtraction in the MUSE data. The panels present two galaxies each hidden under the PSF of a different QSO, and their retrieved spectra: Q1211+1030--top and Q1232-0224--bottom. Galaxies were identified by detecting the [\OII] emission line (first panel from the left) at the redshift of the absorber. To remove the quasar signature we subtracted the continuum around the detected line. The second column presents the absorption lines detected in the normalized quasar spectrum in velocity space with zero velocity set to the systematic redshift of the galaxy. The third and fourth columns show the continuum-subtracted zoom-in MUSE image of the galaxy around the [\OII] emission line and the zoom-in of the HST F702W image of the quasar on the same scale. The red cross represents the centre of the quasar and green circle the position of the detected galaxy.}
\label{fig:qsores}
\end{figure*}

\begin{figure} 
 \includegraphics[width=\columnwidth]{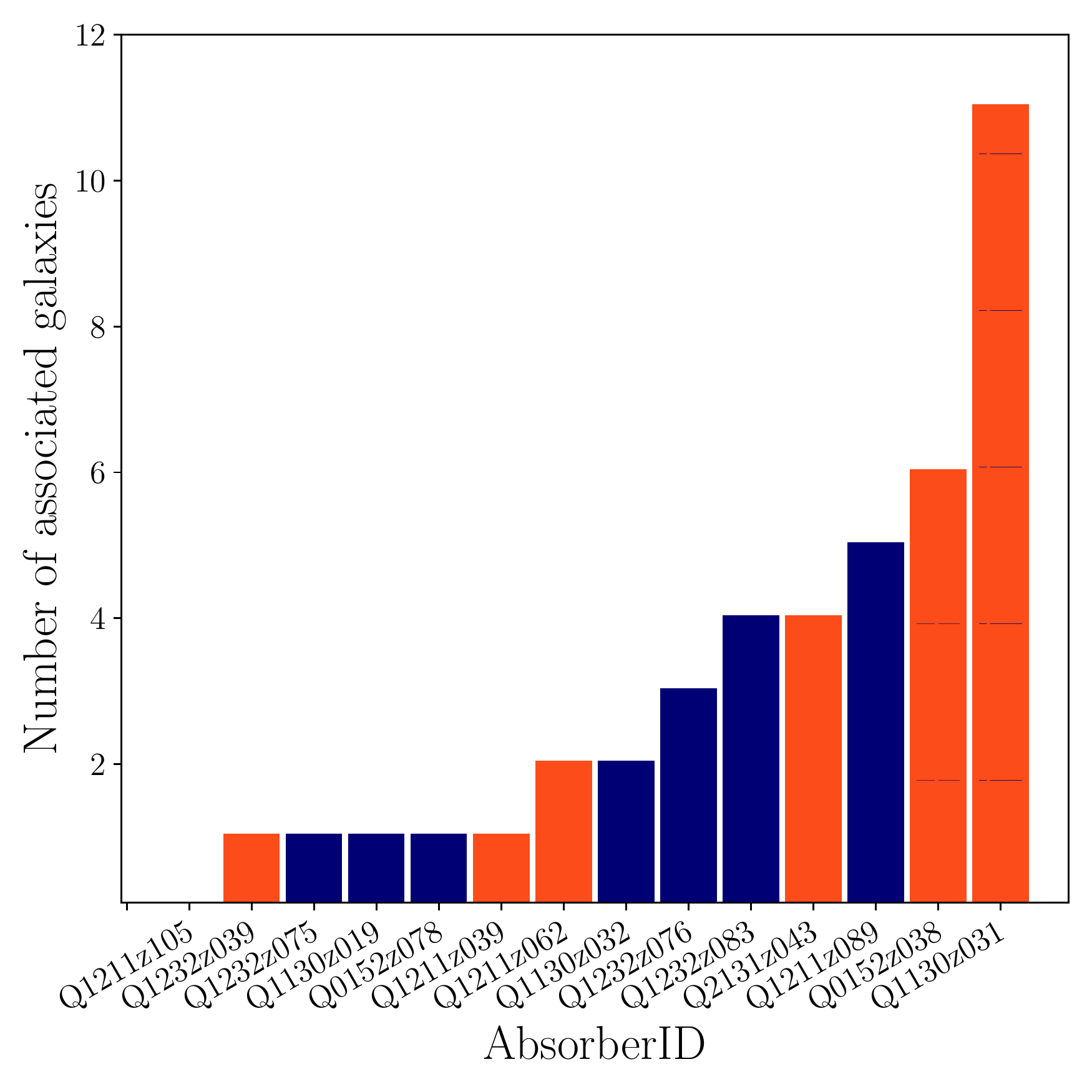}
 \caption{Number of galaxies associated with each absorber in our sample. \HI\ selected galaxies are indicated in orange, metal-selected ones in blue. Most of the absorbers have two or more galaxies at the associated redshifts. Q1130z031 is an exceptionally rich galaxy group and also has the highest \HI\ column density \citep{peroux19}. For Q1211z105 we do not detect any galaxy at the redshift of the absorber down to 0.1 M$_{\odot}$yr$^{-1}$.}
 \label{fig:nrgals}
 \end{figure}
\begin{figure*}
\centering
\includegraphics[width=2\columnwidth]{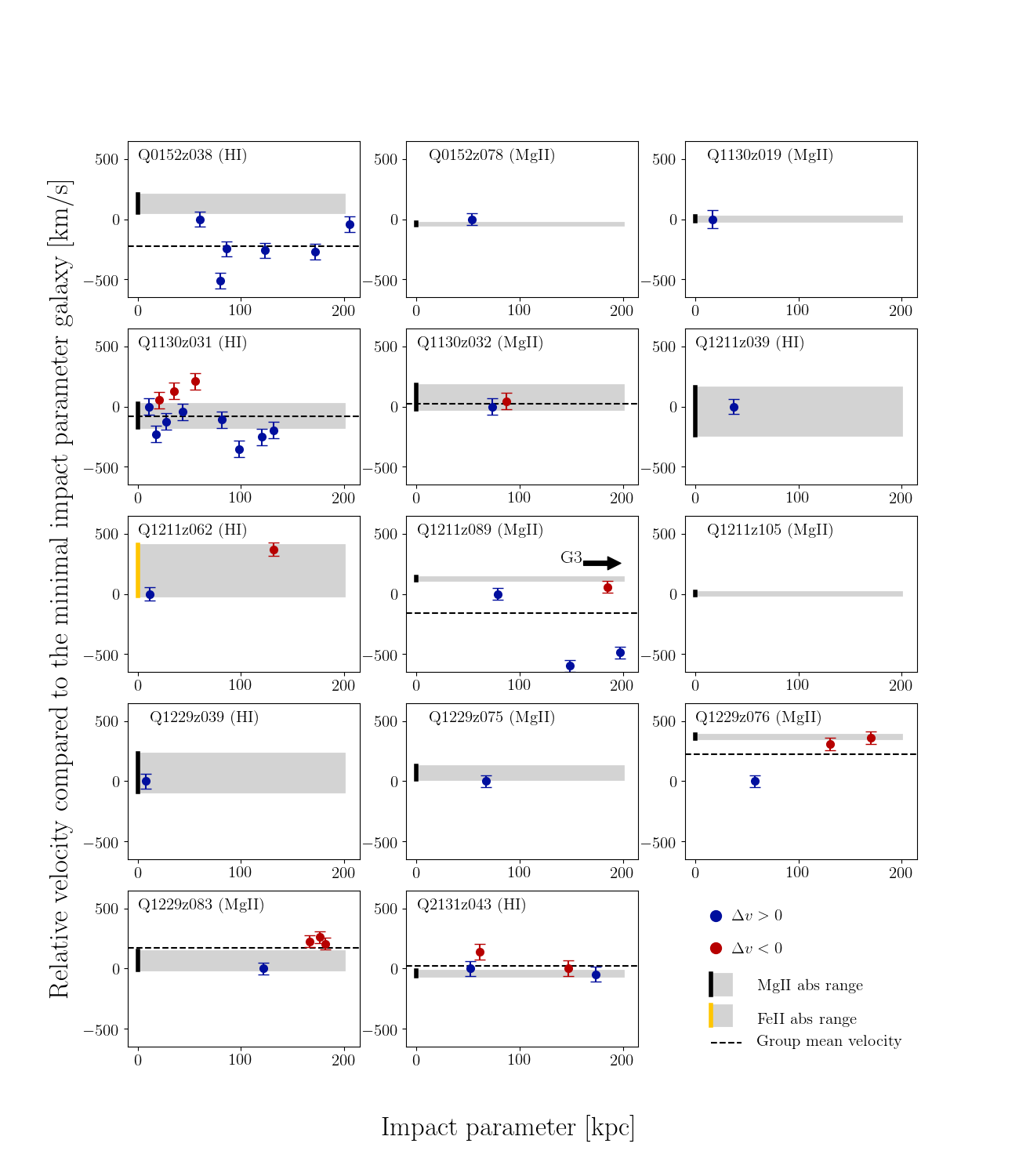}
\caption{Impact parameter - velocity diagram of the MUSE-ALMA Halos sample. The velocity of the galaxy at the lowest impact parameter is set to zero and the velocities of the remaining galaxies and absorption lines are calculated with respect to that galaxy.  Galaxies with $\Delta$v $>$ 0 are marked in red and with $\Delta$v $\leq$ 0 in blue. The black line (and the further grey area) indicates the velocity span of all \MgII $\lambda$2796 absorption line components. For the Q1211z062 system (marked in yellow) \MgII $\lambda$2796 was not covered and \FeII $\lambda$2600 was used instead. Dashed black like marks the mean velocity of the galaxy group. Note that galaxies with the smallest impact parameters are not always closest to the gas in velocity.}
\label{fig:vel}
\end{figure*}

\section{Analysis and Results}
\subsection{Identification of \HI\ and metal absorbers in high-resolution UV quasar spectra}
High-resolution UV quasar spectra are used for identification of the metal absorbers and measurements of the metal line column densities. Together with far-UV measurements of \HI\ column density, this allowed us to measure metallicities of the observed gas. The primary sample is comprised of known strong \HI\ absorbers, we used the high-resolution UV spectra to identify the metal lines associated with these absorbers. Specifically, we identified \FeII\ (\FeII\ $\lambda$ 2382, 2344, 2586, 2600), which we use for the metallicity measurements. We then expanded the sample to other metal absorbers by identification of the characteristic lines such as \MgII\ or \CIV\ doublets down to the equivalent width limit of W$_{\rm r}(2796)$ > 0.02. For the redshift determination, we used \MgII $\lambda$2796; in case of multiple components, we used the strongest one to measure the central wavelength. We limited the sample to systems with redshift $z < 1.4$, so that the [\OII] emission line is within the MUSE wavelength coverage, resulting in a list of 14 \MgII\ absorbers, of which some have been presented elsewhere: Q0152z038 \citep{rahmani18disk}, Q0152z078 \citep{rahmani18wind}, Q1130z031 \citep{peroux19} and Q2131z042 \citep{peroux17}. We used VPFIT\footnote{https://www.ast.cam.ac.uk/$\sim$rfc/vpfit.html}v10.0 to fit the Voigt profiles to unsaturated metal absorption lines. In particular, we fitted \FeII\ of systems whose metallicities were not known from the literature. We referred to the measurements of \HI\ column densities (N(\HI)) in the literature, and for missing cases (Q1211z089, Q1211z105, Q1232z076, Q1232z083) we measured it used the archival HST FOS UV spectra. We derived [Fe/H] metallicities of the absorbing gas using a solar metallicity value of [Fe/H]$_{\odot}$  = -4.5 from \citet{asplund}. However, Fe is not a perfect tracer of the metallicity since it is well known that this element can be depleted onto dust. 

Dust in the gas associated with the absorber could cause reddening of the quasar spectrum. To test if there is a significant amount of dust associated with the absorbing gas, we fit the spectral templates to quasars optical spectra extracted from MUSE cubes and archival optical and near-infrared photometry. The details of the method can be found in the Appendix \ref{appendix:dust}. We estimate the reddening to be $E(B-V)$ = 0.02 - 0.03 for all quasars in the sample, indicating that the that Fe depletion onto dust in absorbers is likely small. Given the remaining uncertainty in the dust correction, we considerably refer to absorption metallicity estimates as lower limits.

\subsection{Ionization correction for low N(\HI) absorbers from Bayesian MCMC approach} 
Apart from the dust, the absorption metallicity measurement can be affected by the H ionization fraction. Most absorbers in our sample are not DLAs (log(N(\HI)) < 20.3 cm$^{-2}$) and in such systems we expect a significant portion of the hydrogen to be ionized, influencing the metallicity derivations. To assess the effect of the ionisation correction on the metallicities derived for such absorbers, we used the radiative transfer calculations from CLOUDY \citep{cloudy}. We performed Bayesian MCMC modelling on the pre-computed grid of Cloudy models from \citet{fumagalli2016}. The grid consisting of multiple radiative transfer models is defined by four parameters: N(\HI) (log(N(\HI)) = 17.0 - 20.5), redshift ($z$ = 0.0 - 4.5), metallicity (log($Z$/$Z_{\odot}$ = -4.0 - 1.0 and $n_{H}$ density (log($n_{H}$) = -4.5 - 0.0). N(\HI) and redshift are constrained from the observations and serve as an input. To determine the most probable model, we use as constraints the measured metal line column densities (\MgII, \MgI\ and \FeII\ in almost all cases). We run the MCMC chain over the grid, using the PyIGM package routines \footnote{PyIGM Package is available at \url{https://github.com/pyigm/pyigm} \citep{prochaska17}}. For the resulting Probability Density Functions (PDF) of each of the parameters, we adopt the median value of the computed [Fe/H] metallicity (see Table \ref{tab:emmiters}). For a detailed description of the method, its extension and assumptions in the UV background we refer the reader to \citet{fumagalli2016} and \citet{wotta2019}. 

\begin{figure} 
\includegraphics[width=\columnwidth]{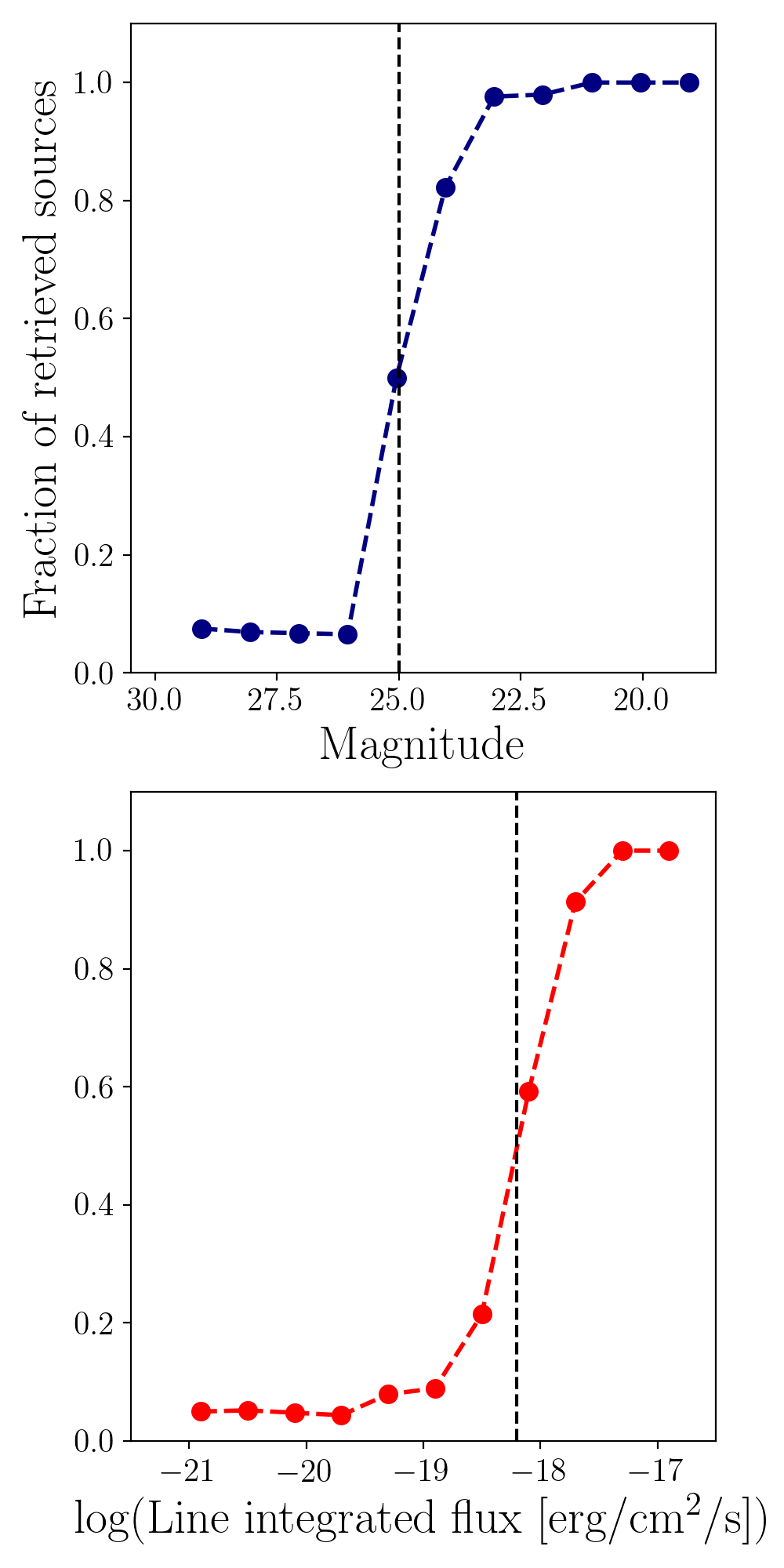}
\caption{Completeness of the MUSELET search for continuum sources and line emitters in MUSE-ALMA Halos MUSE cubes. Detection fraction of line emitters with continuum as a function of their magnitude are on the top panel (blue line) and line emitter without detected continuum as a function of a integrated line flux are on the bottom panel (red line). The black dashed line marks the 50 percent completeness level.}
\label{fig:completness}
\end{figure}

\subsection{Identification of galaxies associated with absorbers}
After identification of all quasar absorbers, we used the MUSE observations of the quasar fields, together with wide-field imaging with HST WFPC2 to search for galaxies associated with these absorbers. We used several analysis techniques to maximize the completeness of the search, including different types of objects (SF-galaxies, passive galaxies, faint, low SFR objects etc.). MUSE datacubes can be viewed as individual narrow-band (NB) images at each wavelength slice or a combined continuum image over the whole observed wavelength range, allowing for two types of object searches in the data: a single spectral-line search at the redshift of the absorber and identification of the sources seen in continuum. We combined 50 (50$\times$ 1.25 \AA) MUSE wavelength planes into a pseudo-narrow-band (NB) images centred at the expected [\OII] observed wavelength at the redshift of the absorbers. The choice of the width of the pseudo-NB corresponds to a velocity difference to the absorber redshift of $\Delta v$ $\pm$ 1000-1500 km s$^{\rm -1}$, depending on the central wavelength of the observed [\OII] line. The slight redshift dependence is a mild effect which is mitigated by the systemic search with MUSELET. Then, we identified all bright sources on the NB images and inspected their full spectra to confirm the redshift by the presence of other emission lines (if detected). This allows us to detect galaxies at a range of impact parameters, from systems close to the quasar (< 1 arcsec) up to the limits imposed by the MUSE field-of-view. Within the spectral coverage of MUSE, [\OII] doublet can be observed up to $z$ = 1.4, however from $z$ = 0.8 it will be the only prominent galactic line covered. We exclude the possibility of the contamination from the low-$z$ sources ( $z$ < 0.8), because at that redshift ranges we would see more then one line (bright [\OIII] or H$\alpha$). For the higher-$z$ case ($z$ > 1.4), the only observable emission line is Ly$\alpha$ (covered by MUSE at $z$ =  2.7 - 6.7). Although theoretically, the detection could be a Ly$\alpha$ within that redshift range, based on the LAE luminosity function and lack of the expected prominent asymmetric shape we exclude that interpretation.

We search for galaxies within $\Delta v \pm$ 1000 km s$^{-1}$ from the absorber. We measure the centre of the lines with a precision below 1 \AA\ (corresponding to an error in the redshift estimation smaller than $\Delta z \pm 0.0003$). We note that the majority of galaxies lie within $\Delta v \pm$ 500 km s$^{-1}$.

For three absorbers, Q1211z062, Q1232z039 and Q1130z031 \citep{peroux19}, we detect galaxies close to the quasar position (within 1 arcsec), blended with the quasar PSF. Spectral PSF subtraction allows for easier detection of such systems, despite the quasar bright contribution. In each case, we detected the [\OII] emission doublet at the redshift of the absorber. We extracted the 1D spectrum of 4 $\times$ 4 spatial pixels around the line detection. The continuum of these objects are not detected. The method is illustrated in Fig.\ref{fig:qsores}, for absorbers from two of the QSOs, where we present a zoom into the extracted spectrum around the [\OII] emission line (in the observed wavelength frame), the main absorption feature of the associated absorber in the galaxy velocity space as well the pseudo-narrow band image alongside the HST image of the same region. 

After narrow-band identification of the sources at the redshift of the absorber, we extracted all continuum sources from the MUSE white light image using the MUSELET source finder, part of the MPDAF package provided by the MUSE GTO team \citep{mpdaf}. MUSELET creates from a cube: (i) a white light image; (ii) R, G, B images corresponding to 1/3 of the wavelength range each; (iii) a set of narrow-band images based on the average of 5 wavelength planes (total width of 6.25 \AA). It then runs SExtractor \citep{sextractor} on these products, resulting in a catalogue of continuum sources. We classified all of the reported sources as either emission line galaxies, absorption line galaxies, stars, unknown continuum sources (without emission or absorption lines) or artifacts. We inspected all extracted spectra and used [\OII] and [\OIII] emission lines or Ca K+H absorption lines to determine the redshift of each source. Among all galaxies associated with absorbers, we detected four passive galaxies (with only Ca K+H absorption lines and red continuum). For completeness, we cross-matched the catalogue of continuum sources detected in MUSE with detections in archival HST WFPC2 images \citep{lebrun}. This search led to no new galaxies discovered at the redshift of the absorbers.
After identification of the galaxies associated with the absorber on the white light continuum MUSE image, we measured their position with respect to the quasar. We fitted a 2D Gaussian to the continuum images of both the quasar and the galaxies to determine their geometrical centres and then measure the angular distance between them. 

To test the completeness of the methods described above, we injected 100 mock sources into each MUSE datacube. The mock sources are single Gaussian emission lines of the integrated flux ranging from 10$^{-21}$ to 10$^{-17}$ erg s$^{-1}$ cm$^{-2}$, represented spatially by a 2D Gaussian of FWHM of 3 pixels, corresponding to the MUSE PSF. We introduce the two types of mock sources: line emitters with continuum flux and without continuum. After running the MUSELET source extraction on these cubes we compared the catalogue of detected sources with the list of mock sources. The completeness of our search is presented in Figure \ref{fig:completness}. 
We are over 90 percent complete for detections brighter then 23.5 mag for continuum objects and line flux greater then 10$^{-17.7}$erg cm$^{-2}$ s$^{-1}$. We can retrieve sources down to the magnitude of 25.7 mag and flux 10$^{-18.5}$erg cm$^{-2}$ s$^{-1}$.
\begin{figure} 

\includegraphics[width=\columnwidth]{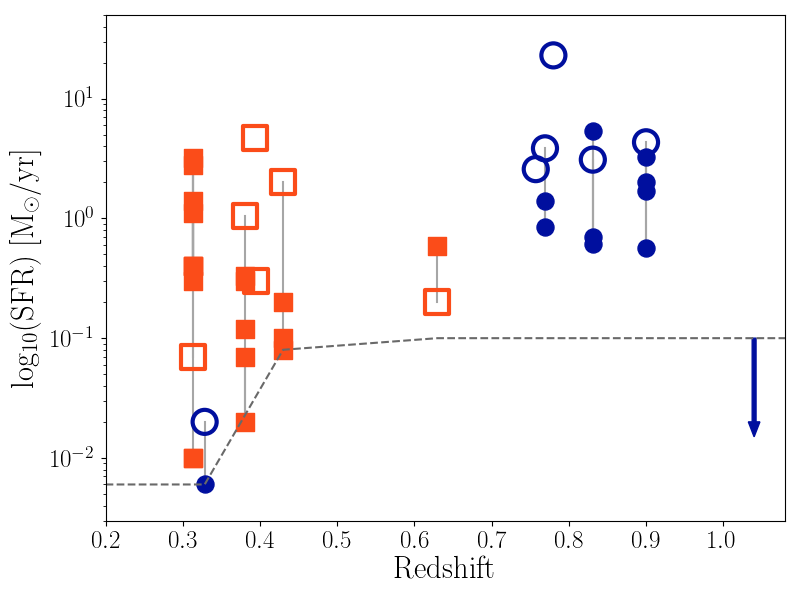}
\caption{The star formation rate (SFR$_{[\OII]}$) of the galaxies associated with absorbers as a function of redshift. Vertical grey lines connect galaxies associated with the same absorber. Open symbols indicate the galaxy at the smallest impact parameter. Absorbers selected based on their \HI\ absorber are marked with orange squares, while metal-selected ones are marked with blue circles. The arrow marks the limits of the non-detection of the galaxy associated with the absorber Q1211z105. The dashed grey line represents the SFR limit at the redshift of the absorbers. We reach star formation rates of 0.1 M$_{\odot}$yr$^{-1}$ for most of the sample and lower (about 0.01 M$_{\odot}$yr$^{-1}$) in case of deeper data (Q1130-1449).}
\label{fig:sfr}
\end{figure}

We detected 43 associated galaxies within a 250 kpc impact parameter from the QSO sight-line. As we summarize in Fig.\ref{fig:nrgals}, we detect systems with multiple galaxies associated with the absorber as well as a few systems where only one galaxy was detected. Only in one case we do not detect any associated galaxy. This results in a high detection rate of 93 per cent. Note that five of the absorbers in the sample were targeted MUSE observations of absorbers with a known counterpart. Nevertheless, among the remaining nine absorbers, we detect at least one galaxy for eight absorbers. Altogether, among the 14 absorbers, eight (57 per cent) have more than one associated galaxy. This high detection rate is likely related to combination of the wide field of view and simultaneous spectral coverage of MUSE. We are probing a large range of impact parameters, from low (< 10 kpc) up to large distances of 200-250 kpc. We detect emission line sources and passive galaxies with no emission lines but solely detected in continuum at the redshift of the absorbers down to an integrated line flux of $ 10^{-18.5}$erg cm$^{-2}$ s$^{-1}$ and magnitude of 25.7 mag at the 20 percent completeness level.

Table \ref{tab:emmiters} summarizes the parameters of all absorbers and their associated galaxies. Spectra of all systems alongside their position marked on MUSE white light images and HST images are shown on the Fig. \ref{fig:example} and in the Appendix.

\begin{figure} 
\includegraphics[width=\columnwidth]{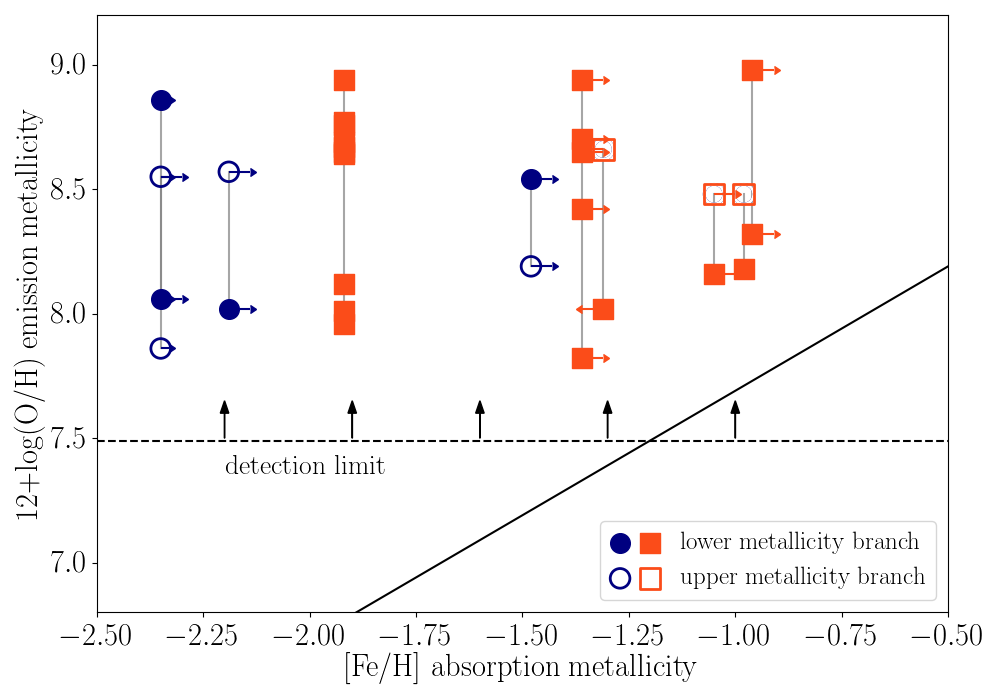}
\caption{Oxygen abundance (emission metallicity) of the galaxies versus the metallicity of the absorbing gas. The solid black line represents the x=y line, the grey dashed line marks the detection limit of our data. Galaxies associated with the same absorber are linked with a vertical line. Filled and hollow circles represent upper and lower metallicity branch solution. Absorbers selected based on their \HI\ absorber are marked with orange squares, while metal-selected ones are marked with blue circles. Measurements from the literature with one solution are also plotted. In all cases where it is measured, the galaxy emission metallicity is larger than that of the absorbing gas.}
\label{fig:metmet}
\end{figure}
\subsection{Measurements of the physical properties of galaxies associated with absorbers}
The majority of the galaxies associated with absorbers in our sample are emission-line objects. We can use the measured line fluxes to calculate the star formation rates of these galaxies as well as their emission-line metallicities. To obtain line fluxes, we fit Gaussian profiles to the [\OII], [\OIII], H$\beta$ and H$\alpha$ emission lines (if detected), using the MPDAF Gaussian profile fitting tool. For the redshift estimation, we use [\OII]$\lambda\lambda$3727,3729 (fitted with double Gaussian profiles) as this line is detected in all galaxies. In the case of galaxies lacking emission lines, we used Ca K+H absorption lines to determine the redshift. 

For each galaxy we compared the velocity offsets between the members of the group and the absorbing gas. In Fig. \ref{fig:vel} we present the relative velocities of the galaxies of each absorbing system calculated with respect to the galaxy at the lowest impact parameter. Therefore, the velocity of the galaxy at the lowest impact parameter is set to zero. Additionally, we calculated the velocity between that galaxy and the absorbing gas. In four cases (out of eight multi-galaxy systems) objects other than the closest galaxy are better aligned in velocity with the gas seen in the absorption. 

For all emission-line galaxies in the sample (except Q1130z019, for which we used H$\alpha$) the [\OII] emission line is covered and detected. We used the flux of the line to calculate the star formation rates of the galaxies following \citet{kobulnicky}. In Fig.\ref{fig:sfr} we present the measured SFR as a function of the redshifts of the absorbers. For sources for which we do not detect emission lines, we calculate the 3$\sigma$ line limits at the expected line positions. 
We detect sources at the redshift of the absorbers down to an SFR of 0.01 M$_{\odot}$yr$^{-1}$ for $z < 0.4$ and 0.1 M$_{\odot}$yr$^{-1}$ for higher redshifts (Fig.\ref{fig:sfr}). We report galaxies with a range of SFR from passive galaxies (< 0.1 M$_{\odot}$yr$^{-1}$) to star-forming galaxies. Except for one case (23 M$_{\odot}$yr$^{-1}$) the SFR of associated galaxies range from 1 to 10 M$_{\odot}$yr$^{-1}$).
These values are not dust-corrected since for objects most H$\alpha$ lays outside the probed wavelength range, thus making measurement of Balmer Decrement, the reddening proxy, impossible. Note that the parameters taken from literature also do not have any dust-correction applied. 

To obtain the galaxy metallicities we chose to use only those galaxies with measurements of the oxygen species [\OII], [\OIII] and H$\beta$. We then used the R23 oxygen abundance calculation from \citet{kobulnicky} as the metallicity proxy. This procedure results in two solutions lower and upper metallicity branch. Ideally, additional parameters based on [\NII] and H$\alpha$ can be used to differentiate between the branches. However, these lines are not covered with MUSE and therefore we present both solutions. For passive galaxies, we use the line detection limits. For Q1211z089 and Q1232z083, the 12+log(O/H) values could not be determined because both [\OIII] and H$\beta$ are outside of the MUSE wavelength coverage. 

A comparison of the metallicity of the galaxies with that of the absorbing gas is shown in Fig.\ref{fig:metmet}. Groups of galaxies associated with single absorbers are connected with vertical lines, and for most of the systems, the two metallicity branches are plotted. We mark the one-to-one relation with a black thick line in the lower part of the diagram. The red dashed line in Fig.\ref{fig:metmet} marks the emission line metallicity limit of our measurements. For all measured systems, the galaxy metallicity is found to be higher than that of the absorbing gas although given the limits, the metallicities still could be consistent with the 1:1 relation. Note, however, that given the metallicity estimate caveats (see Section 3.1) these result must be taken with some caution. Finally, since we are not sensitive to metallicities below 12+log(O/H) = 7.5, we might be missing faint low-metallicity systems in our sample of galaxies associated with absorbers.

\begin{figure} 
\includegraphics[width=\columnwidth]{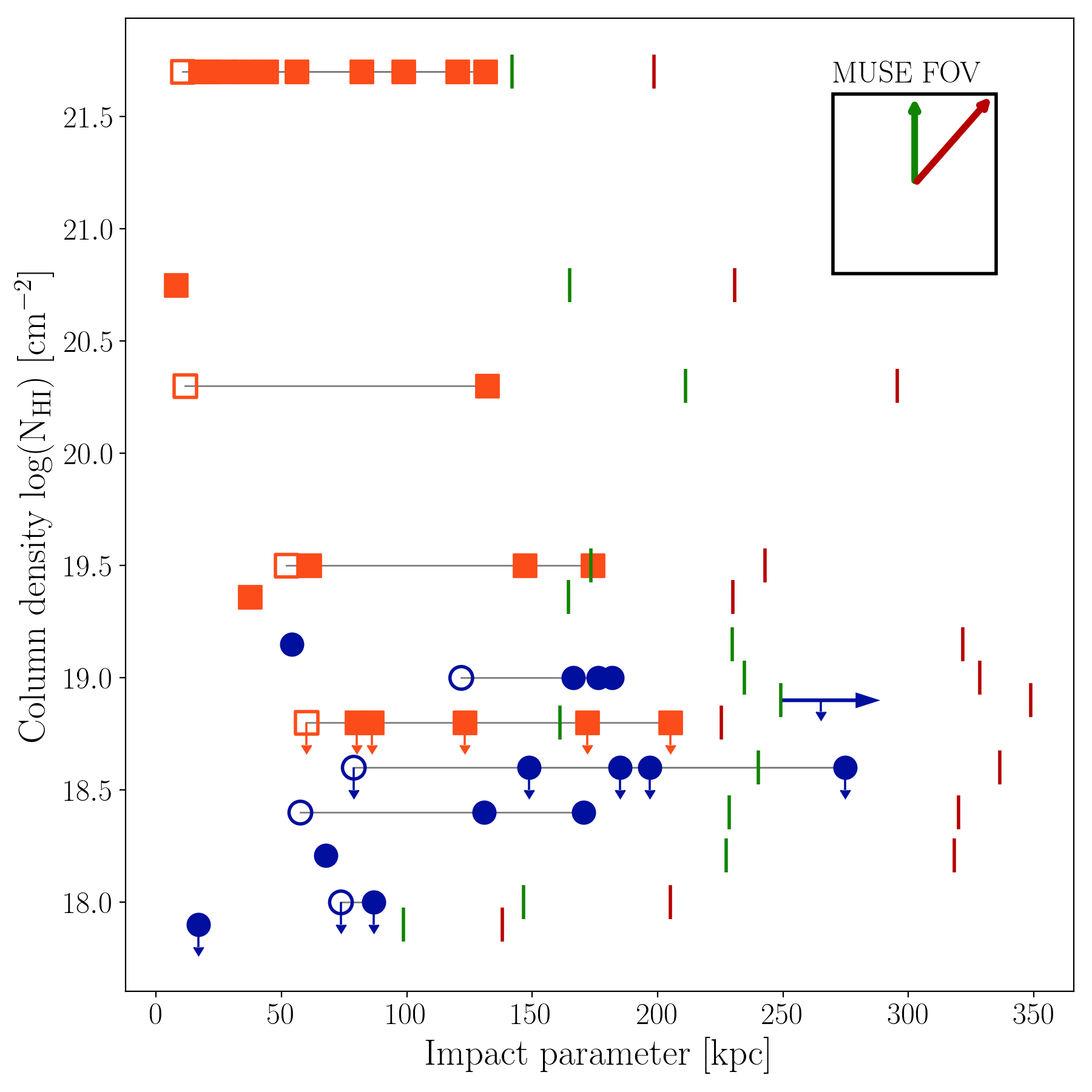}
\caption{\HI\ column density as a function of impact parameter. Absorbers selected based on their \HI\ feature are marked with orange squares, while metal-selected ones are marked with blue circles. The blue, horizontal arrow marks the limits of the Q1211z105 associated galaxy impact parameter non-detection. The hollow symbols represent the lowest impact parameter object in groups. Red and green lines mark the limit of the impact parameter at the redshift of the absorber as defined by the MUSE field of view. Because the MUSE field of view is a square, we can reach larger impact parameters on the diagonal (red line) than along the square's side (green line; see the inset in the right upper corner). Horizontal grey lines connect galaxies associated with a single absorber. We find associated galaxies up to the impact parameters of 250 kpc.}
\label{fig:HIbSFR}
\end{figure}

\section{Discussion}
\subsection{Absorbers are associated with multiple galaxies}
The MUSE IFU proves to be an excellent instrument to search for galaxies associated with absorbers. The MUSE field of view of 1 arcminute translates to 200-250 kpc at redshift $z\sim$1. Since each object comes with a spectrum, we can immediately classify all objects in the field of view, including passive systems, increasing the extent of the search. 

Furthermore, identification of low impact parameter galaxies (< 1 arcsec) is also possible. Low impact parameter sources are the primary candidates for the absorbers hosts but can be potentially hidden under the extended PSF of bright quasars. Using spectral PSF subtraction, we can remove the quasar contamination and detect the emission lines from the source hidden under the PSF of the quasar (the method is described in Section 3.2). We report three such cases in our sample: Q1211z062, Q1232z039 (Fig. \ref{fig:qsores}) and Q1130z031 \citep{peroux19}. 

Candidates for galaxies associated with absorbers were traditionally identified on wide-field images and later confirmed using long-slit spectroscopy. A number of presumably unique galaxy-absorber pairs at different redshifts were discovered this way, supporting the picture of a CGM tracing the halo of a single galaxy. Wide-field IFU spectroscopy improved the completeness of the absorber counterpart search, resulting in discoveries of multiple galaxies associated with absorbers \citep{bielby17,klitsch,bielby19, fossati}. These findings also extend to higher redshifts with $z \sim$ 3 DLAs associated with multiple Ly $\alpha$ Emitters (LAE), as in \citet{mackenzie}. Discoveries of multiple galaxies associated with low-$z$ absorbers challenge the classical view of the CGM connected to the single halo of an isolated galaxy. 

Cuts in both impact parameter and velocity difference between absorber and galaxy differ between published MUSE absorber studies, making the comparison of results not straightforward. MusE GAs FLOw and WInd (MEGAFLOW), a MUSE survey of \MgII\ absorbers, \citep{schroetter} allows for a 1000 km s$^{-1}$ velocity difference between the galaxy and the absorber, and in case of multiple detections, the lowest-impact-parameter object is selected. Quasar Sightline And Galaxy Evolution (QSAGE), an OVI absorption oriented MUSE program \citep{bielby19}, supported by a larger impact parameter coverage with HST grisms, uses a tight limit of 400 km s$^{-1}$ (expected outflow velocity). For sources covered only by HST, the velocity range is extended to 700 km s$^{-1}$ to account for the lower resolution of the HST grism spectra compared to MUSE. In a targeted single-system study, \citet{bielby17} 
find a galaxy group tightly aligned within $\Delta v =$ 140 km s$^{-1}$ around the absorber redshift, and within an impact parameter of 300 kpc, while in the case of a CO pre-selected absorber, \citet{klitsch} looked for counterparts within $\pm$ 2000 km s$^{-1}$. In MUSE-ALMA Halos, we select galaxies to be associated with the absorber if their velocities does not differ by more than 1000 km s$^{-1}$ from the absorber. Impact parameters are only limited by the MUSE field of view (up to 250 kpc at $z\sim$1). We mostly detect systems with multiple galaxies associated with the absorber in velocity range between few km s$^{-1}$ to $\pm$ 750 km s$^{-1}$. We used the luminosity function of [\OII] emitters from \citet{ly} to calculate the expected number of galaxies in the volume at the redshift of the absorbers. We find the expected number of detectable sources to be 0.45 for the higher redshift absorber and around 0.1 for the rest of the sample. Therefore we conclude, that we detect over-densities of galaxies associated with quasar absorbers, compared to random field galaxies.  

Groups of galaxies typically include between 3--30 members and have a total dynamical mass of (10$^{12.5}$--10$^{14}$)$h^{-1}$ M$_{\odot}$ with sizes between (0.1--1)$h^{-1}$ Mpc \citep{bahcall96}. Depending on mass, we would expect a velocity dispersion of group members between 100 and 500 km s$^{-1}$ \citep{bahcall96}. For the groups of absorption-selected galaxies in our sample, the velocity dispersion spans 160--350 km s$^{-1}$. Such values are typical of groups environment.

Detecting multiple galaxies associated with the absorber introduces a complexity to the CGM--galaxy identification. Typically, absorbers at the lowest impact parameter would be considered the most likely counterpart for the absorber. Our analysis clearly demonstrates that impact parameters alone are not conclusive for identifying a unique galaxy associated with the quasar absorber. In Fig.\ref{fig:vel} we present the velocity - impact parameter diagrams for all absorbers in our sample. In several cases, the galaxy at the lowest impact parameter is offset from the gas velocity by more than the typical galaxy rotation velocity (150--200 km s$^{-1}$). By contrast, some galaxies with larger impact parameter are better aligned to the absorbers velocity. This does not point directly to the gas origin since we expect also to see tracers of outflows in the CGM and wind velocities can be as high as 600 kms$^{-1}$. To link the absorbing gas with a particular galaxy, a detailed component-by-component analysis of the absorbing gas profile alongside resolved kinematic studies of the galaxy are needed. 

To put our sample in the context of other known absorbers, we test if the galaxies follow the canonical relations. In Fig.\ref{fig:HIbSFR} we present the \HI\ column density versus impact parameter. Studies so far, mostly for high \HI\ column density absorbers \citep{krogager17,augustin}, showed an anti-correlation between N(\HI) and impact parameter, which is also reproduced by the OWLS hydro-dynamical simulations \citep{rahmatyi}: the larger the N(\HI) column densities the smaller the impact parameter. Including all galaxies associated with absorbers, we are retrieving this relation also at lower \HI\ column densities (from 10$^{\rm18}$ to 10$^{\rm21.7}$ cm$^{-2}$). Multiple associated galaxies, as in our sample, introduce a significant scatter into this relation. 

\begin{figure}
    \centering
    \includegraphics[width=\columnwidth]{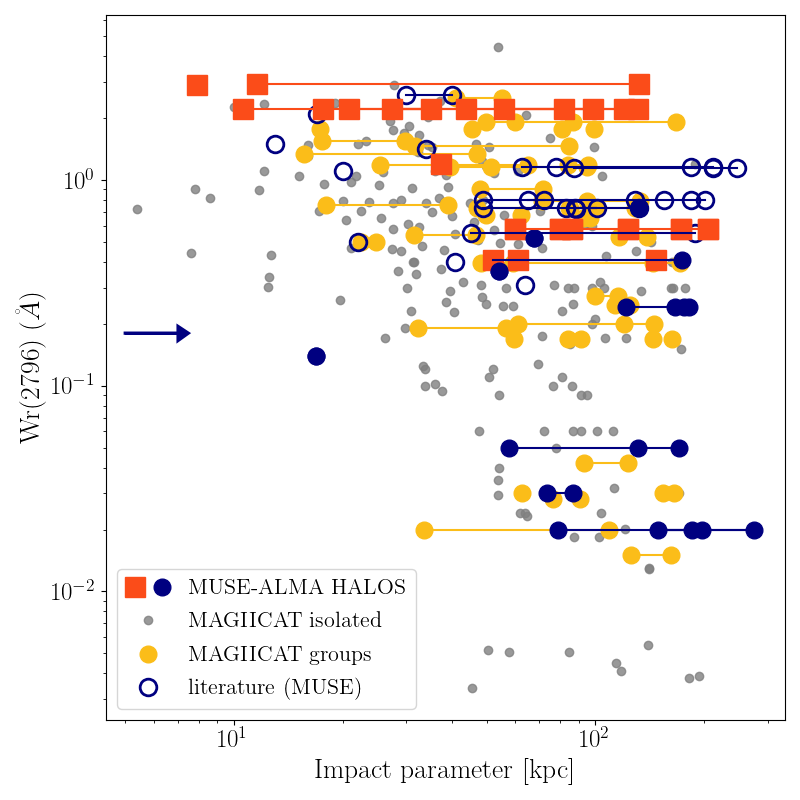}
    \caption{\MgII $\lambda$2796 absorption line equivalent width versus the impact parameter. The filled blue circles are the galaxies identified in MUSE-ALMA Halos (groups linked with a line, absorbers selected based on their \HI\ feature are marked with orange squares), empty blue circles mark systems from the literature \citep{schroetter,bielby17,klitsch,zabl19}, groups are linked with lines. The blue, horizontal arrow marks the limits of the Q1211z105 with no detected associated galaxy. The underlying points come from the MAG\II CAT catalogues of isolated galaxies and groups catalogues in grey and yellow, respectively. Groups of associated galaxies alongside isolated systems define a region in parameter space below which no absorbers are detected. }
    \label{fig:eqw}
\end{figure}

All \HI\ absorbers in our sample are also \MgII\ absorbers. Having both Ly$\alpha$ and \MgII\ for all systems is a unique quality of the sample. Typically, \MgII\ absorbers follow an W$_r$(2796) (\MgII\ $\lambda$2796 equivalent width) and impact parameter anti-correlation \citep[e.g.][]{lanzetta,berbo,bouche06,chen}. We compared our groups to several other MUSE studies from the literature and the \MgII\ absorbers compilation MAG\II CAT \citep{nielsen13} of isolated galaxies as well as galaxy groups \citep{nielsen18}. Fig.\ref{fig:eqw} shows that MUSE-ALMA Halos systems are in agreement with this relation, especially with other absorption-related groups. Together with the group compilation from MAG\II CAT, our data points seem to be offset from the "individual" galaxies. Similar findings were reported by \citet{bordoloi} in the COS-Halos comparison of \MgII\ absorbers in groups and isolated galaxies. The W$_{r}$(2796)--$b$ relation for multiple associated galaxies define an envelope of possible parameters for the absorbers. Despite being sensitive to both low \MgII\ equivalent widths and low impact parameters, our observations do not populate the part of the plot with low W$_r$(2786) and low impact parameters. 

Lowest impact parameter systems can lie far from the absorbing gas in velocity space introducing the question of the nature of the gas detected in absorption and its physical connection the galaxies detected in emission. Detailed component-by-component analysis of the metal absorption lines, together with resolved kinematic and metallicity maps of the galaxies will likely provide information to identify the origin of the absorbing gas. Intra-group gas can be the result of the interactions between the group members, with possible higher density tidal streams or be filled with clouds of material infalling onto the centre of the group dark matter halo. This intra-group medium proves to be more complicated to describe and understand than a single halo picture, and is less straightforward to connect the inflow/outflow features with particular members of the group.
 
\subsection{Metallicities of absorbers and associated galaxies}
\begin{figure} 
\includegraphics[width=\columnwidth]{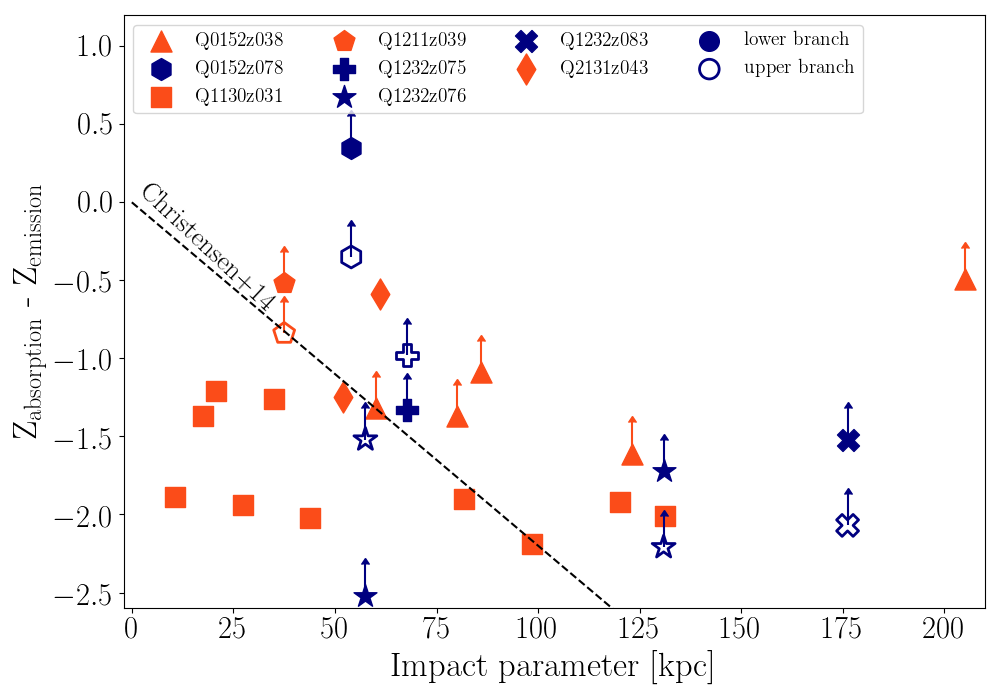}
\caption{Difference in metallicity of the QSO absorption line and the metallicity from emission lines of the galaxies associated in redshift. Different symbols distinguish groups of galaxies associated with various absorbers,absorbers selected based on their \HI\ feature are marked in orange. Filled symbols mark the lower branch metallicities or unequivocal values from the literature, hollow symbols mark the upper metallicity branch. The dashed black line marks the CGM gradient measured by \citet{christensen}.}
\label{fig:diffmet}
\end{figure}

Measurements of \HI\ and metal column densities of the absorbers in our sample give a unique opportunity to compare directly the metallicity of the gas and the integrated metallicity of the associated galaxies. Our absorption metallicity measurements suffer from several caveats which can affect the results presented below (for the detailed discussion see Section 3.1) and due to the lack of dust correction should be treated as lower limits. In Fig.\ref{fig:diffmet} we present the difference of the metallicity of the gas and the galaxies as a function of the impact parameter. We find that in all cases the galaxy metallicity is higher than the gas metallicity lower limits. However, our detection threshold of emission metallicities goes down to 12+log(O/H) = 7.5, which corresponds to the absorbing gas metallicities. 

Measuring the metallicity in two different spatial positions (in the associated galaxy and in the halo) can be interpreted as measuring a metallicity gradient. Resolved studies of nearby galaxies (like CALIFA or MaNGA) report negative ISM metallicity gradients in nearby star-forming galaxies \citep{sanchez14,ho16,belfiore}. Similar studies of high-redshift lensed galaxies have shown a diversity of gradients in star-forming galaxies \citep{lee16, queyrel,jones13, jones15}. Positive metal gradients can point to strong outflows expelling metals to the CGM or suggest inflows of a large portion of metal-poor gas.

Expanding the metallicity gradients from the ISM to larger scales likely probes the physics of the CGM. Gradients between absorbing gas and galaxies were so far measured for systems with impact parameter below 100 kpc \citep{peroux12,christensen} and have an average value of -0.022 dex kpc$^{-1}$. Combining results from absorption spectroscopy of gravitationally lensed quasars with measurements for star-forming galaxies, isolated spirals, quasar-galaxy pairs, and gravitationally lensed galaxies, \citet{Kulkarni} also reported a tentative anti-correlation between the metallicity gradient and the metallicity at the galaxy centre: objects with low metallicity at the centre have positive metallicity gradients going outward. 

We looked at the difference between the absorber and galaxy metallicity as a function of impact parameter (Fig.\ref{fig:diffmet}). Note that our absorption metallicity measurements do not have detailed ionization correction, and additionally, we compare the values based on different spices: Fe for absorption and O for emission. We find no dependence of the metallicity difference on impact parameter and a larger scatter than \citet{christensen} at redshift 0.1 < $z$ < 3.2. This is explained in part by the multiple galaxies associated with each absorber.

Hydrodynamical simulations like FIRE or IllustrisTNG \citep{muratov15, muratov17, nelson}, predict metals to be a tracer of the CGM structure following the flows of gas in and out of the galaxy. Metal-rich outflows are expected to originate in the centres of galaxies, introducing the metallicity differences with respect to the azimuthal angle between the quasar line of sight and the projected major axis of the galaxy. Future work will look these spacial variations of metallicity in more detail \citep[see][]{peroux16}.\\
To relate the gas probed in absorption with the galaxies in the group, a detailed component-by-component analysis provides a promising avenue \citep[][Rahmani et al. in prep.]{rahmani18disk}.

\section{Conclusions}
We report the first results of MUSE-ALMA Halos, a unique multi-wavelength data set aimed at understanding the complexity of the Circum-Galactic Medium, the interactions with the galaxies and the environmental effect on the high \HI\ column density absorbers. The survey consists of five quasar fields for which we identified 14 absorbers at redshift $z < $ 1.4. We searched for the galaxies associated with the absorber within the MUSE field of view (covering 200-250 kpc in impact parameter, depending of the system's redshift) and within $\Delta v$ = 1000 km s$^{-1}$ from the absorber redshift. Each absorber has an \HI\ column density measured either from the literature or from archival FOS spectra. We used MUSE IFU observations to identify galaxies associated with the gas seen in absorption, finding altogether 43 such systems yielding a 89 per cent detection rate among the non-targeted absorbers. We find that most absorbers are associated with multiple galaxies or groups, tracing overdensities of at least a factor of 10 relative to the field galaxies. The detection of multiple galaxies is directly related to the ability to map a large field centered on the QSO, which is a unique capability of MUSE covering impact parameters up to 250 kpc. In addition, the use of spectral PSF subtraction for low impact parameter cases and the sensitivity of MUSE resulting in the detection of low SFR object (SFR > 0.01--0.1 M$_{\odot}$yr$^{-1}$) allow for much more complete study of the galaxy environment linked to the absorber. The observed absorbing gas may then probe not only a single CGM halo but most likely the intra-group material affected by intergalactic interactions. 
By studying the relative velocity differences between the gas and the associated galaxies, we find that the traditional way of assigning the absorber to the lowest impact parameter galaxy should be challenged. We find large dispersion in velocities between galaxies and absorbing gas, sometimes pointing to a more distant galaxy as the main candidate for the host. 
We compare the parameters of the systems in our sample with canonical relations of the absorbers and their host galaxies, N(\HI)--$b$ and W$_{r}$(2796)--$b$. We find that the W$_{r}$(2796)--$b$ relation follows the general trend while defining an envelope of parameters below which no further objects are detected.
Thanks to our multi-wavelength dataset we can compare the lower limits of metallicities of the absorbing gas with the metallicity of the galaxies as measured in the emission. We find that the overall metallicity is always higher in the associated galaxies than in the absorbing gas probed by the QSO. However, we are not sensitive to the low metallicity systems below  12+(O/H) = 7.5, and gas metallicities are lower limits. 
Given the complexity of the environment associated with these absorbers, we advocate the use of the kinematics information that could assist in associating individual absorbing component with each galaxy member. Comparison to the resolved kinematic and ISM metallicity of the galaxies could improve our interpretation of the absorption lines.
With this study, MUSE-ALMA Halos, we show that combining absorption and emission studies of CGM and associated galaxies is a powerful tool to study the evolution of gas inside and outside galaxies. In forthcoming work we will study the molecular gas content of galaxies associated with the absorbers in our sample, using completed ALMA observations of all the quasar fields described in this work.
\section*{Acknowledgements}
We thank the referee for the useful comments. CP thanks the Alexander von Humboldt Foundation for the granting of a Bessel Research Award held at MPA. AK acknowledges STFC grant ST/P000541/1 and Durham University. RA thanks CNRS and CNES (Centre National d'Etudes Spatiales) for support for her PhD. Authors thank C. Howk and N. Lehner for their comments and suggestions. AH thanks A. Bittner, M. Kozubal and D. Wylezalek for comments helping to improve this manuscript.



\bibliographystyle{mnras}
\bibliography{biblio}

\begin{thebibliography}{}
\makeatletter
\relax
\def\mn@urlcharsother{\let\do\@makeother \do\$\do\&\do\#\do\^\do\_\do\%\do\~}
\def\mn@doi{\begingroup\mn@urlcharsother \@ifnextchar [ {\mn@doi@}
  {\mn@doi@[]}}
\def\mn@doi@[#1]#2{\def\@tempa{#1}\ifx\@tempa\@empty \href
  {http://dx.doi.org/#2} {doi:#2}\else \href {http://dx.doi.org/#2} {#1}\fi
  \endgroup}
\def\mn@eprint#1#2{\mn@eprint@#1:#2::\@nil}
\def\mn@eprint@arXiv#1{\href {http://arxiv.org/abs/#1} {{\tt arXiv:#1}}}
\def\mn@eprint@dblp#1{\href {http://dblp.uni-trier.de/rec/bibtex/#1.xml}
  {dblp:#1}}
\def\mn@eprint@#1:#2:#3:#4\@nil{\def\@tempa {#1}\def\@tempb {#2}\def\@tempc
  {#3}\ifx \@tempc \@empty \let \@tempc \@tempb \let \@tempb \@tempa \fi \ifx
  \@tempb \@empty \def\@tempb {arXiv}\fi \@ifundefined
  {mn@eprint@\@tempb}{\@tempb:\@tempc}{\expandafter \expandafter \csname
  mn@eprint@\@tempb\endcsname \expandafter{\@tempc}}}

\bibitem[\protect\citeauthoryear{{Arrigoni Battaia}, {Hennawi}, {Prochaska},
  {O{\~n}orbe}, {Farina}, {Cantalupo}  \& {Lusso}}{{Arrigoni Battaia}
  et~al.}{2019}]{fabMUSE}
{Arrigoni Battaia} F.,  {Hennawi} J.~F.,  {Prochaska} J.~X.,  {O{\~n}orbe} J.,
  {Farina} E.~P.,  {Cantalupo} S.,   {Lusso} E.,  2019, \mn@doi [\mnras]
  {10.1093/mnras/sty2827}, \href
  {https://ui.adsabs.harvard.edu/abs/2019MNRAS.482.3162A} {482, 3162}

\bibitem[\protect\citeauthoryear{{Asplund}, {Grevesse}, {Sauval}  \&
  {Scott}}{{Asplund} et~al.}{2009}]{asplund}
{Asplund} M.,  {Grevesse} N.,  {Sauval} A.~J.,   {Scott} P.,  2009, \mn@doi
  [\araa] {10.1146/annurev.astro.46.060407.145222}, \href
  {https://ui.adsabs.harvard.edu/abs/2009ARA&A..47..481A} {47, 481}

\bibitem[\protect\citeauthoryear{{Augustin} et~al.,}{{Augustin}
  et~al.}{2018}]{augustin}
{Augustin} R.,  et~al., 2018, \mn@doi [\mnras] {10.1093/mnras/sty1287}, \href
  {https://ui.adsabs.harvard.edu/abs/2018MNRAS.478.3120A} {478, 3120}

\bibitem[\protect\citeauthoryear{Augustin et~al.,}{Augustin
  et~al.}{2019}]{augustin19}
Augustin R.,  et~al., 2019, \mn@doi [Monthly Notices of the Royal Astronomical
  Society] {10.1093/mnras/stz2238}

\bibitem[\protect\citeauthoryear{{Bacon} et~al.,}{{Bacon} et~al.}{2006}]{muse}
{Bacon} R.,  et~al., 2006, in Society of Photo-Optical Instrumentation
  Engineers (SPIE) Conference Series. p. 62690J (\mn@eprint {arXiv}
  {astro-ph/0606329}), \mn@doi{10.1117/12.669772}

\bibitem[\protect\citeauthoryear{{Bacon}, {Piqueras}, {Conseil}, {Richard}  \&
  {Shepherd}}{{Bacon} et~al.}{2016}]{mpdaf}
{Bacon} R.,  {Piqueras} L.,  {Conseil} S.,  {Richard} J.,   {Shepherd} M.,
  2016, {MPDAF: MUSE Python Data Analysis Framework}, Astrophysics Source Code
  Library (\mn@eprint {ascl} {1611.003})

\bibitem[\protect\citeauthoryear{{Bahcall}}{{Bahcall}}{1996}]{bahcall96}
{Bahcall} N.~A.,  1996, arXiv e-prints, \href
  {https://ui.adsabs.harvard.edu/abs/1996astro.ph.11148B} {pp
  astro--ph/9611148}

\bibitem[\protect\citeauthoryear{{Belfiore} et~al.,}{{Belfiore}
  et~al.}{2017}]{belfiore}
{Belfiore} F.,  et~al., 2017, \mn@doi [\mnras] {10.1093/mnras/stx789}, \href
  {https://ui.adsabs.harvard.edu/abs/2017MNRAS.469..151B} {469, 151}

\bibitem[\protect\citeauthoryear{{Bergeron} \& {Boiss{\'e}}}{{Bergeron} \&
  {Boiss{\'e}}}{1991}]{berbo}
{Bergeron} J.,  {Boiss{\'e}} P.,  1991, \aap, \href
  {https://ui.adsabs.harvard.edu/abs/1991A&A...243..344B} {243, 344}

\bibitem[\protect\citeauthoryear{{Bergeron}, {Boulade}, {Kunth}, {Tytler},
  {Boksenberg}  \& {Vigroux}}{{Bergeron} et~al.}{1988}]{bergeron}
{Bergeron} J.,  {Boulade} O.,  {Kunth} D.,  {Tytler} D.,  {Boksenberg} A.,
  {Vigroux} L.,  1988, \aap, \href
  {https://ui.adsabs.harvard.edu/abs/1988A&A...191....1B} {191, 1}

\bibitem[\protect\citeauthoryear{{Bertin} \& {Arnouts}}{{Bertin} \&
  {Arnouts}}{1996}]{sextractor}
{Bertin} E.,  {Arnouts} S.,  1996, \mn@doi [\aaps] {10.1051/aas:1996164}, \href
  {https://ui.adsabs.harvard.edu/abs/1996A%26AS..117..393B} {117, 393}

\bibitem[\protect\citeauthoryear{{Bielby}, {Crighton}, {Fumagalli}, {Morris},
  {Stott}, {Tejos}  \& {Cantalupo}}{{Bielby} et~al.}{2017}]{bielby17}
{Bielby} R.,  {Crighton} N.~H.~M.,  {Fumagalli} M.,  {Morris} S.~L.,  {Stott}
  J.~P.,  {Tejos} N.,   {Cantalupo} S.,  2017, \mn@doi [\mnras]
  {10.1093/mnras/stx528}, \href
  {https://ui.adsabs.harvard.edu/abs/2017MNRAS.468.1373B} {468, 1373}

\bibitem[\protect\citeauthoryear{{Bielby} et~al.,}{{Bielby}
  et~al.}{2019}]{bielby19}
{Bielby} R.~M.,  et~al., 2019, \mn@doi [\mnras] {10.1093/mnras/stz774}, \href
  {https://ui.adsabs.harvard.edu/abs/2019MNRAS.486...21B} {486, 21}

\bibitem[\protect\citeauthoryear{{Boisse}, {Le Brun}, {Bergeron}  \&
  {Deharveng}}{{Boisse} et~al.}{1998}]{boise98}
{Boisse} P.,  {Le Brun} V.,  {Bergeron} J.,   {Deharveng} J.-M.,  1998, \aap,
  \href {https://ui.adsabs.harvard.edu/abs/1998A&A...333..841B} {333, 841}

\bibitem[\protect\citeauthoryear{{Bordoloi} et~al.,}{{Bordoloi}
  et~al.}{2011}]{bordoloi}
{Bordoloi} R.,  et~al., 2011, \mn@doi [\apj] {10.1088/0004-637X/743/1/10},
  \href {https://ui.adsabs.harvard.edu/abs/2011ApJ...743...10B} {743, 10}

\bibitem[\protect\citeauthoryear{{Borisova} et~al.,}{{Borisova}
  et~al.}{2016}]{borisova}
{Borisova} E.,  et~al., 2016, \mn@doi [\apj] {10.3847/0004-637X/831/1/39},
  \href {https://ui.adsabs.harvard.edu/abs/2016ApJ...831...39B} {831, 39}

\bibitem[\protect\citeauthoryear{{Bouch{\'e}}, {Murphy}, {P{\'e}roux}, {Csabai}
   \& {Wild}}{{Bouch{\'e}} et~al.}{2006}]{bouche06}
{Bouch{\'e}} N.,  {Murphy} M.~T.,  {P{\'e}roux} C.,  {Csabai} I.,   {Wild} V.,
  2006, \mn@doi [\mnras] {10.1111/j.1365-2966.2006.10685.x}, \href
  {https://ui.adsabs.harvard.edu/abs/2006MNRAS.371..495B} {371, 495}

\bibitem[\protect\citeauthoryear{{Bouch{\'e}}, {Murphy}, {P{\'e}roux},
  {Davies}, {Eisenhauer}, {F{\"o}rster Schreiber}  \& {Tacconi}}{{Bouch{\'e}}
  et~al.}{2007}]{bouche}
{Bouch{\'e}} N.,  {Murphy} M.~T.,  {P{\'e}roux} C.,  {Davies} R.,  {Eisenhauer}
  F.,  {F{\"o}rster Schreiber} N.~M.,   {Tacconi} L.,  2007, \mn@doi [\apjl]
  {10.1086/523594}, \href
  {https://ui.adsabs.harvard.edu/abs/2007ApJ...669L...5B} {669, L5}

\bibitem[\protect\citeauthoryear{{Bouch{\'e}} et~al.,}{{Bouch{\'e}}
  et~al.}{2016}]{bouche16}
{Bouch{\'e}} N.,  et~al., 2016, \mn@doi [\apj] {10.3847/0004-637X/820/2/121},
  \href {https://ui.adsabs.harvard.edu/abs/2016ApJ...820..121B} {820, 121}

\bibitem[\protect\citeauthoryear{{Cantalupo}, {Arrigoni-Battaia}, {Prochaska},
  {Hennawi}  \& {Madau}}{{Cantalupo} et~al.}{2014}]{cantalupoNature}
{Cantalupo} S.,  {Arrigoni-Battaia} F.,  {Prochaska} J.~X.,  {Hennawi} J.~F.,
  {Madau} P.,  2014, \mn@doi [\nat] {10.1038/nature12898}, \href
  {https://ui.adsabs.harvard.edu/abs/2014Natur.506...63C} {506, 63}

\bibitem[\protect\citeauthoryear{{Chen}, {Wild}, {Tinker}, {Gauthier},
  {Helsby}, {Shectman}  \& {Thompson}}{{Chen} et~al.}{2010}]{chen}
{Chen} H.-W.,  {Wild} V.,  {Tinker} J.~L.,  {Gauthier} J.-R.,  {Helsby} J.~E.,
  {Shectman} S.~A.,   {Thompson} I.~B.,  2010, \mn@doi [\apjl]
  {10.1088/2041-8205/724/2/L176}, \href
  {https://ui.adsabs.harvard.edu/abs/2010ApJ...724L.176C} {724, L176}

\bibitem[\protect\citeauthoryear{{Christensen}, {M{\o}ller}, {Fynbo}  \&
  {Zafar}}{{Christensen} et~al.}{2014}]{christensen}
{Christensen} L.,  {M{\o}ller} P.,  {Fynbo} J.~P.~U.,   {Zafar} T.,  2014,
  \mn@doi [\mnras] {10.1093/mnras/stu1726}, \href
  {https://ui.adsabs.harvard.edu/abs/2014MNRAS.445..225C} {445, 225}

\bibitem[\protect\citeauthoryear{{Christensen} et~al.,}{{Christensen}
  et~al.}{2017}]{LisaGrb}
{Christensen} L.,  et~al., 2017, \mn@doi [\aap] {10.1051/0004-6361/201731382},
  \href {https://ui.adsabs.harvard.edu/abs/2017A&A...608A..84C} {608, A84}

\bibitem[\protect\citeauthoryear{{Corlies}, {Peeples}, {Tumlinson}, {O'Shea},
  {Lehner}, {Howk}  \& {O'Meara}}{{Corlies} et~al.}{2018}]{corlies}
{Corlies} L.,  {Peeples} M.~S.,  {Tumlinson} J.,  {O'Shea} B.~W.,  {Lehner} N.,
   {Howk} J.~C.,   {O'Meara} J.~M.,  2018, arXiv e-prints, \href
  {https://ui.adsabs.harvard.edu/abs/2018arXiv181105060C} {p. arXiv:1811.05060}

\bibitem[\protect\citeauthoryear{{Eisenhauer} et~al.,}{{Eisenhauer}
  et~al.}{2003}]{sinfoni}
{Eisenhauer} F.,  et~al., 2003, in {Iye} M.,  {Moorwood} A. F.~M.,  eds,
  Society of Photo-Optical Instrumentation Engineers (SPIE) Conference Series
  Vol. 4841, Instrument Design and Performance for Optical/Infrared
  Ground-based Telescopes. pp 1548--1561 (\mn@eprint {arXiv}
  {astro-ph/0306191}), \mn@doi{10.1117/12.459468}

\bibitem[\protect\citeauthoryear{{Ferland} et~al.,}{{Ferland}
  et~al.}{2013}]{cloudy}
{Ferland} G.~J.,  et~al., 2013, \rmxaa, \href
  {https://ui.adsabs.harvard.edu/abs/2013RMxAA..49..137F} {49, 137}

\bibitem[\protect\citeauthoryear{{Fossati} et~al.,}{{Fossati}
  et~al.}{2019}]{fossati}
{Fossati} M.,  et~al., 2019, \mn@doi [\mnras] {10.1093/mnras/stz2693}, \href
  {https://ui.adsabs.harvard.edu/abs/2019MNRAS.490.1451F} {490, 1451}

\bibitem[\protect\citeauthoryear{{Frank} et~al.,}{{Frank} et~al.}{2012}]{frank}
{Frank} S.,  et~al., 2012, \mn@doi [\mnras] {10.1111/j.1365-2966.2011.20172.x},
  \href {https://ui.adsabs.harvard.edu/abs/2012MNRAS.420.1731F} {420, 1731}

\bibitem[\protect\citeauthoryear{{Fraternali}}{{Fraternali}}{2017}]{fraternali}
{Fraternali} F.,  2017, in {Fox} A.,  {Dav{\'e}} R.,  eds,  Astrophysics and
  Space Science Library Vol. 430, Gas Accretion onto Galaxies. p.~323
  (\mn@eprint {arXiv} {1612.00477}), \mn@doi{10.1007/978-3-319-52512-9_14}

\bibitem[\protect\citeauthoryear{{Fumagalli}, {O'Meara}  \&
  {Prochaska}}{{Fumagalli} et~al.}{2016}]{fumagalli2016}
{Fumagalli} M.,  {O'Meara} J.~M.,   {Prochaska} J.~X.,  2016, \mn@doi [\mnras]
  {10.1093/mnras/stv2616}, \href
  {https://ui.adsabs.harvard.edu/abs/2016MNRAS.455.4100F} {455, 4100}

\bibitem[\protect\citeauthoryear{{Fynbo} et~al.,}{{Fynbo}
  et~al.}{2010}]{fynbo10}
{Fynbo} J.~P.~U.,  et~al., 2010, \mn@doi [\mnras]
  {10.1111/j.1365-2966.2010.17294.x}, \href
  {https://ui.adsabs.harvard.edu/abs/2010MNRAS.408.2128F} {408, 2128}

\bibitem[\protect\citeauthoryear{{Fynbo} et~al.,}{{Fynbo}
  et~al.}{2013}]{fynbo13}
{Fynbo} J.~P.~U.,  et~al., 2013, \mn@doi [\mnras] {10.1093/mnras/stt1579},
  \href {https://ui.adsabs.harvard.edu/abs/2013MNRAS.436..361F} {436, 361}

\bibitem[\protect\citeauthoryear{{Ho}}{{Ho}}{2016}]{ho16}
{Ho} I.~T.,  2016, PhD thesis, University of Hawai'i at Manoa

\bibitem[\protect\citeauthoryear{{Husemann}, {Bennert}, {Scharw{\"a}chter},
  {Woo}  \& {Choudhury}}{{Husemann} et~al.}{2016}]{brend}
{Husemann} B.,  {Bennert} V.~N.,  {Scharw{\"a}chter} J.,  {Woo} J.~H.,
  {Choudhury} O.~S.,  2016, \mn@doi [\mnras] {10.1093/mnras/stv2478}, \href
  {https://ui.adsabs.harvard.edu/abs/2016MNRAS.455.1905H} {455, 1905}

\bibitem[\protect\citeauthoryear{{Jones}, {Ellis}, {Richard}  \&
  {Jullo}}{{Jones} et~al.}{2013}]{jones13}
{Jones} T.,  {Ellis} R.~S.,  {Richard} J.,   {Jullo} E.,  2013, \mn@doi [\apj]
  {10.1088/0004-637X/765/1/48}, \href
  {https://ui.adsabs.harvard.edu/abs/2013ApJ...765...48J} {765, 48}

\bibitem[\protect\citeauthoryear{{Jones} et~al.,}{{Jones}
  et~al.}{2015}]{jones15}
{Jones} T.,  et~al., 2015, \mn@doi [\aj] {10.1088/0004-6256/149/3/107}, \href
  {https://ui.adsabs.harvard.edu/abs/2015AJ....149..107J} {149, 107}

\bibitem[\protect\citeauthoryear{{Kacprzak}, {Murphy}  \&
  {Churchill}}{{Kacprzak} et~al.}{2010}]{kacprzak10}
{Kacprzak} G.~G.,  {Murphy} M.~T.,   {Churchill} C.~W.,  2010, \mn@doi [\mnras]
  {10.1111/j.1365-2966.2010.16667.x}, \href
  {https://ui.adsabs.harvard.edu/abs/2010MNRAS.406..445K} {406, 445}

\bibitem[\protect\citeauthoryear{{Kacprzak}, {Churchill}, {Evans}, {Murphy}  \&
  {Steidel}}{{Kacprzak} et~al.}{2011}]{kacprzak11}
{Kacprzak} G.~G.,  {Churchill} C.~W.,  {Evans} J.~L.,  {Murphy} M.~T.,
  {Steidel} C.~C.,  2011, \mn@doi [\mnras] {10.1111/j.1365-2966.2011.19261.x},
  \href {https://ui.adsabs.harvard.edu/abs/2011MNRAS.416.3118K} {416, 3118}

\bibitem[\protect\citeauthoryear{{Klitsch}, {P{\'e}roux}, {Zwaan}, {Smail},
  {Oteo}, {Biggs}, {Popping}  \& {Swinbank}}{{Klitsch} et~al.}{2018}]{klitsch}
{Klitsch} A.,  {P{\'e}roux} C.,  {Zwaan} M.~A.,  {Smail} I.,  {Oteo} I.,
  {Biggs} A.~D.,  {Popping} G.,   {Swinbank} A.~M.,  2018, \mn@doi [\mnras]
  {10.1093/mnras/stx3184}, \href
  {https://ui.adsabs.harvard.edu/abs/2018MNRAS.475..492K} {475, 492}

\bibitem[\protect\citeauthoryear{{Kobulnicky}, {Kennicutt}  \&
  {Pizagno}}{{Kobulnicky} et~al.}{1999}]{kobulnicky}
{Kobulnicky} H.~A.,  {Kennicutt} Robert~C. J.,   {Pizagno} J.~L.,  1999,
  \mn@doi [\apj] {10.1086/306987}, \href
  {https://ui.adsabs.harvard.edu/abs/1999ApJ...514..544K} {514, 544}

\bibitem[\protect\citeauthoryear{{Krogager}, {M{\o}ller}, {Fynbo}  \&
  {Noterdaeme}}{{Krogager} et~al.}{2017}]{krogager17}
{Krogager} J.~K.,  {M{\o}ller} P.,  {Fynbo} J.~P.~U.,   {Noterdaeme} P.,  2017,
  \mn@doi [\mnras] {10.1093/mnras/stx1011}, \href
  {https://ui.adsabs.harvard.edu/abs/2017MNRAS.469.2959K} {469, 2959}

\bibitem[\protect\citeauthoryear{{Kulkarni}, {Cashman}, {Lopez}, {Ellison},
  {Som}  \& {Jos{\'e} Maureira}}{{Kulkarni} et~al.}{2019}]{Kulkarni}
{Kulkarni} V.~P.,  {Cashman} F.~H.,  {Lopez} S.,  {Ellison} S.~L.,  {Som} D.,
  {Jos{\'e} Maureira} M.,  2019, arXiv e-prints, \href
  {https://ui.adsabs.harvard.edu/abs/2019arXiv191010759K} {p. arXiv:1910.10759}

\bibitem[\protect\citeauthoryear{{Lane}, {Briggs}, {Turnshek}  \& {Rao}}{{Lane}
  et~al.}{1998}]{lane19}
{Lane} W.~M.,  {Briggs} F.~H.,  {Turnshek} D.~A.,   {Rao} S.~M.,  1998, in
  American Astronomical Society Meeting Abstracts. p. 04.09

\bibitem[\protect\citeauthoryear{{Lanzetta} \& {Bowen}}{{Lanzetta} \&
  {Bowen}}{1990}]{lanzetta}
{Lanzetta} K.~M.,  {Bowen} D.,  1990, \mn@doi [\apj] {10.1086/168922}, \href
  {https://ui.adsabs.harvard.edu/abs/1990ApJ...357..321L} {357, 321}

\bibitem[\protect\citeauthoryear{{Larkin} et~al.,}{{Larkin}
  et~al.}{2006}]{osiris}
{Larkin} J.,  et~al., 2006, in Society of Photo-Optical Instrumentation
  Engineers (SPIE) Conference Series. p. 62691A, \mn@doi{10.1117/12.672061}

\bibitem[\protect\citeauthoryear{{Le Brun}, {Bergeron}, {Boisse}  \&
  {Deharveng}}{{Le Brun} et~al.}{1997}]{lebrun}
{Le Brun} V.,  {Bergeron} J.,  {Boisse} P.,   {Deharveng} J.~M.,  1997, \aap,
  \href {https://ui.adsabs.harvard.edu/abs/1997A&A...321..733L} {321, 733}

\bibitem[\protect\citeauthoryear{{Leethochawalit}, {Jones}, {Ellis}, {Stark},
  {Richard}, {Zitrin}  \& {Auger}}{{Leethochawalit} et~al.}{2016}]{lee16}
{Leethochawalit} N.,  {Jones} T.~A.,  {Ellis} R.~S.,  {Stark} D.~P.,  {Richard}
  J.,  {Zitrin} A.,   {Auger} M.,  2016, \mn@doi [\apj]
  {10.3847/0004-637X/820/2/84}, \href
  {https://ui.adsabs.harvard.edu/abs/2016ApJ...820...84L} {820, 84}

\bibitem[\protect\citeauthoryear{{Lehner} et~al.,}{{Lehner}
  et~al.}{2013}]{lehner}
{Lehner} N.,  et~al., 2013, \mn@doi [\apj] {10.1088/0004-637X/770/2/138}, \href
  {https://ui.adsabs.harvard.edu/abs/2013ApJ...770..138L} {770, 138}

\bibitem[\protect\citeauthoryear{{Lofthouse} et~al.,}{{Lofthouse}
  et~al.}{2019}]{magg}
{Lofthouse} E.~K.,  et~al., 2019, arXiv e-prints, \href
  {https://ui.adsabs.harvard.edu/abs/2019arXiv191013458L} {p. arXiv:1910.13458}

\bibitem[\protect\citeauthoryear{{Lusso} et~al.,}{{Lusso} et~al.}{2019}]{lusso}
{Lusso} E.,  et~al., 2019, \mn@doi [\mnras] {10.1093/mnrasl/slz032}, \href
  {https://ui.adsabs.harvard.edu/abs/2019MNRAS.485L..62L} {485, L62}

\bibitem[\protect\citeauthoryear{{Ly} et~al.,}{{Ly} et~al.}{2007}]{ly}
{Ly} C.,  et~al., 2007, \mn@doi [\apj] {10.1086/510828}, \href
  {https://ui.adsabs.harvard.edu/abs/2007ApJ...657..738L} {657, 738}

\bibitem[\protect\citeauthoryear{{Mackenzie} et~al.,}{{Mackenzie}
  et~al.}{2019}]{mackenzie}
{Mackenzie} R.,  et~al., 2019, arXiv e-prints, \href
  {https://ui.adsabs.harvard.edu/abs/2019arXiv190407254M} {p. arXiv:1904.07254}

\bibitem[\protect\citeauthoryear{{Martin}, {Shapley}, {Coil}, {Kornei},
  {Bundy}, {Weiner}, {Noeske}  \& {Schiminovich}}{{Martin}
  et~al.}{2012}]{martin}
{Martin} C.~L.,  {Shapley} A.~E.,  {Coil} A.~L.,  {Kornei} K.~A.,  {Bundy} K.,
  {Weiner} B.~J.,  {Noeske} K.~G.,   {Schiminovich} D.,  2012, \mn@doi [\apj]
  {10.1088/0004-637X/760/2/127}, \href
  {https://ui.adsabs.harvard.edu/abs/2012ApJ...760..127M} {760, 127}

\bibitem[\protect\citeauthoryear{{M{\o}ller} \& {Warren}}{{M{\o}ller} \&
  {Warren}}{1993}]{moller93}
{M{\o}ller} P.,  {Warren} S.~J.,  1993, \aap, \href
  {https://ui.adsabs.harvard.edu/abs/1993A&A...270...43M} {270, 43}

\bibitem[\protect\citeauthoryear{{M{\o}ller}, {Fynbo}  \& {Fall}}{{M{\o}ller}
  et~al.}{2004}]{moller04}
{M{\o}ller} P.,  {Fynbo} J.~P.~U.,   {Fall} S.~M.,  2004, \mn@doi [\aap]
  {10.1051/0004-6361:20040194}, \href
  {https://ui.adsabs.harvard.edu/abs/2004A&A...422L..33M} {422, L33}

\bibitem[\protect\citeauthoryear{{Muratov}, {Kere{\v{s}}},
  {Faucher-Gigu{\`e}re}, {Hopkins}, {Quataert}  \& {Murray}}{{Muratov}
  et~al.}{2015}]{muratov15}
{Muratov} A.~L.,  {Kere{\v{s}}} D.,  {Faucher-Gigu{\`e}re} C.-A.,  {Hopkins}
  P.~F.,  {Quataert} E.,   {Murray} N.,  2015, \mn@doi [\mnras]
  {10.1093/mnras/stv2126}, \href
  {https://ui.adsabs.harvard.edu/abs/2015MNRAS.454.2691M} {454, 2691}

\bibitem[\protect\citeauthoryear{{Muratov} et~al.,}{{Muratov}
  et~al.}{2017}]{muratov17}
{Muratov} A.~L.,  et~al., 2017, \mn@doi [\mnras] {10.1093/mnras/stx667}, \href
  {https://ui.adsabs.harvard.edu/abs/2017MNRAS.468.4170M} {468, 4170}

\bibitem[\protect\citeauthoryear{{Muzahid}, {Kacprzak}, {Charlton}  \&
  {Churchill}}{{Muzahid} et~al.}{2016}]{muzahid16}
{Muzahid} S.,  {Kacprzak} G.~G.,  {Charlton} J.~C.,   {Churchill} C.~W.,  2016,
  \mn@doi [\apj] {10.3847/0004-637X/823/1/66}, \href
  {https://ui.adsabs.harvard.edu/abs/2016ApJ...823...66M} {823, 66}

\bibitem[\protect\citeauthoryear{{Muzahid} et~al.,}{{Muzahid}
  et~al.}{2019}]{muzahid19}
{Muzahid} S.,  et~al., 2019, arXiv e-prints, \href
  {https://ui.adsabs.harvard.edu/abs/2019arXiv191003593M} {p. arXiv:1910.03593}

\bibitem[\protect\citeauthoryear{{Nelson} et~al.,}{{Nelson}
  et~al.}{2019}]{nelson}
{Nelson} D.,  et~al., 2019, arXiv e-prints, \href
  {https://ui.adsabs.harvard.edu/abs/2019arXiv190205554N} {p. arXiv:1902.05554}

\bibitem[\protect\citeauthoryear{{Nielsen}, {Churchill}, {Kacprzak}  \&
  {Murphy}}{{Nielsen} et~al.}{2013}]{nielsen13}
{Nielsen} N.~M.,  {Churchill} C.~W.,  {Kacprzak} G.~G.,   {Murphy} M.~T.,
  2013, \mn@doi [\apj] {10.1088/0004-637X/776/2/114}, \href
  {https://ui.adsabs.harvard.edu/abs/2013ApJ...776..114N} {776, 114}

\bibitem[\protect\citeauthoryear{{Nielsen}, {Kacprzak}, {Pointon}, {Churchill}
  \& {Murphy}}{{Nielsen} et~al.}{2018}]{nielsen18}
{Nielsen} N.~M.,  {Kacprzak} G.~G.,  {Pointon} S.~K.,  {Churchill} C.~W.,
  {Murphy} M.~T.,  2018, \mn@doi [\apj] {10.3847/1538-4357/aaedbd}, \href
  {https://ui.adsabs.harvard.edu/abs/2018ApJ...869..153N} {869, 153}

\bibitem[\protect\citeauthoryear{{P{\'e}roux}, {Bouch{\'e}}, {Kulkarni}, {York}
   \& {Vladilo}}{{P{\'e}roux} et~al.}{2011}]{peroux11}
{P{\'e}roux} C.,  {Bouch{\'e}} N.,  {Kulkarni} V.~P.,  {York} D.~G.,
  {Vladilo} G.,  2011, \mn@doi [\mnras] {10.1111/j.1365-2966.2010.17598.x},
  \href {https://ui.adsabs.harvard.edu/abs/2011MNRAS.410.2237P} {410, 2237}

\bibitem[\protect\citeauthoryear{{P{\'e}roux}, {Bouch{\'e}}, {Kulkarni}, {York}
   \& {Vladilo}}{{P{\'e}roux} et~al.}{2012}]{peroux12}
{P{\'e}roux} C.,  {Bouch{\'e}} N.,  {Kulkarni} V.~P.,  {York} D.~G.,
  {Vladilo} G.,  2012, \mn@doi [\mnras] {10.1111/j.1365-2966.2011.19947.x},
  \href {https://ui.adsabs.harvard.edu/abs/2012MNRAS.419.3060P} {419, 3060}

\bibitem[\protect\citeauthoryear{{P{\'e}roux} et~al.,}{{P{\'e}roux}
  et~al.}{2016}]{peroux16}
{P{\'e}roux} C.,  et~al., 2016, \mn@doi [\mnras] {10.1093/mnras/stw016}, \href
  {https://ui.adsabs.harvard.edu/abs/2016MNRAS.457..903P} {457, 903}

\bibitem[\protect\citeauthoryear{{P{\'e}roux} et~al.,}{{P{\'e}roux}
  et~al.}{2017}]{peroux17}
{P{\'e}roux} C.,  et~al., 2017, \mn@doi [\mnras] {10.1093/mnras/stw2444}, \href
  {https://ui.adsabs.harvard.edu/abs/2017MNRAS.464.2053P} {464, 2053}

\bibitem[\protect\citeauthoryear{{P{\'e}roux} et~al.,}{{P{\'e}roux}
  et~al.}{2019}]{peroux19}
{P{\'e}roux} C.,  et~al., 2019, \mn@doi [\mnras] {10.1093/mnras/stz202}, \href
  {https://ui.adsabs.harvard.edu/abs/2019MNRAS.485.1595P} {485, 1595}

\bibitem[\protect\citeauthoryear{{Prochaska} et~al.,}{{Prochaska}
  et~al.}{2017}]{prochaska17}
{Prochaska} J.~X.,  et~al., 2017, \mn@doi [\apj] {10.3847/1538-4357/aa6007},
  \href {https://ui.adsabs.harvard.edu/abs/2017ApJ...837..169P} {837, 169}

\bibitem[\protect\citeauthoryear{{Queyrel} et~al.,}{{Queyrel}
  et~al.}{2012}]{queyrel}
{Queyrel} J.,  et~al., 2012, \mn@doi [\aap] {10.1051/0004-6361/201117718},
  \href {https://ui.adsabs.harvard.edu/abs/2012A&A...539A..93Q} {539, A93}

\bibitem[\protect\citeauthoryear{{Quiret} et~al.,}{{Quiret}
  et~al.}{2016}]{quiret}
{Quiret} S.,  et~al., 2016, \mn@doi [\mnras] {10.1093/mnras/stw524}, \href
  {https://ui.adsabs.harvard.edu/abs/2016MNRAS.458.4074Q} {458, 4074}

\bibitem[\protect\citeauthoryear{{Rahmani} et~al.,}{{Rahmani}
  et~al.}{2016}]{rahmani16}
{Rahmani} H.,  et~al., 2016, \mn@doi [\mnras] {10.1093/mnras/stw1965}, \href
  {https://ui.adsabs.harvard.edu/abs/2016MNRAS.463..980R} {463, 980}

\bibitem[\protect\citeauthoryear{{Rahmani} et~al.,}{{Rahmani}
  et~al.}{2018a}]{rahmani18disk}
{Rahmani} H.,  et~al., 2018a, \mn@doi [\mnras] {10.1093/mnras/stx2726}, \href
  {https://ui.adsabs.harvard.edu/abs/2018MNRAS.474..254R} {474, 254}

\bibitem[\protect\citeauthoryear{{Rahmani} et~al.,}{{Rahmani}
  et~al.}{2018b}]{rahmani18wind}
{Rahmani} H.,  et~al., 2018b, \mn@doi [\mnras] {10.1093/mnras/sty2216}, \href
  {https://ui.adsabs.harvard.edu/abs/2018MNRAS.480.5046R} {480, 5046}

\bibitem[\protect\citeauthoryear{{Rahmati} \& {Schaye}}{{Rahmati} \&
  {Schaye}}{2014}]{rahmatyi}
{Rahmati} A.,  {Schaye} J.,  2014, \mn@doi [\mnras] {10.1093/mnras/stt2235},
  \href {https://ui.adsabs.harvard.edu/abs/2014MNRAS.438..529R} {438, 529}

\bibitem[\protect\citeauthoryear{{Rao}, {Turnshek}  \& {Nestor}}{{Rao}
  et~al.}{2006}]{rao06}
{Rao} S.~M.,  {Turnshek} D.~A.,   {Nestor} D.~B.,  2006, \mn@doi [\apj]
  {10.1086/498132}, \href
  {https://ui.adsabs.harvard.edu/abs/2006ApJ...636..610R} {636, 610}

\bibitem[\protect\citeauthoryear{{Rao}, {Belfort-Mihalyi}, {Turnshek},
  {Monier}, {Nestor}  \& {Quider}}{{Rao} et~al.}{2011}]{rao}
{Rao} S.~M.,  {Belfort-Mihalyi} M.,  {Turnshek} D.~A.,  {Monier} E.~M.,
  {Nestor} D.~B.,   {Quider} A.,  2011, \mn@doi [\mnras]
  {10.1111/j.1365-2966.2011.19119.x}, \href
  {https://ui.adsabs.harvard.edu/abs/2011MNRAS.416.1215R} {416, 1215}

\bibitem[\protect\citeauthoryear{{Rubin}, {Prochaska}, {Koo}  \&
  {Phillips}}{{Rubin} et~al.}{2012}]{rubin}
{Rubin} K.~H.~R.,  {Prochaska} J.~X.,  {Koo} D.~C.,   {Phillips} A.~C.,  2012,
  \mn@doi [\apjl] {10.1088/2041-8205/747/2/L26}, \href
  {https://ui.adsabs.harvard.edu/abs/2012ApJ...747L..26R} {747, L26}

\bibitem[\protect\citeauthoryear{{S{\'a}nchez} et~al.,}{{S{\'a}nchez}
  et~al.}{2014}]{sanchez14}
{S{\'a}nchez} S.~F.,  et~al., 2014, \mn@doi [\aap]
  {10.1051/0004-6361/201322343}, \href
  {https://ui.adsabs.harvard.edu/abs/2014A&A...563A..49S} {563, A49}

\bibitem[\protect\citeauthoryear{{Schroetter} et~al.,}{{Schroetter}
  et~al.}{2016}]{schroetter}
{Schroetter} I.,  et~al., 2016, \mn@doi [\apj] {10.3847/1538-4357/833/1/39},
  \href {https://ui.adsabs.harvard.edu/abs/2016ApJ...833...39S} {833, 39}

\bibitem[\protect\citeauthoryear{{Schroetter} et~al.,}{{Schroetter}
  et~al.}{2019}]{schro19}
{Schroetter} I.,  et~al., 2019, arXiv e-prints, \href
  {https://ui.adsabs.harvard.edu/abs/2019arXiv190709967S} {p. arXiv:1907.09967}

\bibitem[\protect\citeauthoryear{{Selsing}, {Fynbo}, {Christensen}  \&
  {Krogager}}{{Selsing} et~al.}{2016}]{Selsing2016}
{Selsing} J.,  {Fynbo} J.~P.~U.,  {Christensen} L.,   {Krogager} J.-K.,  2016,
  \mn@doi [\aap] {10.1051/0004-6361/201527096}, \href
  {http://adsabs.harvard.edu/abs/2016A%26A...585A..87S} {585, A87}

\bibitem[\protect\citeauthoryear{{Shull}, {Danforth}  \& {Tilton}}{{Shull}
  et~al.}{2014}]{shull}
{Shull} J.~M.,  {Danforth} C.~W.,   {Tilton} E.~M.,  2014, \mn@doi [\apj]
  {10.1088/0004-637X/796/1/49}, \href
  {https://ui.adsabs.harvard.edu/abs/2014ApJ...796...49S} {796, 49}

\bibitem[\protect\citeauthoryear{{Steidel}, {Dickinson}  \&
  {Persson}}{{Steidel} et~al.}{1994}]{Steidel}
{Steidel} C.~C.,  {Dickinson} M.,   {Persson} S.~E.,  1994, \mn@doi [\apjl]
  {10.1086/187686}, \href
  {https://ui.adsabs.harvard.edu/abs/1994ApJ...437L..75S} {437, L75}

\bibitem[\protect\citeauthoryear{{Tumlinson}, {Peeples}  \& {Werk}}{{Tumlinson}
  et~al.}{2017}]{tumlinson}
{Tumlinson} J.,  {Peeples} M.~S.,   {Werk} J.~K.,  2017, \mn@doi [\araa]
  {10.1146/annurev-astro-091916-055240}, \href
  {https://ui.adsabs.harvard.edu/abs/2017ARA%26A..55..389T} {55, 389}

\bibitem[\protect\citeauthoryear{{Umehata} et~al.,}{{Umehata}
  et~al.}{2019}]{umehata}
{Umehata} H.,  et~al., 2019, \mn@doi [Science] {10.1126/science.aaw5949}, \href
  {https://ui.adsabs.harvard.edu/abs/2019Sci...366...97U} {366, 97}

\bibitem[\protect\citeauthoryear{{Weilbacher}, {Streicher}  \&
  {Palsa}}{{Weilbacher} et~al.}{2016}]{musepipe}
{Weilbacher} P.~M.,  {Streicher} O.,   {Palsa} R.,  2016, {MUSE-DRP: MUSE Data
  Reduction Pipeline} (\mn@eprint {ascl} {1610.004})

\bibitem[\protect\citeauthoryear{{Whiting}, {Webster}  \& {Francis}}{{Whiting}
  et~al.}{2006}]{whiting}
{Whiting} M.~T.,  {Webster} R.~L.,   {Francis} P.~J.,  2006, \mn@doi [\mnras]
  {10.1111/j.1365-2966.2006.10101.x}, \href
  {https://ui.adsabs.harvard.edu/abs/2006MNRAS.368..341W} {368, 341}

\bibitem[\protect\citeauthoryear{{Wisotzki} et~al.,}{{Wisotzki}
  et~al.}{2018}]{wisotzki}
{Wisotzki} L.,  et~al., 2018, \mn@doi [\nat] {10.1038/s41586-018-0564-6}, \href
  {https://ui.adsabs.harvard.edu/abs/2018Natur.562..229W} {562, 229}

\bibitem[\protect\citeauthoryear{{Wotta}, {Lehner}, {Howk}, {O'Meara},
  {Oppenheimer}  \& {Cooksey}}{{Wotta} et~al.}{2019a}]{wotta}
{Wotta} C.~B.,  {Lehner} N.,  {Howk} J.~C.,  {O'Meara} J.~M.,  {Oppenheimer}
  B.~D.,   {Cooksey} K.~L.,  2019a, \mn@doi [\apj] {10.3847/1538-4357/aafb74},
  \href {https://ui.adsabs.harvard.edu/abs/2019ApJ...872...81W} {872, 81}

\bibitem[\protect\citeauthoryear{{Wotta}, {Lehner}, {Howk}, {O'Meara},
  {Oppenheimer}  \& {Cooksey}}{{Wotta} et~al.}{2019b}]{wotta2019}
{Wotta} C.~B.,  {Lehner} N.,  {Howk} J.~C.,  {O'Meara} J.~M.,  {Oppenheimer}
  B.~D.,   {Cooksey} K.~L.,  2019b, \mn@doi [\apj] {10.3847/1538-4357/aafb74},
  \href {https://ui.adsabs.harvard.edu/abs/2019ApJ...872...81W} {872, 81}

\bibitem[\protect\citeauthoryear{{Yanny}, {Hamilton}, {Schommer}, {Williams}
  \& {York}}{{Yanny} et~al.}{1987}]{yanny87}
{Yanny} B.,  {Hamilton} D.,  {Schommer} R.~A.,  {Williams} T.~B.,   {York}
  D.~G.,  1987, \mn@doi [\apjl] {10.1086/185049}, \href
  {https://ui.adsabs.harvard.edu/abs/1987ApJ...323L..19Y} {323, L19}

\bibitem[\protect\citeauthoryear{{Yanny}, {York}  \& {Gallagher}}{{Yanny}
  et~al.}{1989}]{yanny89}
{Yanny} B.,  {York} D.~G.,   {Gallagher} J.~S.,  1989, \mn@doi [\apj]
  {10.1086/167231}, \href
  {https://ui.adsabs.harvard.edu/abs/1989ApJ...338..735Y} {338, 735}

\bibitem[\protect\citeauthoryear{{Yanny}, {York}  \& {Williams}}{{Yanny}
  et~al.}{1990a}]{yanny90}
{Yanny} B.,  {York} D.~G.,   {Williams} T.~B.,  1990a, \mn@doi [\apj]
  {10.1086/168473}, \href
  {https://ui.adsabs.harvard.edu/abs/1990ApJ...351..377Y} {351, 377}

\bibitem[\protect\citeauthoryear{{Yanny}, {Barden}, {Gallagher}  \&
  {York}}{{Yanny} et~al.}{1990b}]{yanny}
{Yanny} B.,  {Barden} S.,  {Gallagher} John~S. I.,   {York} D.~G.,  1990b,
  \mn@doi [\apj] {10.1086/168547}, \href
  {https://ui.adsabs.harvard.edu/abs/1990ApJ...352..413Y} {352, 413}

\bibitem[\protect\citeauthoryear{{Zabl} et~al.,}{{Zabl} et~al.}{2019}]{zabl19}
{Zabl} J.,  et~al., 2019, \mn@doi [\mnras] {10.1093/mnras/stz392}, \href
  {https://ui.adsabs.harvard.edu/abs/2019MNRAS.485.1961Z} {485, 1961}

\bibitem[\protect\citeauthoryear{{Zafar}, {Popping}  \& {P{\'e}roux}}{{Zafar}
  et~al.}{2013}]{zafar13}
{Zafar} T.,  {Popping} A.,   {P{\'e}roux} C.,  2013, \mn@doi [\aap]
  {10.1051/0004-6361/201321153}, \href
  {https://ui.adsabs.harvard.edu/abs/2013A&A...556A.140Z} {556, A140}

\makeatother
\end{thebibliography}



\appendix
\section{Dust extinction}
\label{appendix:dust}
To test if dust depletion of metals in absorbers in our sample is significant we fit a spectral template to the quasar spectra extracted from the MUSE cubes. We furthermore include optical ($g$, $r$, $i$, $z$, and $y$ and near-infrared photometry from PANSTARRS and 2MASS. We exclude photometric bands that are affected by strong emission lines (such as \CIV\ and H$\alpha$). Similarly, we exclude from the fit regions of the spectra that are affected by broad emission lines.
The spectra are fitted using the quasar template spectrum by \citet{Selsing2016} including two free parameters: the rest-frame V-band reddening, A(V), and a variable intrinsic power-law index applied as an offset, $\Delta\beta$, relative to the template power-law index of $\beta_{\lambda}=-1.7$. This change in power-law index allows us to assess the impact of the unknown intrinsic quasar shape and propagate this uncertainty into the reddening estimate. We apply the reddening in the rest-frame of the lowest-redshift absorber (if more than one is present along the line-of-sight) as this will give the most conservative estimate of the reddening; less reddening is required for higher redshift absorbers. Lastly, the spectral template is normalized to match the observed flux level by including the nuisance parameter, $f_0$. 
In most cases the best-fit models are in good agreement with the data. We conclude that a low amount of dust reddening is the most probable solution.

\section{Galaxy properties}
\label{appendix:spectra}

We present HST WFPC2 images with positions of galaxies associated with the absorbers described in this work (Figures A1 - A6), as well as their spectra and continuum detection on MUSE white light images (A7 - A14). MUSE white light and HST images for each quasar field present all detected galaxies associated with absorbers in our sample, differentiated by colours. Both images are in the same scale frame and oriented North up, East to the left. For clarity, we show the Q1130z031 rich group members on a separate figure \citep[for details on this system see][]{peroux19}.

Each spectrum panel gives the plot of the MUSE spectra of the galaxies from 4800 \AA\ to 9300 \AA\ except for the already published MUSE studies of Q0152z038 \citep{rahmani18disk}, Q0152z078 \citep{rahmani18wind}, Q1130z031 \citep{peroux19} and Q2131z043 \citep{peroux17}. Emission lines in the galaxy are marked in red, as are the absorption lines of interstellar \CaII H\&K from the galaxy. The spectrum of sky lines before substraction is plotted in red. In the upper left-hand corner of each galaxy spectrum frame appears the identifier of the galaxy (first four digits of QSO right ascension, two decimal points of the absorber redshift (eg. z019 for z=0.19), and the galaxy ID (e.g., G1)) with which the galaxy spectrum is associated; and the impact parameter (in kpc) of the QSO field images (Figures A1 to A6) projected onto the plane perpendicular to the sightline of the QSO. Two postage stamps of the galaxy appear to the right of this label, above the spectra (from MUSE and from the WFPC on HST). If the HST inset is not present, the particular galaxy is outside the field of view of HST WFPC2 camera. To the right of the galaxy spectrum is the spectrum of \MgII\ $\lambda$2796 or \FeII\ $\lambda$2600 for the absorption system noted in the galaxy spectrum frame (in rest-frame velocity units). The blue line is the normalized UVES/HIRES quasar spectrum, red is a Voigt profile fit to the line. Because of the velocity span of some of the galaxies, not all panels are presented on the same velocity scale. We do not present among the spectrum figures the spectra of the galaxies extracted from the quasar PSF (Q1211z062-G1 and Q1232z039-G1), because we were only able to retrieve the [\OII] emission line with false continuum; they are presented in Fig.\ref{fig:qsores}. 

In the QSO fields (Figures A1 to A6), each galaxy associated with an absorber is marked with a circle, colour-coded by the absorber ID. The name of each absorber and a matching coloured dot matching appears in the upper-left-hand or lower-right-hand corner of the images of the QSO fields. 

\begin{figure*}
\subfloat{\includegraphics[width=\columnwidth]{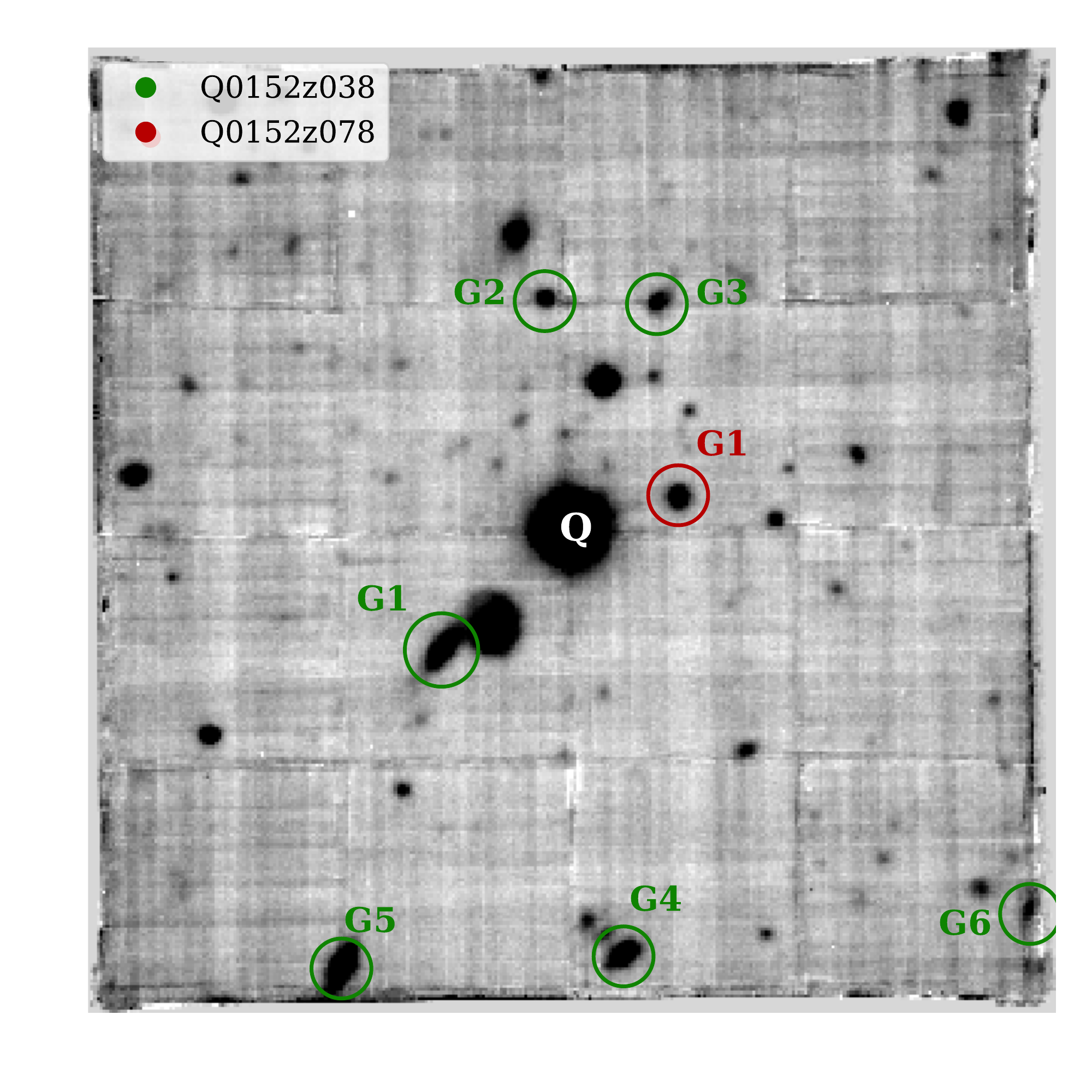}}
\subfloat{\includegraphics[width=\columnwidth]{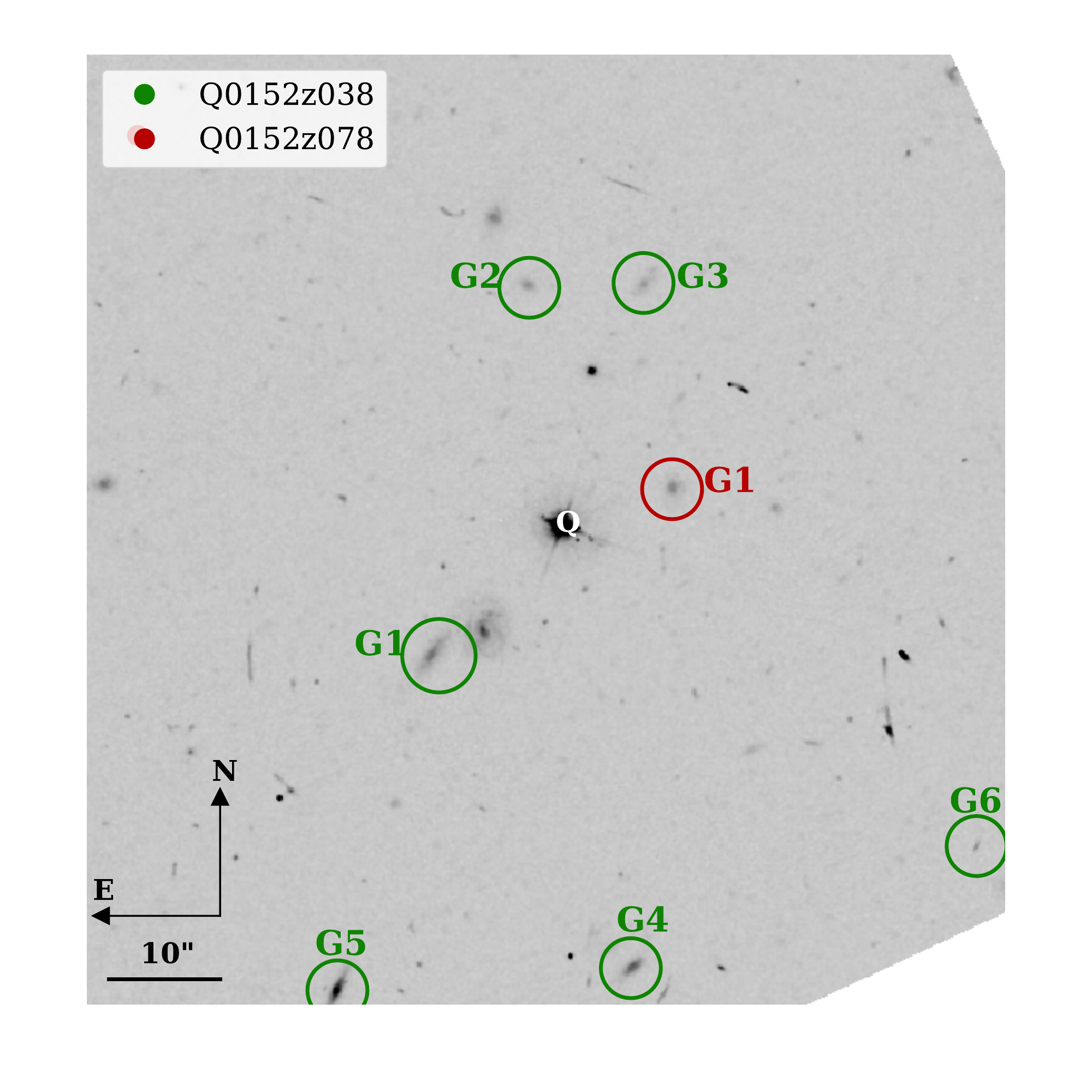}}
\caption{MUSE white light (left) and HST F702W (right) images of Q0152-2001 quasar field with galaxies associated to all absorbers marked. Green circles mark galaxies associated with Q0152z038 absorber, red - Q0152z078 absorber. "Q" indicates the quasar.}
\label{fig:Q0152}
\end{figure*}

\begin{figure*} 
\includegraphics[width=2\columnwidth]{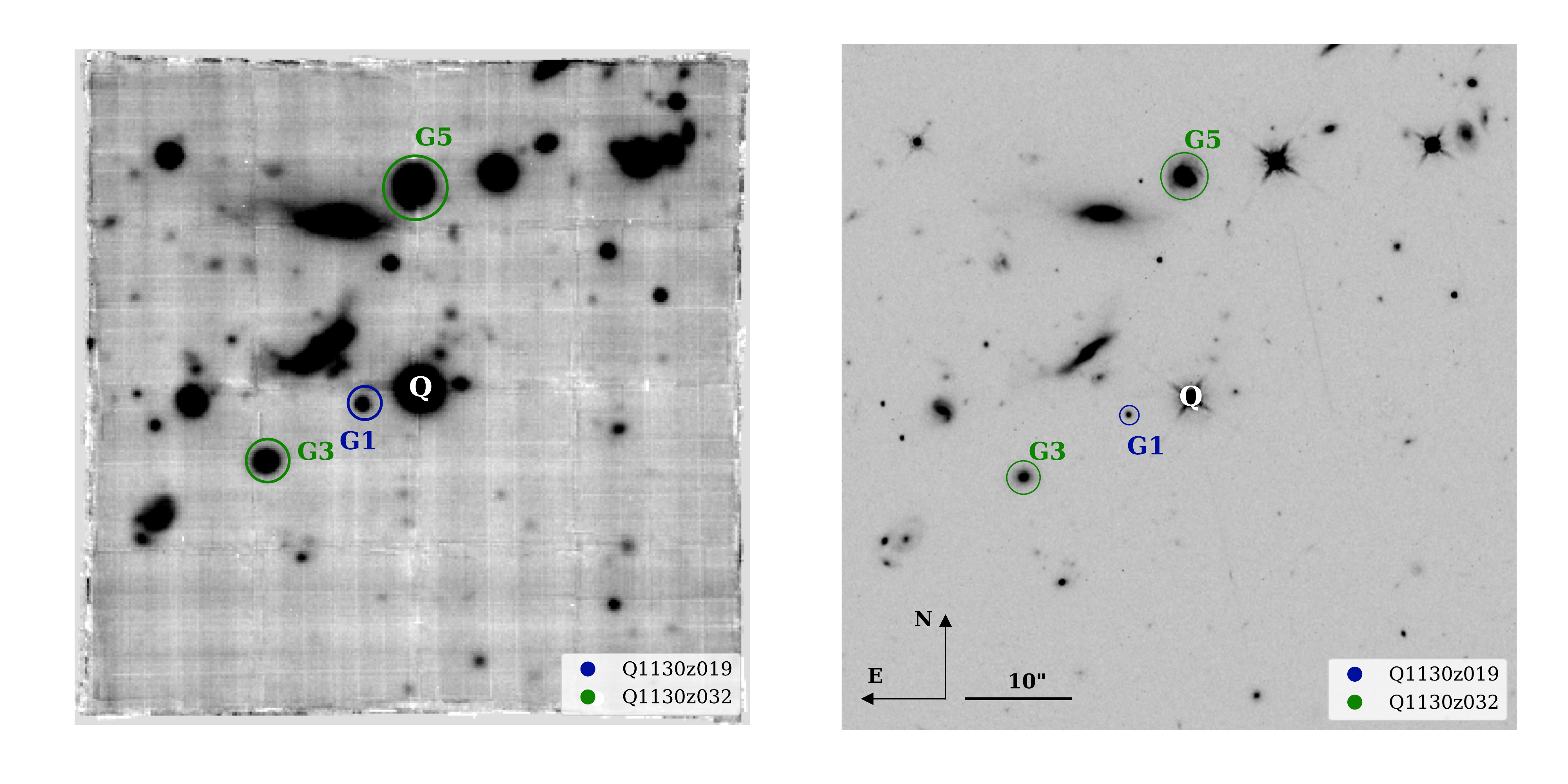}
\caption{MUSE white light (left) and HST F104W IR (right) images of Q1130-1449 quasar field with galaxies associated to two absorbers marked. The blue circle marks the galaxy associated with Q1130z019 absorber, green circle with Q1130z032 absorber."Q" indicates the quasar. The third absorber in this QSO, Q1130z031, is shown on Fig.\ref{fig:Q1130z031}}
\label{fig:Q1130}
\end{figure*}

\begin{figure*} 
\includegraphics[width=2\columnwidth]{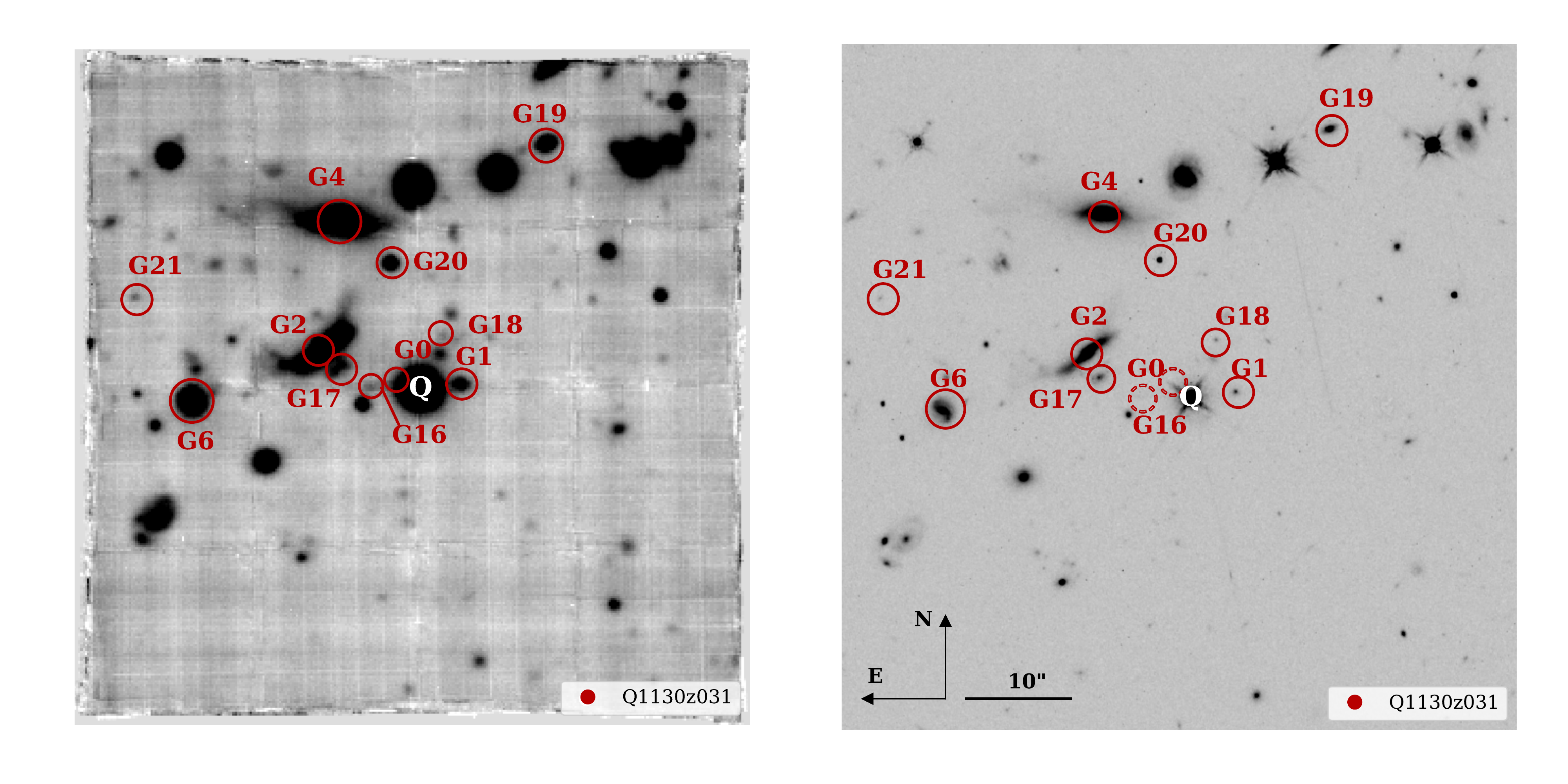}
\caption{MUSE white light (left) and HST F104W IR (right) images of Q1130-1449 quasar field with galaxies associated to Q1130z031. Positions of galaxies not detected on the HST image (G0, G16) are marked with dashed circles."Q" indicates the quasar.}
\label{fig:Q1130z031}
\end{figure*}

\begin{figure*}
\subfloat{\includegraphics[width=\columnwidth]{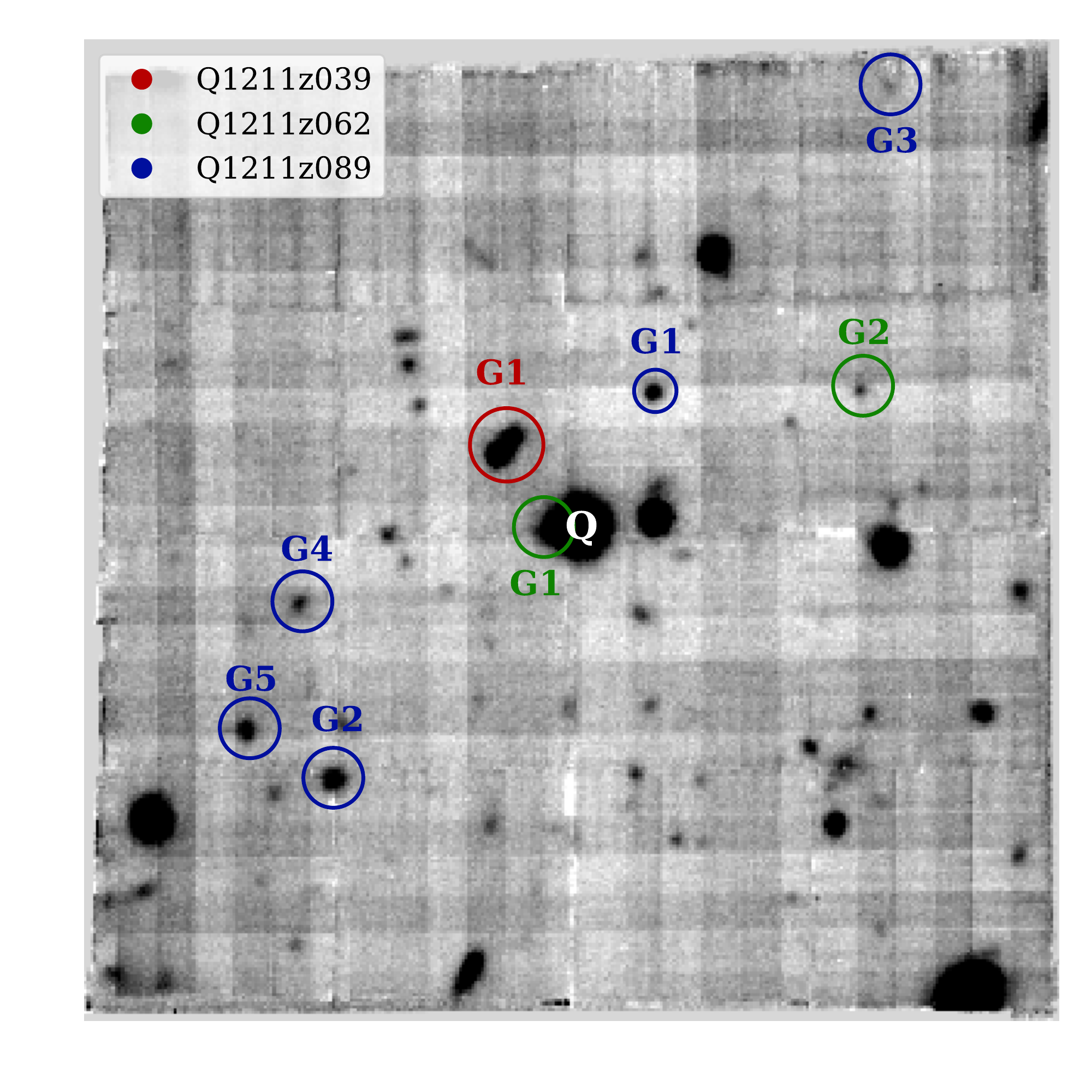}}
\subfloat{\includegraphics[width=\columnwidth]{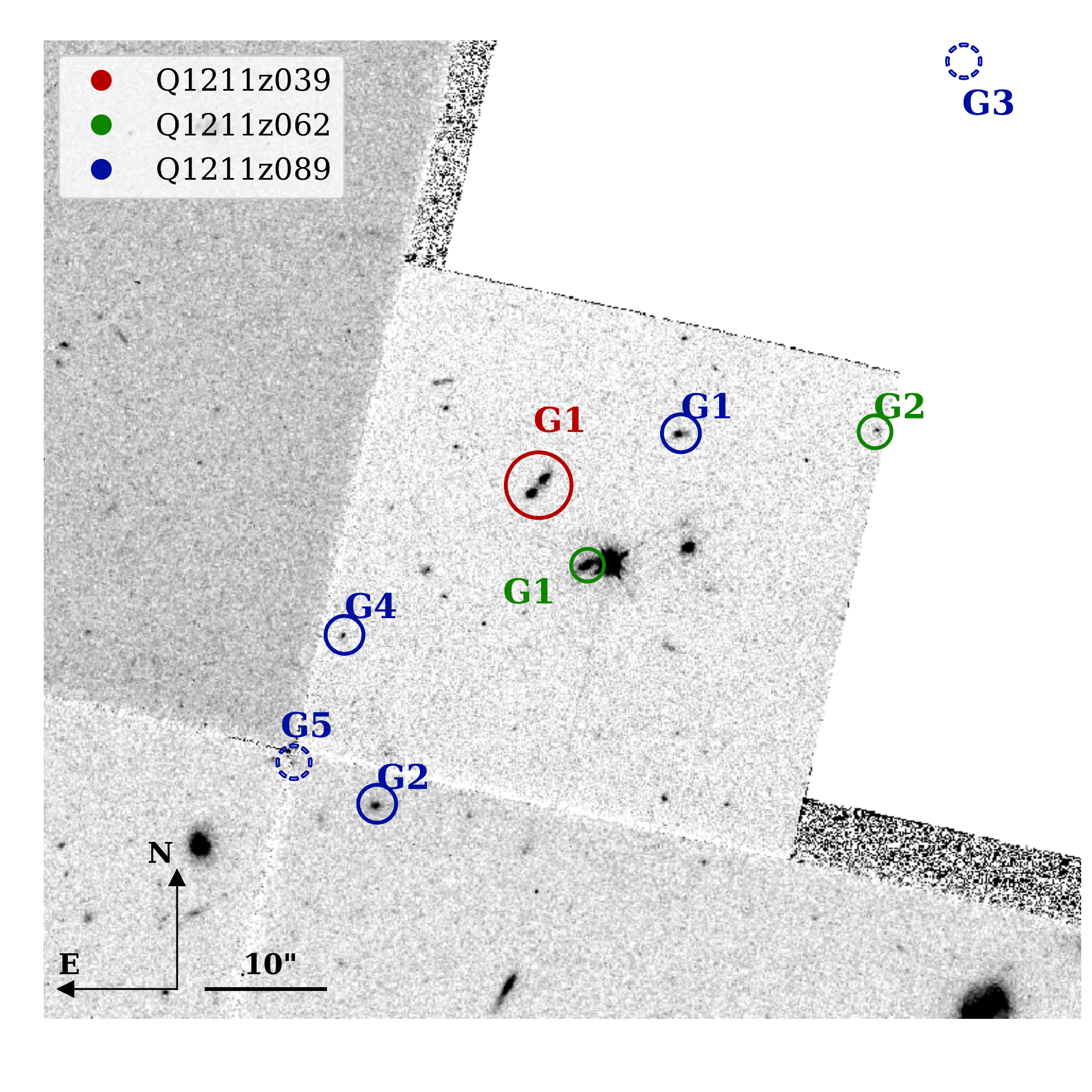}}
\caption{MUSE white light (left) and HST F702W (right) images of Q1211+1030 quasar field with galaxies associated to all absorbers marked. Red circles mark galaxies associated with Q1211z039 absorber, green - Q1211z062 absorber, blue -  Q1211z089 absorber. "Q" indicates the quasar.}
\label{fig:Q1211}

\end{figure*}

\begin{figure*} 
\includegraphics[width=2\columnwidth]{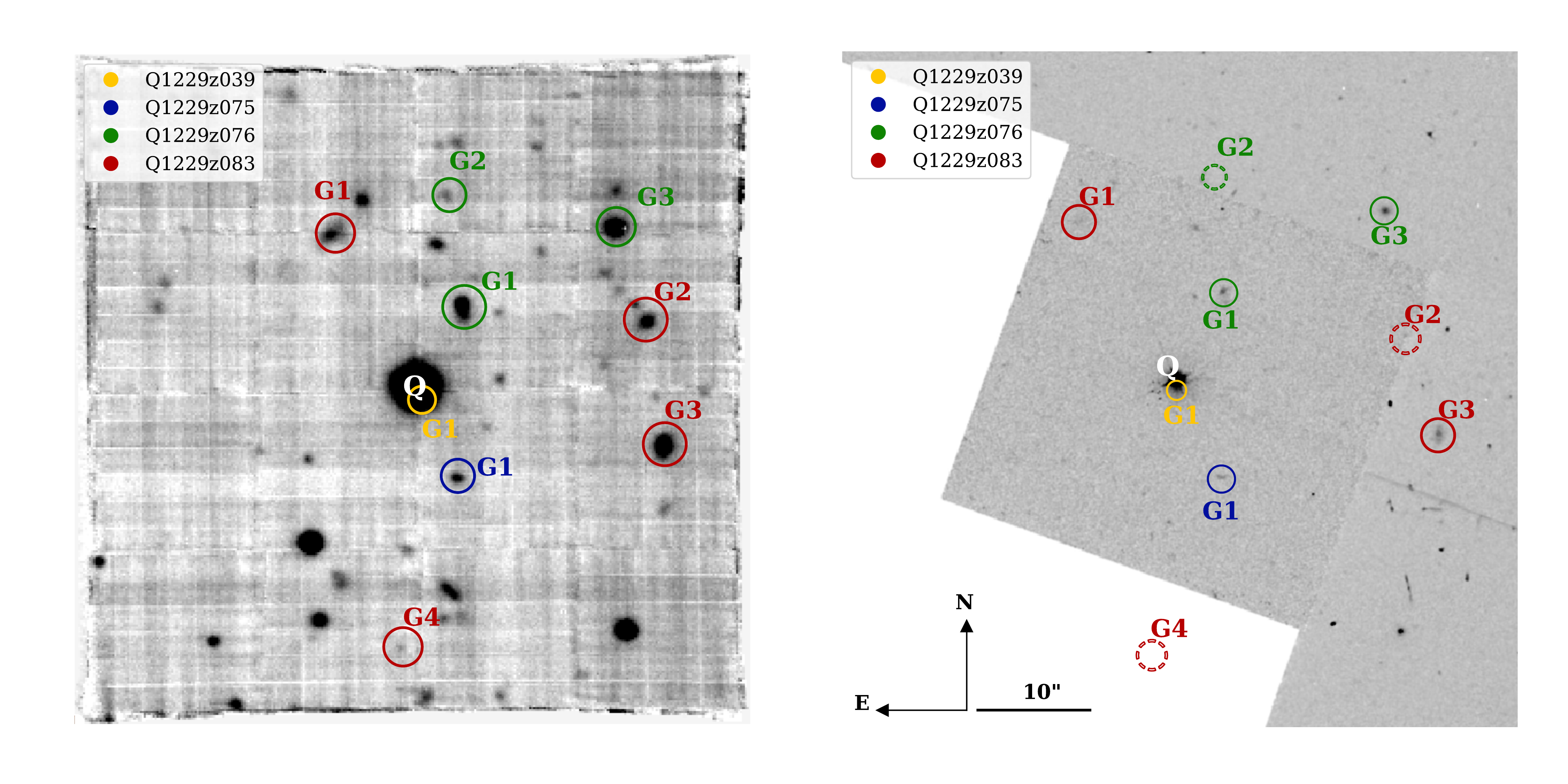}
\caption{MUSE white light (left) and HST F702W (right) images of Q1232-0224 quasar field with galaxies associated to all absorbers marked. Yellow circles mark galaxies associated with Q1232z039 absorber, blue - Q1232z075 absorber, green -  Q1229z076 absorber and red with Q1232z083 absorber. "Q" indicates the quasar.}
\label{fig:Q1229}
\end{figure*}

\begin{figure*} 
\includegraphics[width=2\columnwidth]{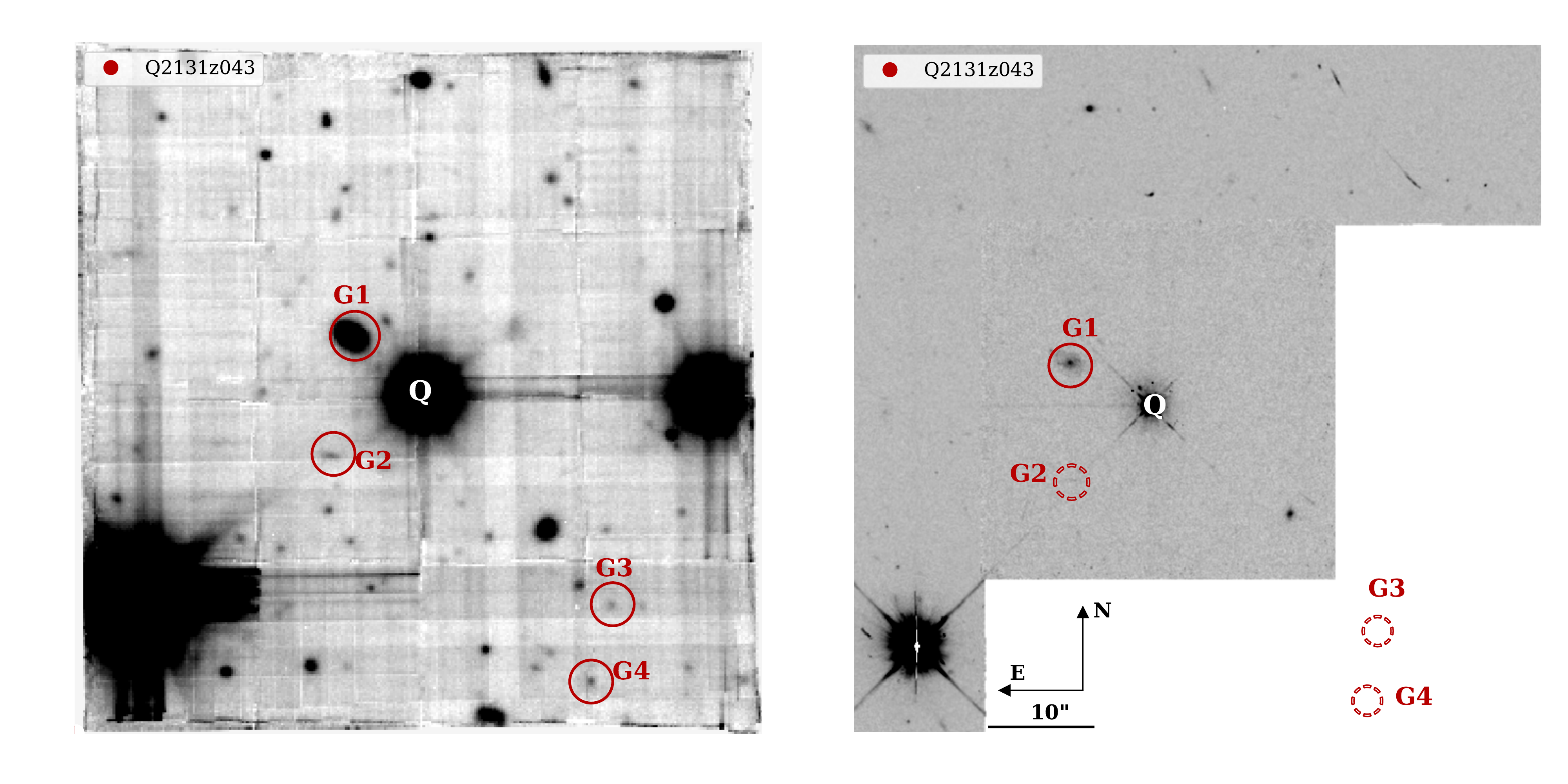}
\caption{MUSE white light (left) and HST F702W (right) images of Q2131-1207 quasar field with galaxies associated to an absorbers marked in red. Dashed circles mark the non-detections. "Q" indicates the quasar.}
\label{fig:Q2131}
\end{figure*}

\begin{figure*} 
\includegraphics[width=2\columnwidth]{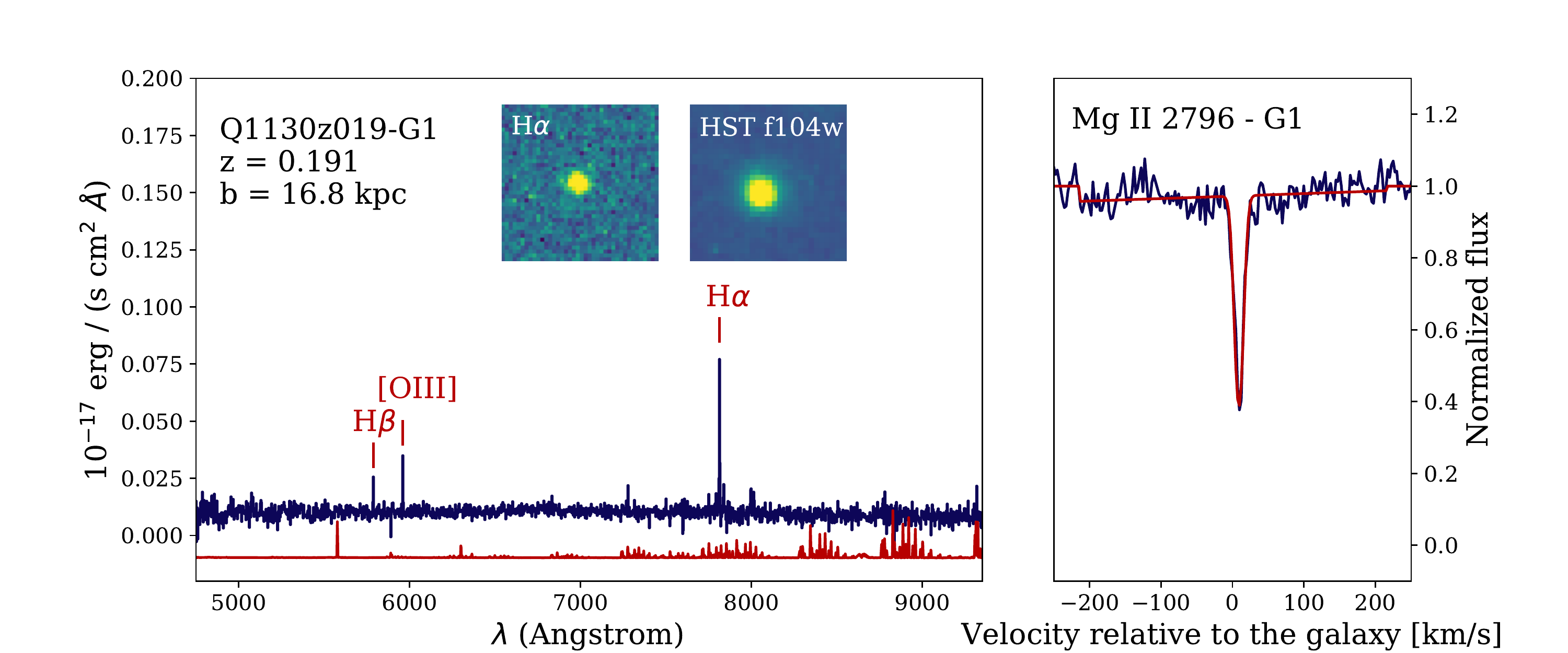}
\caption{Q1130z019 absorber and associated galaxy (continuum images \ref{fig:Q1130}).}
\label{fig:Q1130z019spec}
\end{figure*}

\begin{figure*} 
\includegraphics[width=2\columnwidth]{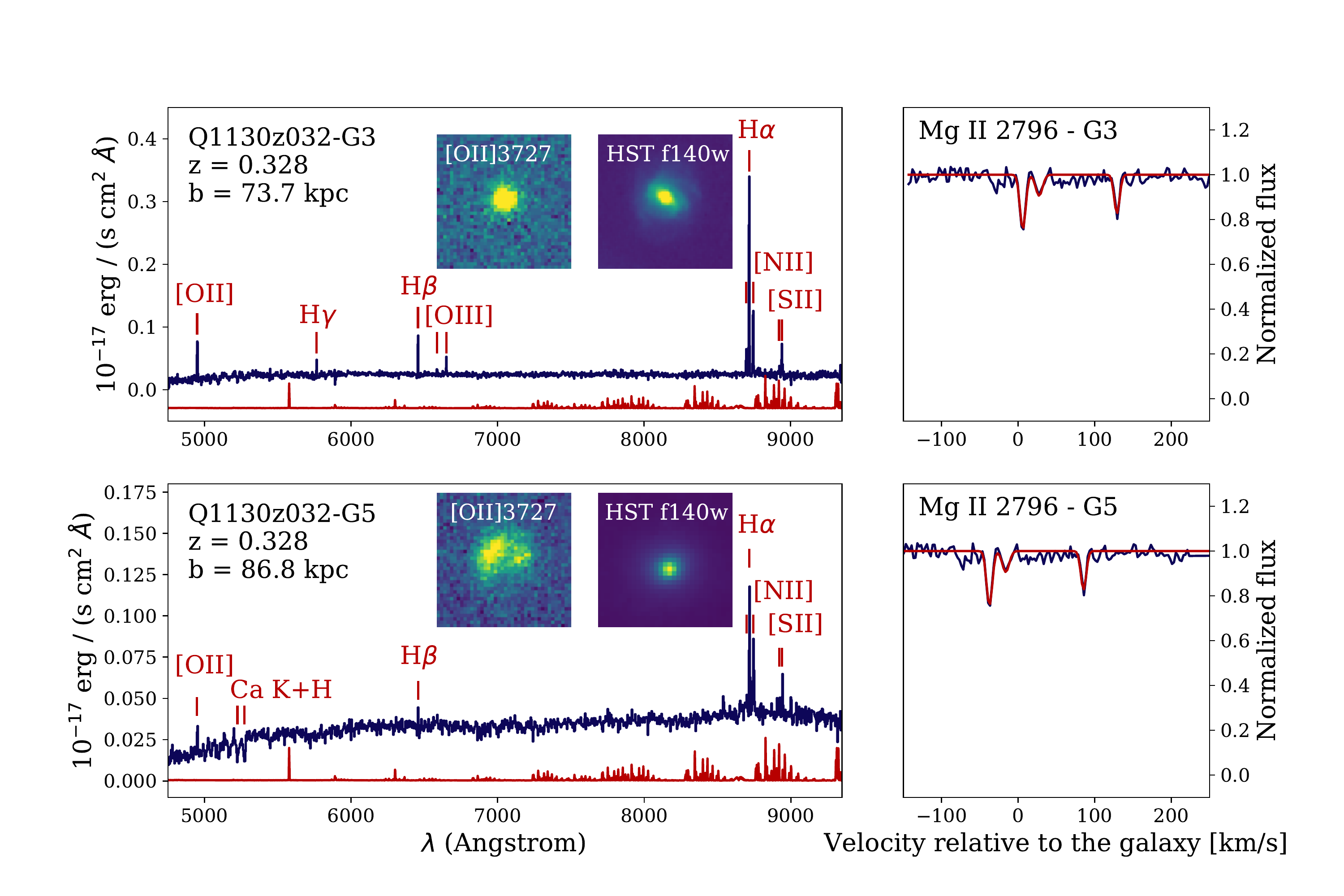}
\caption{Q1130z032 absorber and associated galaxies (continuum images \ref{fig:Q1130}).}
\label{fig:Q1130z032spec}
\end{figure*}

\begin{figure*} 
\includegraphics[width=2\columnwidth]{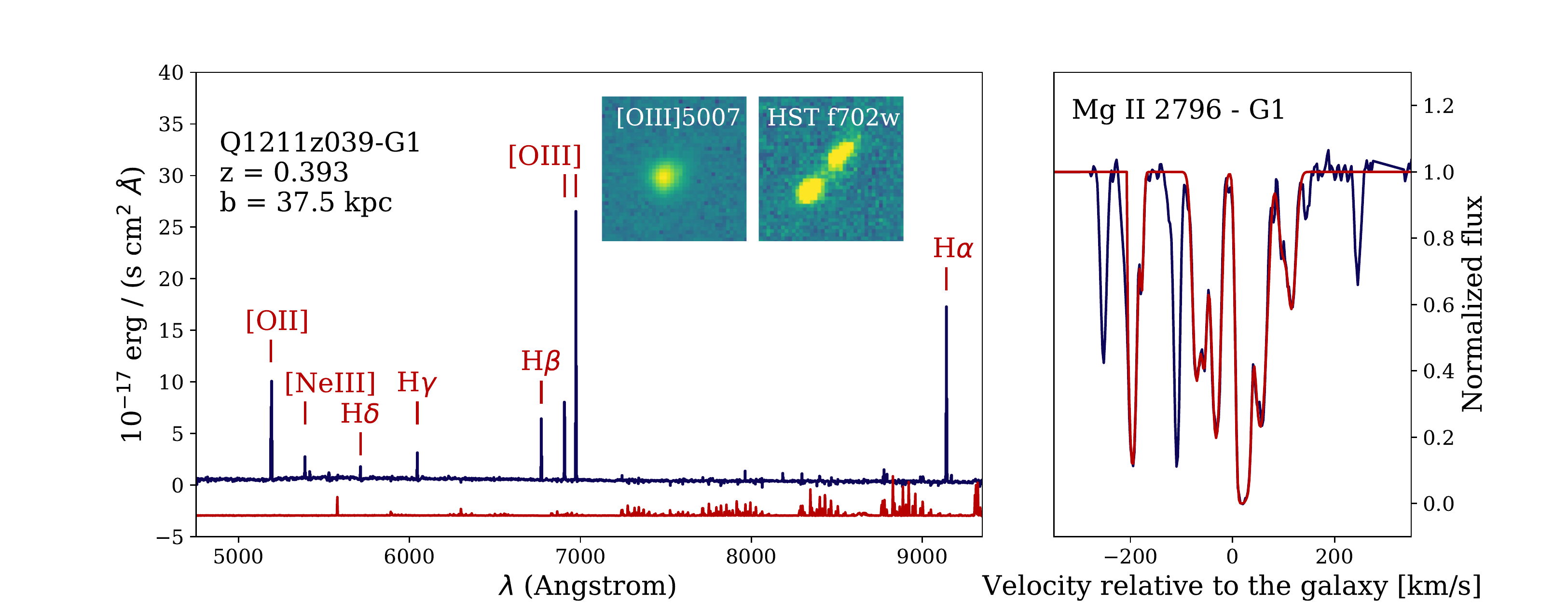}
\caption{Q1211z039 absorber and associated galaxy (continuum images \ref{fig:Q1211}).}
\label{fig:Q1211z039spec}
\end{figure*}

\begin{figure*} 
\includegraphics[width=2\columnwidth]{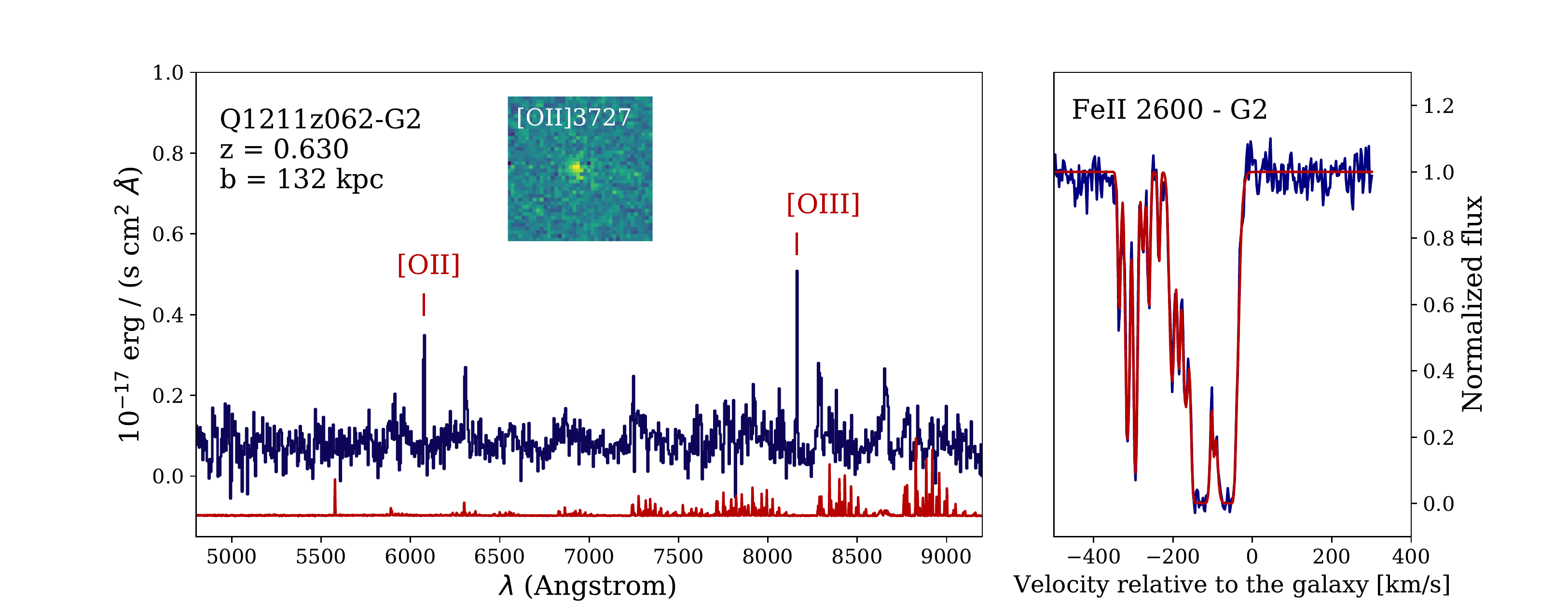}
\caption{Q1211z062 absorber; G1 is a low impact parameter associated galaxy, and only [\OII] emission line was extracted form the quasar PSF (presented on the Fig. \ref{fig:qsores}). Here we present only the spectrum of the second associated galaxy (G2). For this absorber \MgII $\lambda$2796 falls in the UVES spectral gap, we plot \FeII $\lambda$2600 profile instead (continuum images \ref{fig:Q1211}).}
\label{fig:Q1211z062spec}
\end{figure*}

\begin{figure*}
\includegraphics[width=2\columnwidth]{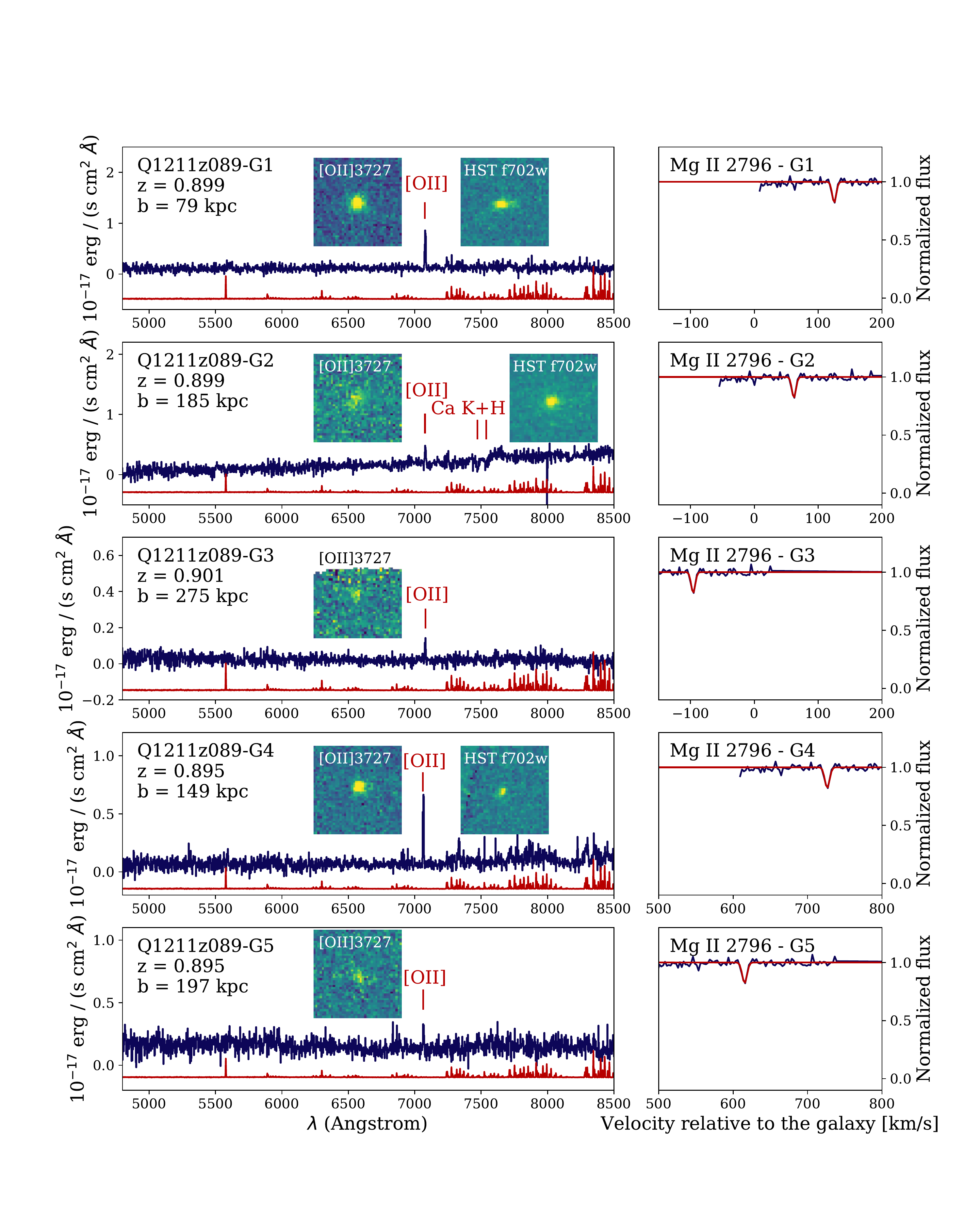}
\caption{Q1211z089 absorber and associated galaxies. G3 is outside HST WFPC2 field of view, G5 lays on the edge of the WF and PC chips of the HST WFPC2 camera and is not detected in continuum (continuum images \ref{fig:Q1211}). Note that due to larger velocity offset, G4 and G5 are presented in different velocity range than G1-G3.}
\label{fig:Q1211z089spec}
\end{figure*}

\begin{figure*} 
\centering
\includegraphics[width=\columnwidth]{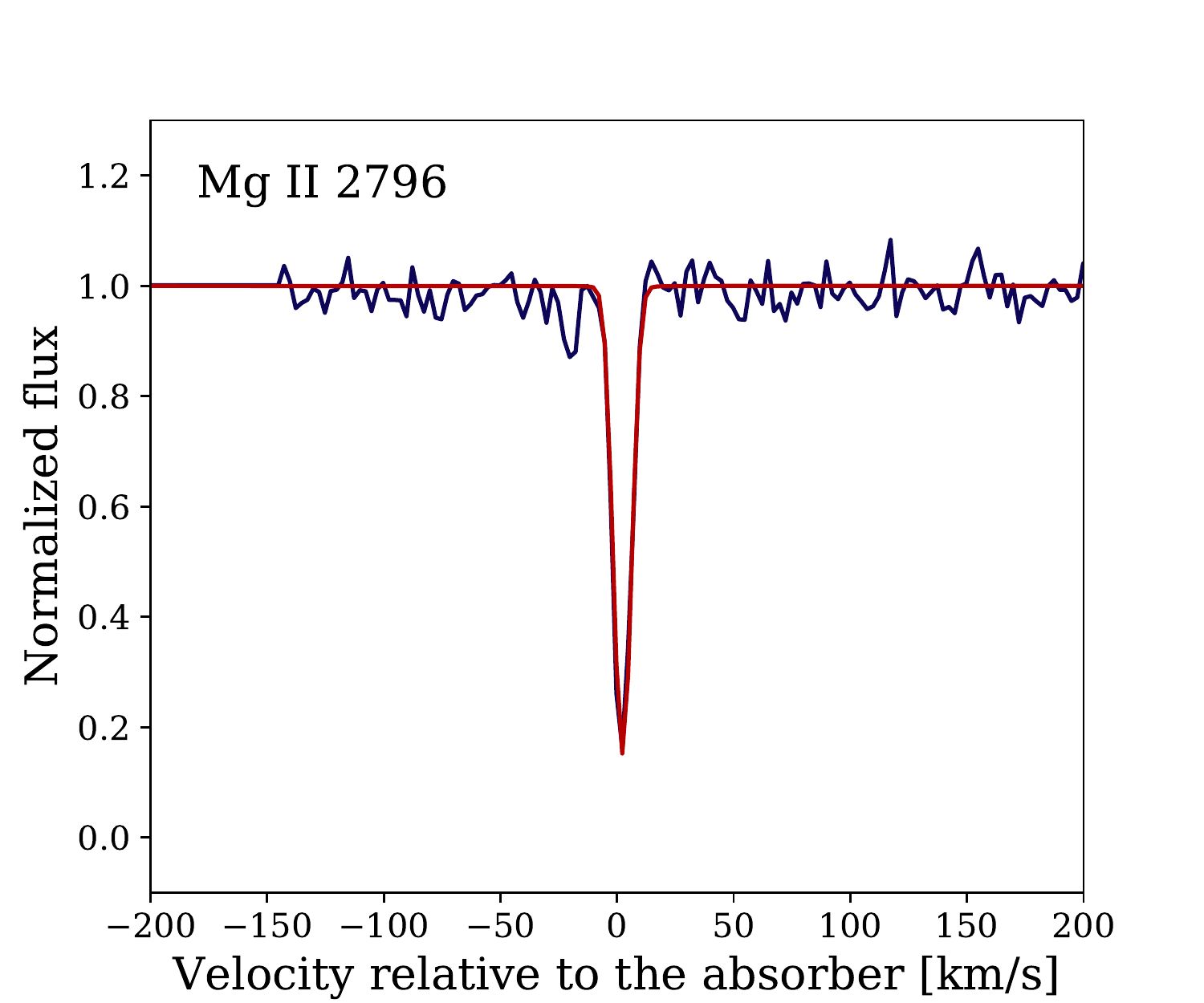}
\caption{Q1211z105 absorber \MgII\ profile. This is the only absorber in the sample for which no associated galaxy was found.}
\label{fig:Q1211z105spec}
\end{figure*}

\begin{figure*} 
\includegraphics[width=2\columnwidth]{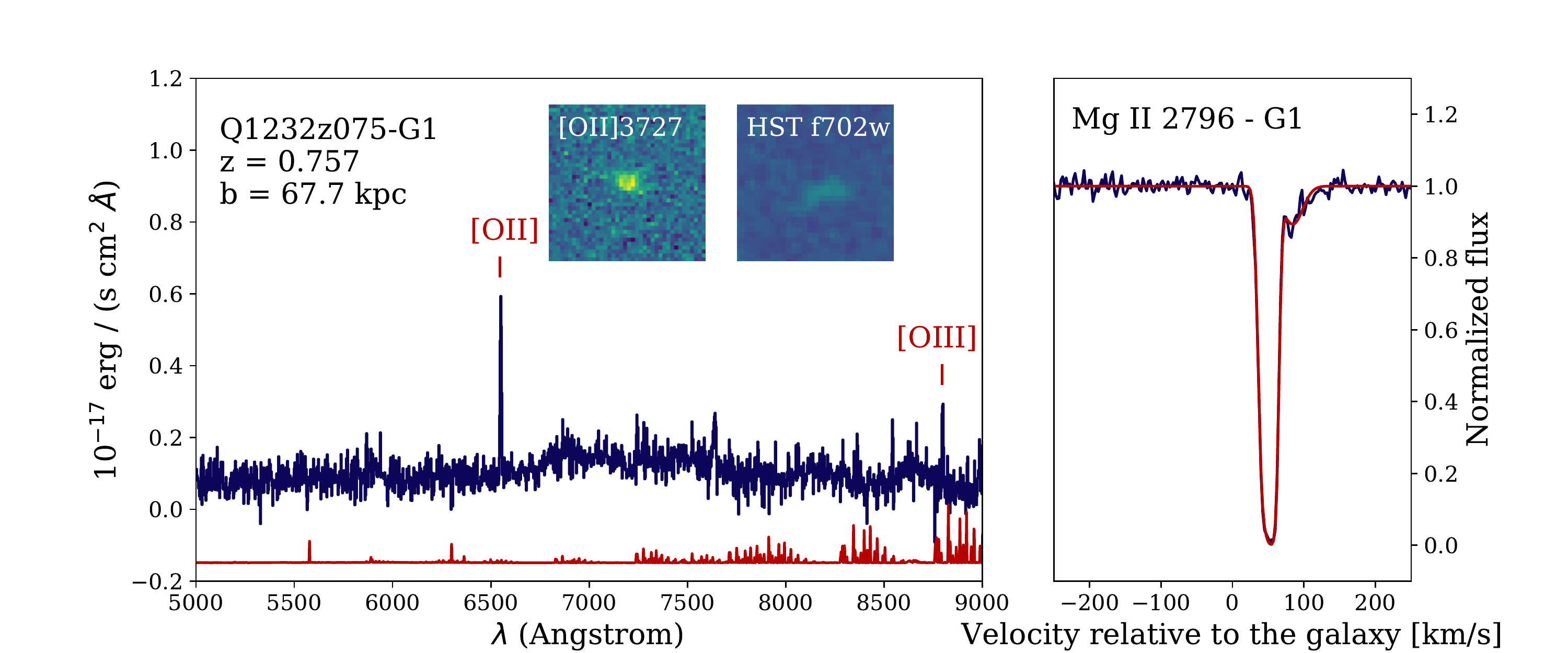}
\caption{Q1232z075 absorber and associated galaxy (continuum images \ref{fig:Q1229}).}
\label{fig:Q1229z075spec}
\end{figure*}

\begin{figure*} 
\includegraphics[width=2\columnwidth]{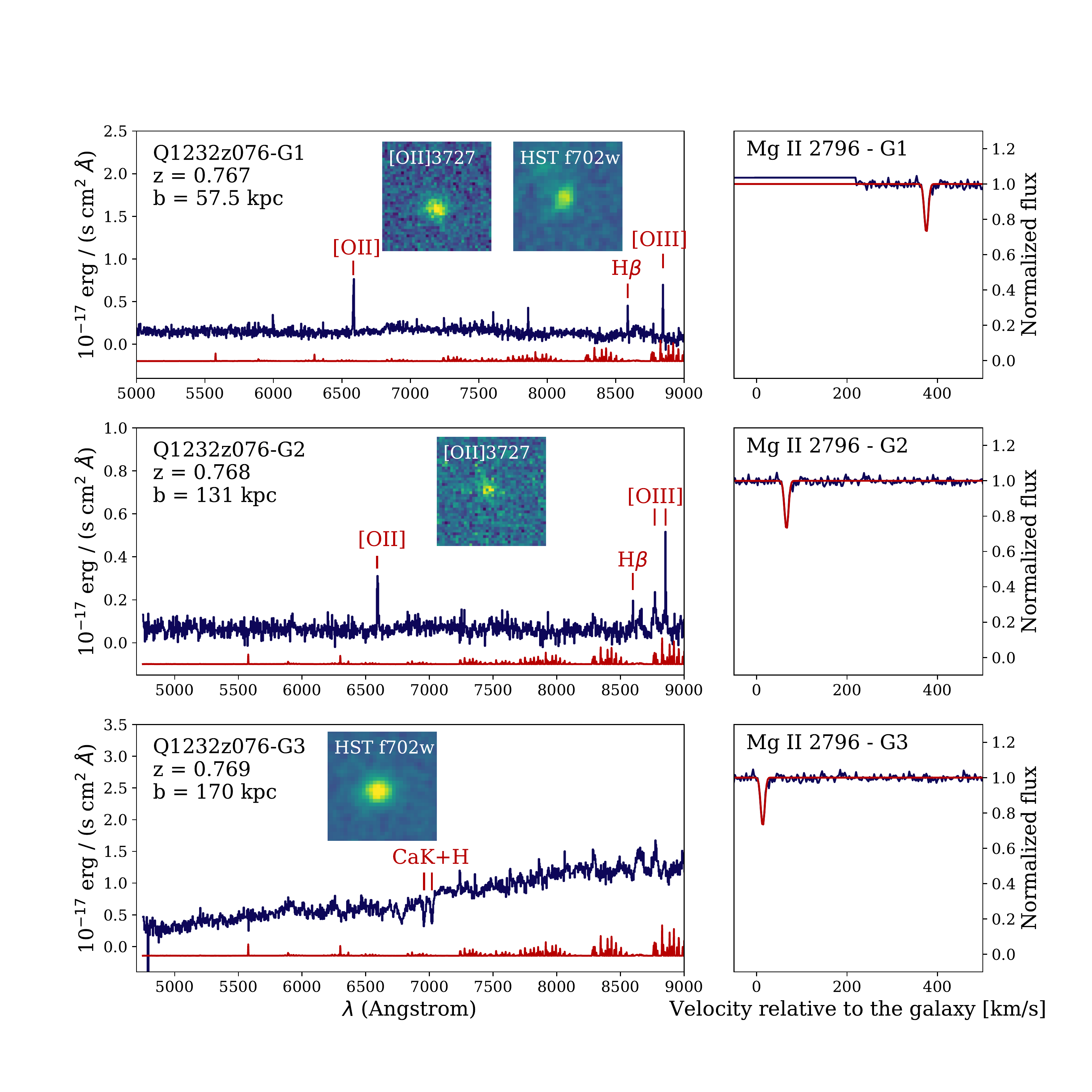}
\caption{Q1232z076 absorber and associated galaxies. G3 is a quiescent galaxy with only absorption line and was detected only in continuum in MUSE whit light image (continuum images \ref{fig:Q1229}).}
\label{fig:Q1229z076spec}
\end{figure*}

\begin{figure*} 
\includegraphics[width=2\columnwidth]{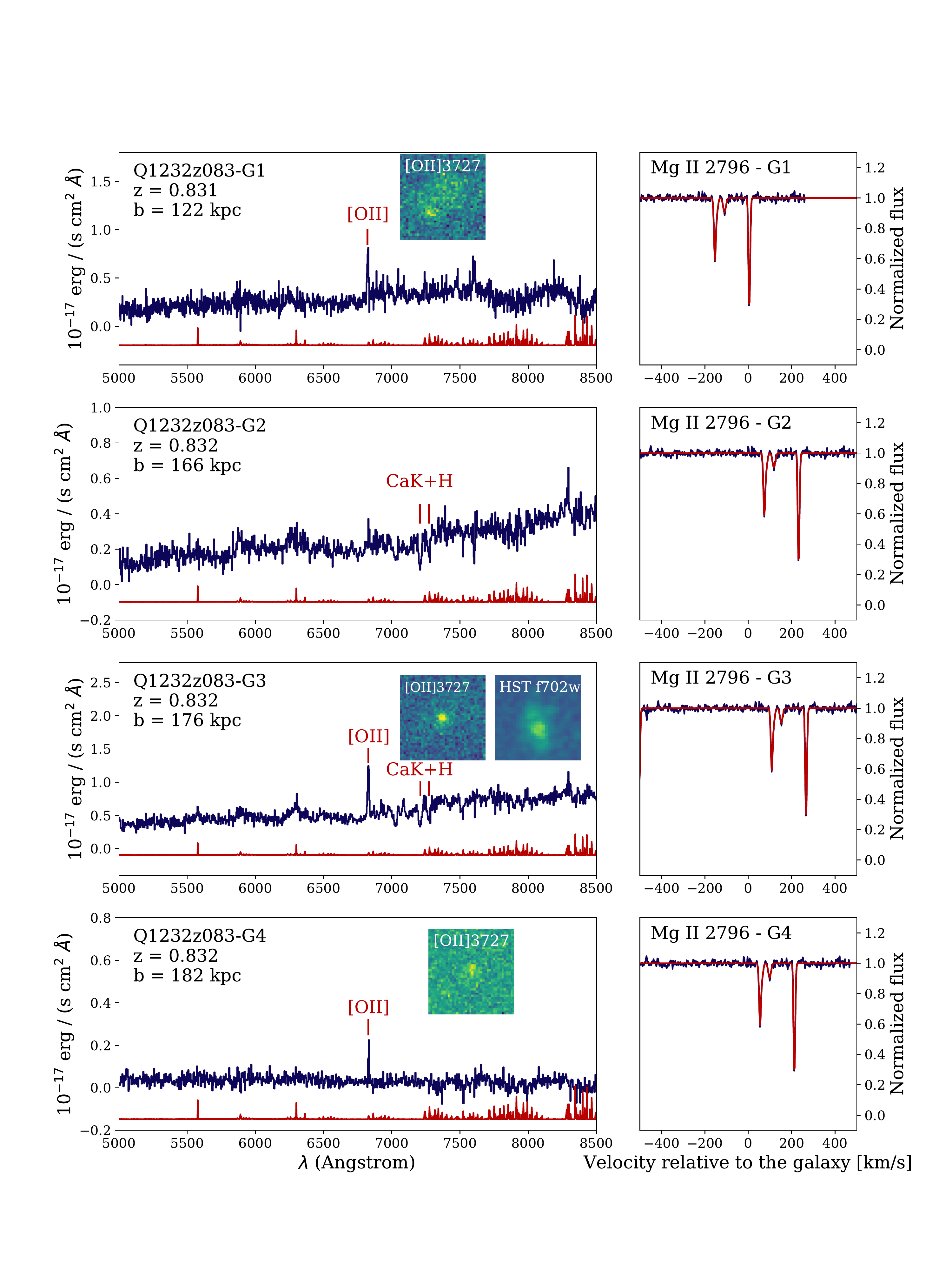}
\caption{Q1232z083 absorber and associated galaxies G1 and G2 were not detected in continuum in HST image. G1 is very diffuse object, G2 is and absorption line system, not detected in continuum in HST image, inset presents only MUSE white light image centered at the galaxy (continuum images \ref{fig:Q1229}).}
\label{fig:Q1229z083spec}
\end{figure*}





	


\bsp	
\label{lastpage}
\end{document}